\documentclass[10pt]{article}
\usepackage{graphicx}
\usepackage{epsf}
\usepackage{epsfig}
\usepackage{amssymb}
\usepackage{amsmath}
\makeatletter
\@addtoreset{equation}{section}
\makeatother

\def\bea{\begin{eqnarray}}
\def\eea{\end{eqnarray}}
\def\be{\begin{equation}}
\def\ee{\end{equation}}
\def\nn{\nonumber\\}
\def\l{\lambda}
\def\r#1{(\ref{#1})}
\begin{document}

\def\wgta#1#2#3#4{\hbox{\rlap{\lower.35cm\hbox{$#1$}}
\hskip.2cm\rlap{\raise.25cm\hbox{$#2$}}
\rlap{\vrule width1.3cm height.4pt}
\hskip.55cm\rlap{\lower.6cm\hbox{\vrule width.4pt height1.2cm}}
\hskip.15cm
\rlap{\raise.25cm\hbox{$#3$}}\hskip.25cm\lower.35cm\hbox{$#4$}\hskip.6cm}}

\def\wgtb#1#2#3#4{\hbox{\rlap{\raise.25cm\hbox{$#2$}}
\hskip.2cm\rlap{\lower.35cm\hbox{$#1$}}
\rlap{\vrule width1.3cm height.4pt}
\hskip.55cm\rlap{\lower.6cm\hbox{\vrule width.4pt height1.2cm}}
\hskip.15cm
\rlap{\lower.35cm\hbox{$#4$}}\hskip.25cm\raise.25cm\hbox{$#3$}\hskip.6cm}}

\def\begeqar{\begin{eqnarray}}
\def\endeqar{\end{eqnarray}}

%---------------------------------------------------------------------
%
%
%-------------------------------------------------------------
%
%

\begin{center}

%%%%%%%%%%%%%%%%%%%%%%%%%%%%%%%%%%%%%%%%%%%%%%%%%%%%%%%%%%%%%%%%%%%%%%%%%%%%%%%
\Large{Continuum Limit of the Integrable $sl(2/1)$ $3-\bar{3}$ Superspin Chain}
%%%%%%%%%%%%%%%%%%%%%%%%%%%%%%%%%%%%%%%%%%%%%%%%%%%%%%%%%%%%%%%%%%%%%%%%%%%%%%%
\vskip 1cm

{\large Fabian H. L. Essler$^{a}$, Holger Frahm$^b$ and
Hubert Saleur$^{c,d}$}
\vspace{1.0em}

%% Address:
%%
{\sl\small $^a$
The Rudolf Peierls Centre for Theoretical Physics\\
University of Oxford, 1Keble Road, Oxford OX1 3NP, UK\\}
{\sl\small $^b$
  Institut f\"ur Theoretische Physik, Universit\"at Hannover,\\
  30167~Hannover, Germany\\}
{\sl\small $^c$ Service de Physique Th\'eorique, CEA Saclay,\\ 
Gif Sur Yvette, 91191, France\\}
{\sl\small $^d$ Department of Physics and Astronomy,
University of Southern California\\
Los Angeles, CA 90089, USA\\}

\end{center}

\begin{abstract}
By a combination of analytical and numerical techniques, we 
analyze the continuum limit of the integrable 
$3\otimes\bar{3}\otimes 3\otimes\bar{3}\ldots$ $sl(2/1)$ superspin 
chain. We discover  
profoundly new features, including a continuous spectrum of conformal 
weights, whose numerical evidence is infinite degeneracies of the 
scaled gaps in the thermodynamic limit. This indicates that the
corresponding conformal field theory has a non compact target 
space (even though our lattice model involves only finite dimensional 
representations). We argue that our results are compatible with this
theory being the level $k=1$, `$SU(2/1)$ WZW model' (whose precise
definition requires some care). In doing so, we establish several new 
results for this model. With regard to potential applications to the
spin quantum Hall effect, we conclude that the continuum limit of the 
$3\otimes\bar{3}\otimes 3\otimes\bar{3}\ldots$ $sl(2/1)$ integrable superspin 
chain is {\sl not} the same as (and is in fact very different from) 
the continuum limit of the corresponding chain with two-superspin
interactions only, which is known to be a model for the spin
quantum Hall effect. The study of possible RG flows between the two
theories is left for further study.

\end{abstract}

\section{Introduction}

The  sigma model approach  \cite{Pruisken} to phase transitions in non interacting 
disordered systems provides a convenient and appealing description of 
the physics at hand. The best known example is the 
transition between plateaux in the integer quantum Hall effect (class 
A), which 
can be  described using a $U(2n)/U(n)\times U(n)$ sigma model in the 
limit $n\rightarrow 0$. More recently, the equivalent `spin quantum 
Hall effect' in d-wave superconductors has been considered (class 
C \cite{Altland}), 
and described with the $SP(2n)/U(n)$ sigma model, in the limit 
$n\rightarrow 0$\cite{SeMF99}. 

In both cases, the existence of a delocalization transition is 
associated with the masslessness  in the IR 
of the sigma model at topological angle $\theta=\pi$. Identification 
of the corresponding conformal field theory would lead to a presumably 
exact determination of critical exponents which have been studied  in 
great details in numerical works \cite{Janssen},\cite{Evers} as well as in experiments.

Unfortunately, the determination of the IR fixed points in these 
problems has proven extremely difficult. The canonical example of 
what `might' happen is the case $n=1$ of either replica model, which 
coincides with the $O(3)$ sigma model at $\theta=\pi$. In this case, 
there is a wealth of evidence \cite{AffleckHaldane}
that the IR fixed point is the $SU(2)$  Wess 
Zumino model at level $k=1$. In the flow, the field is promoted from 
being coset valued to group valued, a highly non perturbative feature 
which might have been hard to believe, were it not for the existence 
of exact solutions \cite{ZamoZamo}. 

The most reasonable way to understand what happens in the disordered 
system is first to trade the replica limit for a super group 
\cite{Efetov},\cite{Weidenmuller}, 
$U(1,1/2)/U(1/1)\times U(1/1)$ in the quantum Hall effect, and 
$OSP(2n/2n)/U(n|n)$ in the spin quantum Hall effect. Once the 
corresponding sigma models have been satisfactorily defined 
\cite{Zirnbauer}, it is then
tempting to conjecture, by analogy with the $O(3)$ example above, what the IR fixed point 
might be. In this way, Zirnbauer  \cite{Zirnbaueri} arrived at a proposal for the quantum 
Hall effect that is related with the WZW model (at level $k=1$)
on $PSU(1,1/2)$. 
Roughly, this is obtained by observing that the base of the 
supersymmetric target space is $H^{2}\times S^{2}$, and then guessing
that this 
gets `promoted' to $H^{3}\times S^{3}$ in the IR. One then tries  to build 
the minimal proposal in agreement with this and known results 
on the exponents. It is important that for the $PSU(1,1/2)$ WZW  model  
the kinetic term is an exactly marginal deformation 
\cite{Berkovits},\cite{Bershadsky}. The 
final proposal of Zirnbauer corresponds to a particular value  of 
this kinetic term, and is not the  WZW 
model.  Hence, the symmetry is a global symmetry $G_{L}\times G_{R}$ 
instead of a local one. 
Unfortunately it is not clear whether this proposal gives  the correct IR fixed point. 
In particular, it is rather unpleasant to have to select a particular 
value for an exactly marginal coupling constant if one hopes to 
describe what seems to be genuine universal physics. 

A less abstract direction of attack  comes from network 
models. These can be shown to correspond, in the proper anisotropic 
limit to quantum spin chains with super group 
symmetries \cite{Superspinchains}. Ideally, the 
problem would then be solved if one were able to apply in this 
situation the formidable arsenal developed over the years in the study 
of quantum spin chains. Unfortunately, there are difficulties here as 
well. For the ordinary as well as the spin quantum Hall effect, the 
Hilbert space of the chain is of type  $V\otimes \bar{V}\otimes V\otimes 
\bar{V}\otimes\cdots$, while the Hamiltonian is the invariant Casimir, the 
proper generalization of $\vec{S}.\vec{S}$. In the ordinary quantum 
Hall effect, $V$ and $\bar{V}$ are infinite dimensional, while in the 
spin quantum Hall effect, $V$ and $\bar{V}$ 
are the two conjugate three 
dimensional representations. 

In either case, the Hamiltonian is not integrable. More precisely, the 
Hamiltonian does not seem to derive from a family of commuting transfer 
matrices obtained via a standard quantum group approach. In the spin 
quantum Hall case, the Hamiltonian can however be diagonalized  
\cite{Gruzberg} (more 
precisely, the zeroes of the characteristic polynomial can be found, 
as the Hamiltonian itself is not fully diagonalizable) using known 
results about the XXZ chain and representation theory of the 
Temperley Lieb algebra \cite{ReadSaleur}. No such approach seems possible in the 
ordinary quantum Hall spin chain.

Since one is after the universality class of the IR fixed point, 
it is natural to wonder \cite{Gade} whether the Hamiltonian might be 
substituted by an integrable version without affecting the exponents.  
Examples are known where such a trick would not work: for instance, 
the higher odd spin integrable $SU(2)$ spin chains flow to level 
$k=2s$ WZW models, while the chains with Heisenberg type couplings 
flow to level $k=1$. But examples are also known where the trick does 
work. Indeed, with unitary groups and finite dimensional 
representations, group symmetry on the lattice plus criticality 
implies that the continuum limit is a WZW model, the stablest of all 
being $k=1$ \cite{Affleck}. Therefore, a spin chain with $SU(n)$ symmetry, if 
critical, will generically be in the same universality class  as  the integrable 
Sutherland chain, that is the level one $SU(n)$ WZW model. 
Criticality is however not guaranteed  in general. For instance, if one 
takes  $SU(n)$ and the alternance  of $n$ and $\bar{n}$ 
representations, the chain whose Hamiltonian is the Casimir has a 
gap, while the integrable one is gapless, and in the universality 
class of the WZW model \cite{Afflecki}. 

Our purpose in this paper is to explore the matter further by 
concentrating on the spin quantum Hall effect where many things are 
known exactly. In  a nutshell, the question we want to explore is whether
the integrable  $3\otimes \bar{3}\otimes 3\otimes \bar{3}\ldots$
  $sl(2/1)$ spin chain is {\sl massless} and if so, in {\sl the same
universality class} as the non integrable$3\otimes \bar{3}\otimes
3\otimes \bar{3}\ldots$ Heisenberg chain, which is known to describe
the physics of the spin quantum Hall effect.

A positive answer would raise hopes for an integrable approach to the 
ordinary quantum Hall effect. Unfortunately, we will see that the 
answer is a resounding no. The reason for this is the proliferation 
of possible fixed points 
that appear when ordinary compact symmetries are replaced 
by super group, maybe non compact, symmetries, and sheds light on 
what should be directions of further research.

We will present evidence from various sources that the {\sl
integrable} alternating $3,\bar{3}$ chain is in the universality class
of the $SU(2/1)$ WZW model at level $1$\cite{Goddard}.  This model has
also been called $OSP(2/2)_{-2}$ in the literature \cite{Andreas}, and
has appeared previously in the study of Dirac fermions in a random
gauge potential\cite{Bhaseen}. In contrast, it is already known
from \cite{ReadSaleur} that the non integrable alternating $3,\bar{3}$
Heisenberg chain does not scale to a WZW model, but to 
a new kind of theory, where the currents have logarithmic OPEs, 
and the symmetry is not of Kac Moody type \cite{ReadSaleuri}. 

It is important to stress here that the WZW model is a {\sl very} 
different theory from the one \cite{ReadSaleur}
associated with the Heisenberg 
Hamiltonian; in particular, it has a continuous spectrum of critical 
exponents, which is not bounded from below. Trying to infer generic 
physical properties from the integrable chain would be a 
considerable mistake in this case, and presumably in the ordinary 
quantum Hall case as well.

The paper is organized as follows. In the next section, we gather 
information, obtained from the literature as well as our own 
calculations, about the WZW model. In sections three and four, we present
the results of a numerical and analytical study of the Bethe ansatz equations of the
integrable spin chain, comparing our results with the expectations for
the WZW model. Section five contains elements of extension to the case
of higher level. Section six goes back to the spin quantum Hall
problem. Technical details are provided in the appendices.

\section{First considerations on the WZW model}

\subsection{Algebraic  generalities}

We start with a short review of the ``base''  $sl(2/1)$ algebra. It
contains a sub $sl(2)$ with generators $J^{\pm}_{0},J^{3}_{0}$, an extra
$U(1)$ generator $B_{0}$, and two pairs of  
fermionic generators $V_{0}^{\pm}$ and $W_{0}^{\pm}$. Its representation theory  
is complicated, although this is one of the best understood cases   in 
the super Lie algebra literature 
\cite{Marcu},\cite{Leites},\cite{Germoni},\cite{Germonithesis}. 
Typical representations are characterized by a pair of numbers 
traditionally called $b,j$ (and denoted $[b,j]$ in what follows). 
Here,
$j$ is a $sl(2)$ spin quantum number and takes values 
$0,1/2,1,\ldots$, whereas $b$ can be any complex number. Typicality 
requires $b\neq \pm j$. Typical representations are irreducible. 
Their dimension is  $8j$, 
and they decompose into $sl(2)$ representations with spin 
$j,j-1/2,j-1/2,j-1$ with, respectively, $b$ numbers 
$b,b+1/2,b-1/2,b$ (for $j=1/2$, the value $j-1$ is omitted).   We 
introduce the notation $\rho_{j}^{b}$ to denote an $sl(2)$ 
representation of spin $j$ for which the other $u(1)$ generator takes 
the constant value $b$, so the former decomposition reads 
$\rho_{j}^{b}\oplus 
\rho_{j-1/2}^{b+1/2}\oplus\rho_{j-1/2}^{b-1/2}\oplus\rho_{j-1}^{b}$. 
The Casimir is diagonal in these representations, 
and proportional to $j^{2}-b^{2}$. 

\begin{figure}
\centering
    \includegraphics[width=3in]{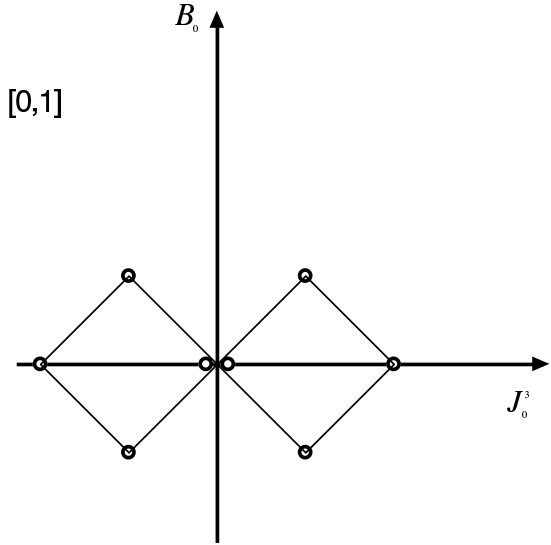}
\caption{Graphical representation of the representation $[0,1/2]$.}
\label{fund}
\end{figure}

A particularly important representation in the WZW model will be the four 
dimensional 
$[b=0,j=1/2]$ which is represented graphically in Fig. \ref{fund}. It is the defining 
representation for $osp(2/2)$. 

There are many more representations. Simple atypical (that is, 
irreducible atypical) occur when $b=\pm j$, and have dimension $4j+1$. 
Examples are the three dimensional representations 
$[1/2,1/2]$ and $[-1/2,1/2]$ that we will use to define the `Hilbert' space for 
the quantum spin chain. Representations $[j,j]$ contain only 
multiplets $\rho_{j}^{j}$ and $\rho_{j-1/2}^{j+1/2}$, so $[1/2,1/2]$ 
contains $\rho_{1/2}^{1/2}$ and $\rho_{0}^{1}$. Similarly, representations 
$[-j,j]$ contain only multiplets with $\rho_{j}^{-j}$ and 
$\rho_{j-1/2}^{j-1/2}$, so $[-1/2,1/2]$ contains $\rho_{1/2}^{-1/2}$ 
 and $\rho_{0}^{-1}$. While typical representations have vanishing 
 superdimension, simple atypical ones have superdimension equal to 
 plus or minus one. 
 
The other atypical representations are not simple: they are 
indecomposable \footnote{We  may use indecomposables as short hand 
for non simple indecomposables, while in the mathematics literature, 
indecomposable encompasses irreducible.}  but partly reducible, and form
what is often called 
`blocks'. The most important example is the one appearing in the 
tensor product $[0,1/2]\otimes [0,1/2]$ which decomposes as the 
adjoint $[0,1]$ and another eight dimensional representation usually 
denoted as 
$[0,-1/2,1/2,0]$. This representation  is the semi direct sum of $[1/2,1/2]$, 
$[-1/2,1/2]$, and of two $[0,0]$ representations. Its $su(2)$ content 
is 
 $\rho_{1/2}^{1/2}$, $\rho_{1/2}^{-1/2}$
, $\rho_{0}^{1}$, $\rho_{0}^{-1}$,  and $\rho_{0}^{0}$ twice. A 
graphical representation is given in figure 2. Note that this 
representation has vanishing superdimension. 

\begin{figure}
\centering
    \includegraphics[width=3in]{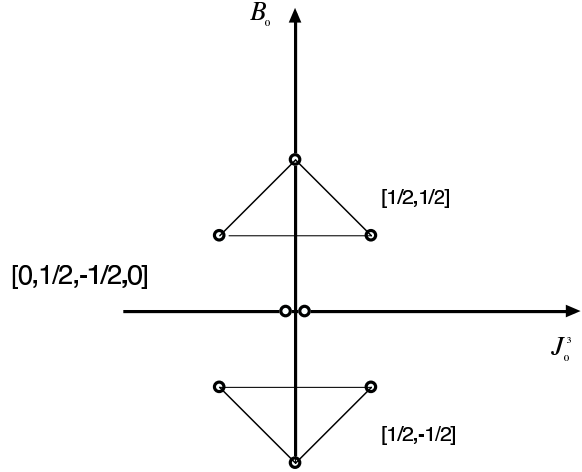}
\caption{Graphical representation of the representation $[0,-1/2,1/2,0]$.}
\label{}
\end{figure}

It is useful to represent such an indecomposable representation by its 
`quiver diagram' \cite{Benson}. Denote by $j$ the representation $[j,j]$ and 
by $-j$ the representation $[-j,j]$ (these simple 
atypical representations can be  obtained by maximum supersymmetrization of the 
    fundamental or its conjugate - $2j$ boxes in a Young diagram 
    representation). Then the eight 
dimensional indecomposable can be represented as
\begin{equation}
    \hbox{$\begin{array}{rcl}&^{0}_{\bullet}\\
\hphantom{-1}\swarrow&&\searrow\\
^{-1/2}_{~~\bullet}\hphantom{\swarrow}&&\hphantom{\swarrow}^{1/2}_{\bullet}\\
\makebox[0pt]{}\searrow&&\swarrow\\&^{0}_{\bullet}\end{array}$}.
\end{equation}
Here,  the bottom part of the 
diagram means that the representation contains a singlet as an 
invariant subspace. There is then two  submodules  of dimension three which are 
not invariant but are (and then, isomorphic to $1/2$ and $-1/2$) modulo 
the singlet, and finally, on top, a submodule of dimension one which is invariant 
modulo all the others.  A generic element of the algebra, written in the  
basis made of the top, middle and bottom components of the quiver in 
this order, will have the form
\begin{equation}
    \rho=\left(\begin{array}{cccc}
    0&0&0&0\\
    a&\rho_{1/2}^{1/2}&0&0\\
    b&0&\rho_{1/2}^{-1/2}&0\\
    c&d&e&0\end{array}\right)
    \end{equation}
Here, the $\rho$ are $3\times 3$ matrices, $a$ and $b$ are $3\times 1$,    
$d$ and $e$ are $1\times 3$, and $c$ is $1\times 1$. 

More generally, replacing $0$ by $j$, and $1/2,-1/2$ by 
$j-1/2,j+1/2$, or the same with the conjugate of all these 
representations, gives the quiver diagram for indecomposables of 
dimension $8$ (for $j=0$) and $16j+4$ ($j>0$).
These representations (denoted $[j,j-1/2,j+1/2,j]$) and 
their conjugates are the only indecomposables whose quiver has a 
`loop'. They are projective representations (see the appendix for some 
algebraic reminders). 
All the other indecomposables \footnote{We restrict here to 
cases where the Cartan subalgebra is diagonalizable, which should be 
the relevant one for our problem.} are not projective, and 
associated with one dimensional 
quivers. They  will be discussed soon. 

We now turn to the current algebra. First, a note about conventions. We 
chose normalizations such that  the $SU(2/1)$ WZW at level $k$ 
contains a sub $SU(2)$ current algebra at level $k$ in the 
standard convention for the latter. Other conventions 
would denote this model as 
the $OSP(2/2)$ level $-2k$ model (the two 
superalgebras are isomorphic).

The basic OPEs read:
\begin{eqnarray}
J^{+}(z)J^{-}(0)&=&{k\over z^{2}}+{2\over z}J^{3}\nonumber\\
J^{3}(z)J^{\pm}(0)&=&\pm {J^{\pm}\over z}\nonumber\\
J^{3}(z)J^{3}(0)={k\over 2z^{2}}~&~&~B(z)B(0)=-{k\over 2z^{2}}\nonumber\\
J^{+}(z)V^{-}(0)={V^{+}\over z}~&~&~ J^{+}(z)W^{-}(0)={W^{+}\over 
z}\nonumber\\
J^{-}(z)V^{+}(0)={V^{-}\over z}~&~&~ J^{-}(z)W^{+}(0)={W^{-}\over 
 z}\nonumber\\
 J^{3}(z)V^{\pm}(0)=\pm {V^{\pm}\over 2z}~&~&~ 
  J^{3}(z)W^{\pm}(0)=\pm {W^{\pm}\over 2z}\nonumber\\
  B(z)V^{\pm}(0)={V^{pm}\over 2z}~&~&~B(z)W^{\pm}(0)=-{W^{\pm}\over 
  2z}\nonumber\\
  W^{+}(z)V^{+}(0)={J^{+}\over z}~&~&~W^{-}(z)V^{-}(0)=-{J^{-}\over 
  z}\nonumber\\
  V^{+}(z)W^{-}(0)&=&-{k\over z^{2}}+{B-J^{3}\over z}\nonumber\\
  V^{-}(z)W^{+}(0)&=&-{k\over z^{2}}+{-B-J^{3}\over z}
 \end{eqnarray}
and the generators of the base algebra obtained by taking the zero 
modes of the currents. The Sugawara construction then leads to vanishing central charge 
independently of the level $k$, as the adjoint has vanishing 
superdimension. 
The conformal weights for affine representations $\widehat{[b,j]}$ based on 
irreducible typical representations $[b,j]$ read
\begin{equation}
    h={j^{2}-b^{2}\over k+1}
\end{equation}
Affine highest weight representations
built on  simple atypical representations have conformal weight equal to zero. 
Finally, for those built on indecomposable atypical representations, 
$L_{0}$ is not diagonalizable.

\subsection{The case $k=1$ and a free field representation}

The simplest example, and the one most likely to be encountered as the 
continuum limit of a lattice model, is the case $k=1$. It is well
known \cite{Goddard, Andreas}
that this theory  admits 
a free field representation, based on a pair of symplectic fermions 
$\eta_{1},\eta_{2}$, and a pair of  bosons $\phi,\phi'$. The boson 
$\phi$ is compact and has the usual metric, the boson $\phi'$ has a 
metric of the opposite sign. The propagators are
\begin{eqnarray}
    \langle\eta_{1}(z,\bar{z})\eta_{2}(w,\bar{w})\rangle&=&-\ln|z-w|^{2}\nonumber\\
    \langle\phi(z)\phi(w)\rangle&=&-\ln(z-w)\nonumber\\
    \langle\phi'(z)\phi'(w)\rangle&=&\ln(z-w)
\end{eqnarray}
The central charge of the model is $-2+1+1=0$, as required for the 
$SU(2/1)$ WZW models.

The currents admit the following free field representation:
\begin{eqnarray}
    J^{+}&=&e^{\sqrt{2}i\phi}\nonumber\\
    J^{-}&=&e^{-i\sqrt{2}i\phi}\nonumber\\
    J^{3}&=&{1\over\sqrt{2}}i\partial\phi\nonumber\\
    B&=&-{1\over\sqrt{2}}i\partial\phi'
    \end{eqnarray}
The fermionic currents consist of two $sl(2)$ doublets with $b=\pm 
1/2$:
\begin{eqnarray}
    V^{\pm}&=&e^{{1\over\sqrt{2}}i(\pm 
    \phi+\phi')}\partial\eta_{1}\nonumber\\
    W^{\pm}&=&e^{{1\over\sqrt{2}}i(\pm 
	 \phi-\phi')}\partial\eta_{2}
	 \end{eqnarray}

The  WZW model has never  (to our knowledge)
been completely understood, even for $k=1$: questions such as the 
operator content or the operator algebra only admit partial answers, 
at best.  Of course, it is tempting to speculate 
that the operator content derives somehow
from the sub $SU(2)$ part, which for $k=1$ contains only the affine 
representations at $j=0,j=1/2$. The $j=1/2$ case corresponds, for the 
fundamental representation of $osp(2/2)$ which has $b=0$, to a 
conformal weight $h={1\over 8}$, and one thus expects to have, for 
$k=1$, at least the representations with $h=0$ and $h={1\over 8}$. 
But it is easy to argue that the operator content is
considerably richer, and in particular
exhibits logarithmic features. Indeed, consider the tensor  
product of the $[0,1/2]$ representation (the multiplet for the 
fundamental field of the theory) with itself;
\begin{equation}
    [0,1/2]\otimes [0,1/2]=[0,1]+[0,1/2,-1/2,0]\label{tensprod}
\end{equation}
The right hand side does not contain the true singlet $[0,0]$, which 
means, jumping from tensor product in the base algebra 
to fusion product in the 
conformal field theory \cite{Andreas},  that the identity field must have logarithmic partners, and fields 
of dimension $h=0$ be organized in several representations of 
$sl(2/1)$. 

To gain further insight on this question,  we can study the OPE 
associated with  (\ref{tensprod}) by using the 
free field representation. The doublet in the $[0,1/2]$ 
representation can be represented by  \cite{Andreas}
\begin{equation}
    v_{\pm}=e^{\pm {i\phi\over\sqrt{2}}} \mu
\end{equation}
where $\mu$ is the $\lambda={1\over 2}$ twist field in the symplectic 
fermion theory, with dimension $-1/8$, which, added to $+1/4$, gives 
the desired $1/8$. The two singlets meanwhile can be represented as 
\cite{Andreas}
\begin{equation}
    w_{\pm}=e^{\pm {i\phi'\over\sqrt{2}}}\nu^{\pm}
\end{equation}
where $\nu^{\pm}$ are the $\lambda=-1/2$ and $\lambda=3/2$ twist 
fields in the symplectic fermion theory. Their dimension is $3/8$, 
which, added to $-1/4$, gives the desired $1/8$. The fields in the 
adjoint representation are  the currents, whose expression was given 
earlier. 
Finally,  for the fields in the indecomposable representation, they are 
given by
\begin{eqnarray}
    &~~~\eta_{1}\eta_{2}~~~&\nonumber\\
   e^{\pm 
    {i\phi\over\sqrt{2}}}e^{-{i\phi'\over\sqrt{2}}}\eta_{1},~~
    e^{-i\sqrt{2}\phi'}\eta_{1}\partial\eta_{1}&&
    e^{\pm 
	{i\phi\over\sqrt{2}}}e^{-{i\phi'\over\sqrt{2}}}\eta_{2},~~
	e^{-i\sqrt{2}\phi'}\eta_{2}
\partial\eta_{2}\nonumber\\
&~~~~~1~~~~~&\label{indecid}
\end{eqnarray}
where the fields are arranged 
exactly as in the quiver for  $[0,-1/2,1/2,0]$ given above. 
All these fields  have vanishing conformal dimension. The 
superdimension of $[0,-1/2,1/2,0]$ vanishes. On the other hand, it is 
known (see below) that the Witten index of the theory is equal to 
one. This means that in the level $k=1$ WZW model there must be 
at least the fields in  (\ref{indecid}) {\sl} and a `true' identity field, 
which is not embedded in a bigger indecomposable representation. The 
free field representations is thus not complete as is, since two 
fields `1' must be introduced.

In fact, it  is possible to build (many) other indecomposable 
representations associated with fields of vanishing conformal weight. 
The simplest  are obtained as representations containing either the field 
$\eta_{1}$ or the field $\eta_{2}$. For instance, the fields 
\begin{equation}
    \eta_{2},~~e^{
    {i\over\sqrt{2}}(\pm \phi+\phi')},~~e^{i\sqrt{2}\phi'}\partial\eta_{1}
    \end{equation}
form a representation one can represent by the diagram
$\stackrel{0}{\bullet}\rightarrow\stackrel{1/2}{\bullet}$. Similarly, 
the fields
\begin{equation}
    \eta_{1},~~e^{
    {i\over\sqrt{2}}(\pm \phi-\phi')},~~
    e^{-i\sqrt{2}\phi'}\partial\eta_{2}
    \end{equation}
can be represented by the diagram
    $\stackrel{-1/2}{\bullet}\leftarrow\stackrel{0}{\bullet}$. 
    The module at the end of the arrow is
      invariant, and quotienting by this module amounts to erasing 
       the dot, that is the remaining module is then invariant.
    For 
    more general such diagrams, 
     the dots will  represent atypical representations of the 
    type $[\pm j,j]$ discussed earlier.
    
These structures do not come alone, 
but are always 
embeddable in larger structures (in mathematical terms, they are not 
`projective' representations). This is easily seen in the free field 
representation. For instance, 
introducing 
the field $e^{i\sqrt{2}(\phi-\phi')}\eta_{2}$ and acting with the 
generators gives rise to 
$\stackrel{-1}{\bullet}\rightarrow\stackrel{-1/2}
{\bullet}\leftarrow\stackrel{0}{\bullet}$. 
where the  $[-1,1]$ submodule is constituted by  the fields 
\begin{equation}
    e^{i\sqrt{2}(\phi-\phi')}\eta_{2},~~e^{{i\over\sqrt{2}}
    (\phi-\phi')}\eta_{2}\eta_{1},~~
    e^{-i\sqrt{2}\phi'}\eta_{2},~~ 
    e^{{i\over\sqrt{2}}(\phi+\phi')}\eta_{2}\eta_{1},~~
    e^{-i\sqrt{2}(\phi+\phi')}\eta_{2}
    \end{equation}
In turn, these fields can be connected to bigger representations, and 
one finds that the conformal fields of vanishing weight built `on top' 
of a single fermion $\eta_{1}$ or $\eta_{2}$  can be arranged into the 
semi-infinite quiver 
\begin{equation}
   \ldots\stackrel{-p}{\bullet}\rightarrow\stackrel{-p+{1\over 2}}{\bullet}
   \leftarrow\stackrel{-p+1}{\bullet}\rightarrow\ldots 
   \stackrel{-1}{\bullet}\rightarrow\stackrel{-1/2}{\bullet}
   \leftarrow\stackrel{0}{\bullet}
   \end{equation}
  or its mirror image.    The
 highest weight  (that is, the field annihilated by the zero modes of 
 all raising operators) of $-n/2$ is the field 
    $e^{{ni\over\sqrt{2}}(\phi-\phi')}\eta_{2}$ for $n$ even, 
    $e^{{ni\over\sqrt{2}}(\phi-\phi')}$   for $n$ odd, and similarly
    for $+n/2$.
    
 Of course, since we are at level $k=1$, these higest weights for 
 $n>1$ are not 
  affine highest 
weights, that is, their OPE with the  $SU(2)$ currents gives poles of degree 
higher than one. For 
instance for $n=2$,
\begin{equation}
    J^{-}(z) e^{i\sqrt{2}(\phi-\phi')}\eta_{2}(w)={1\over (z-w)^{2}}
    :e^{-i\sqrt{2}\phi(z)}e^{i\sqrt{2}(\phi-\phi')}\eta_{2}(w):
    \end{equation}
 Therefore, applying modes with positive indices will produce fields 
 with {\sl negative dimension}. In this example we get the field 
 $e^{-i\sqrt{2}\phi'}\eta_{2}$ of dimension $-1$. This field belongs 
 to the representation $[b=-3/2,j=1/2]$, for which 
 $e^{i/\sqrt{2}(\phi-3\phi')}\partial\eta_{2}$ is an affine highest 
 weight. The pattern generalizes, and one finds that for any $n>1$, 
 the fields at weight $h=0$ in the semi-infinite quiver  for node 
 $-n/2$ are affine 
 descendents of the field 
 \begin{equation}
     \exp\left[{i\over\sqrt{2}}(\phi-(2n-1)\phi')\right]\partial\eta_{2}\ldots\partial^{n-1}\eta_{2}
     \end{equation}
 which is the highest weight of $[-n+1/2,1/2]$, with $h=-n(n-1)/ 2$. Concatenating 
 $[-1/2,1/2]$ and $[0,0]$ in a four dimensional representation we will 
 still denote (abusively) by $[-1/2,1/2]$, the set of conformal fields associated 
 with the semi-infinite quiver seems to be made of all affine highest 
 weight representations with base $[b=-n+1/2,j=1/2]$, 
 $n=1,\ldots,\infty$. Similarly for the mirror image, we get all the 
 representations $[b=n-1/2,1/2]$, $n=1,\ldots,\infty$. 
 
 Note that as we increase the value of $n$ we get larger and larger
 negative dimensions, and the semi-infinite quiver corresponds to a 
 theory whose spectrum is not bounded from below. This feature is
 rather characteristic of supergroup WZW models. It  does not seem to 
 make physical applications impossible however, as we will discuss 
 below. 
 
The algebraic structure of the representations of the affine algebra 
associated with the semi-infinite quivers is very interesting of course, 
and deserves much further study,  beyond the scope of this 
paper. We shall refer to them, whenever necessary, as   `infinite blocks'
representations of the current algebra. 
 
 Note that in a physical theory, left and right sectors have to be 
 combined. While we have treated the fermions so far as chiral 
 objects, they are not, and when completing with the left moving 
 parts of the boson operators, the left moving part of the fermions 
 should also be included. 
 
 Let us now backtrack a bit and wonder what of this algebraic zoo
will appear in the WZW model. 
 First, observe that the free field representation of the currents uses only the 
derivatives of the fermions, which are well defined conformal (chiral) 
fields. One might have hoped that maybe a theory can be defined that 
does not use the fermion fields themselves (and thus is based only on 
the ``small algebra''\cite{kausch}). However, consideration of the four point 
functions and the Knizhnik Zamolodchikov equation seem to exclude 
this possibility, as the identity clearly must have logarithmic 
partners, and the fermions themselves are necessary to reproduce this 
feature in the free field representations \cite{Andreas}. One could 
then
have hoped 
that fields of dimension zero are reduced to 
the true singlet and the eight dimensional indecomposable 
$[0,1/2,-1/2,0]$.
This seems also incorrect. For instance, it is known 
that the fusion of the twist fields with dimension $h=-1/8$ and 
$h=3/8$ in the symplectic fermions theory does produce the fermionic 
fields themselves \cite{kausch}. Also, we shall see later that  
modular transformations of characters suggests the presence of more 
fields of weight zero.  If so, then the non projective 
indecomposables must appear, and then there is no reason to restrict 
them. We in fact speculate  that the infinite blocks defined 
previously 
appear in the WZW 
model at level one. This in turn implies apppearance of arbitrarily 
large negative conformal weights.

Indecomposable representations of $sl(2/1)$ lead to affine highest weight 
representations with vanishing conformal weights. To have a non 
vanishing conformal weight for an affine highest weight 
representation requires dealing with a typical,  irreducible representation of 
$sl(2/1)$. There, the presence of a sub $SU(2)$ at level one 
presumably allows only spins $j=0,1/2$ to occur. So a minimal guess 
for the WZW model would be to add to 
the infinite blocks  just discussed, the affine highest weight representation
based on
$[0,j=1/2]$.  But we have already seen appearance of 
the representations 
$[-n+1/2,1/2]$  in relation with  the infinite blocks. Thus it seems 
to us that the  spectrum of the 
WZW model should at least include 
\begin{equation}
    \widehat{[0,0]}\otimes \widehat{[0,0]}~~
    \bigoplus_{n=-\infty}^\infty~~ \widehat{[-n+1/2,1/2]}\otimes 
    \widehat{[n-1/2,1/2]}\label{opcont}
    \end{equation}

\subsection{Relevance to lattice models?}

The appearance of  arbitrarily large negative conformal weights
is related with the fact that the invariant metric on 
the supergroup is not definite positive, and thus the functional 
integral ill defined - a feature  directly at the origin of the `wrong sign' 
in the propagator of the free field  $\phi'$. This might lead one to 
think that the conformal field theory we have tried to define through 
the free field representation of the $SU(2/1)$ current algebra cannot 
in any case appear as the continuum limit of a lattice model, for 
which (see below) one can easily establish the existence of a unique 
ground state. But things are more subtle. It might be 
for instance that the arbitrarily large negative conformal weights
\footnote{Note that these are not necessarily excluded on "physical 
grounds'. It might well be that in some 
systems, these arbirarily large negative conformal weights
{\sl do} describe physical  observables, like powers of the wave function 
\cite{Chamon}.}
correspond to non normalizable states, and thus  are absent 
from the spectrum and the partition function. This would be 
the situation for a lattice model whose continuum limit is 
a 
non compact boson \cite{JacobsenSaleur} . It might also be that the arbitrarily large 
negative conformal weights {\sl do} correspond to normalizable 
states, but that the correspondence between the spectrum and the 
partition function is more complicated than for unitary conformal 
field theories say, and proceeds through some sort of analytic 
continuation.  This  seems to be the case for the $\beta\gamma$ 
system for instance \cite{Lesage}, where instead of the naive characters, the 
modular invariant partition function is made of character functions, 
which encode the arbitrarily large negative conformal weights in a 
subtle way. We will argue that this is the situation in the $SU(2/1)$ 
model as well. In other words, the {\sl spectrum deduced from the 
lattice model may be related to the spectrum of the conformal field 
theory only through some sort of analytical continuation}. We will 
discuss below a very naive version of this continuation. 

Finally, and getting  a bit ahead of ourselves, one might recall that 
Zirnbauer  (see  \cite{Bocquet} for 
a thorough discussion) has suggested to 
 replace the target space from (in this case) $SU(2/1)$
with its indefinite metric by a  
{\sl non compact target space} with a positive metric - a  Riemannian 
symmetric superspace. The resulting  functional integral would then  be well 
defined, and  the spectrum bounded from below. It is not 
clear to us how one would study this model in practice. But it is 
tempting to conjecture that its spectrum would coincide with the 
analytical continuation of the naive spectrum deduced from the 
algebraic study of the current algebra, that is the spectrum observed 
in the lattice model. If so, the question of `what is the continuum 
limit of the lattice model' may have, in a way, two answers. 
This obviously requires more thinking.

To proceed, we now turn to characters and their  modular 
transformations. This will give us a handle on the mysterious 
analytical continuation at play, and allow us to make contact with 
the conjecture (\ref{opcont}).

\section{Some results about the WZW model from characters}

\subsection{Integrable characters.}

\subsubsection{Ramond sector}

At level $k=1$,
four {\sl admissible} 
characters \cite{Taormina} have been identified. In the Ramond sector 
(periodic fermions on the cylinder)
one has the field $j=1/2,b=0$ with character
\begin{equation}
\chi^{R,I}_{h=1/8}(\tau,\sigma,\nu)=q^{1/8}z^{1/2}F^{R}(\tau,\sigma,\nu)
\sum_{a\in Z}q^{2a^{2}+a}
(z^{2a}-z^{-2a-1})
\end{equation}
Here we use the definition
$\chi=\hbox{Tr } q^{L_{0}-c/24} e^{2i\pi\sigma J_{0}^{3}}e^{2i\pi\nu 
B_{0}}$, $J_{0}^{3}$ and $B_{0}$ being the zero modes of the $J^{3}$ 
field (the third component of the spin) and the $B$ field. 
Normalizations  are such that for  the highest weight state in  the 
$h=1/8$ R sector, $J^{3}_{0}=j=1/2$, $B_{0}=b=0$. We have set 
$z=e^{2i\pi\sigma}$ and $\zeta=e^{2i\pi\nu}$. Finally, 
\begin{equation}
    F^{R}(q,z,\zeta)=\prod_{n=1}^{\infty}
    {(1+z^{1/2}\zeta^{1/2}q^{n})(1+z^{-1/2}\zeta^{1/2}q^{n-1})
    (1+z^{1/2}\zeta^{-1/2}q^{n})(1+z^{-1/2}\zeta^{-1/2}q^{n-1})
    \over (1-q^{n})^{2}(1-zq^{n})(1-z^{-1}q^{n-1})}
\end{equation}
It is convenient to rewrite this character as \cite{Taormina}
\begin{equation}
    \chi^{R,I}_{h=1/8}(\tau,\sigma,\nu)={1\over \eta(\tau)}
    \left[\chi_{0}(\tau,\sigma)\chi_{1/2}(\tau,\nu)+
    \chi_{1/2}(\tau,\sigma)\chi_{0}(\tau,\nu)\right]\label{firstchar}
\end{equation}
where the $\chi$'s  in  (\ref{firstchar}) are characters of the $SU(2)_{1}$ WZW 
model, and are given 
by the following well known expressions \cite{yellowpages}
\begin{eqnarray}
    \chi_{0}(\tau,\sigma)&=&{1\over \eta(\tau)}\sum_{a\in Z} 
    q^{a^{2}}z^{a} \nonumber\\
    \chi_{1/2}(\tau,\sigma)&=&{1\over \eta(\tau)}\sum_{a\in Z} 
	q^{(a+1/2)^{2}}z^{a+1/2}
\end{eqnarray}
(note that these characters are even functions of $\sigma$, as 
expected by symmetry of the $J^{3}_{0}$ spectrum). 
and we have set  $z=e^{2i\pi\sigma}$; similarly we set 
$\zeta=e^{2i\pi\nu}$. The  specialized Ramond character  $\sigma=\nu=0$ 
for the $h=1/8$ field reads 
therefore
\begin{equation}
    \chi^{R,I}_{h=1/8}(\tau)={2\over 
    \eta}\chi_{0}\chi_{1/2}(\tau)=4q^{1/8}\left(1+5q+22q^{2}+\ldots\right)
\end{equation}
where we recognize the overall multiplicity $4$, coinciding with the 
dimension of the representation $[0,1/2]$. The latter decomposes as 
one spin $1/2$ and two spin $0$  $SU(2)$ representations.

The Ramond character for the identity field is considerably more 
complicated \cite{Taormina}:
\begin{equation}
    \chi^{R,IV}_{h=0}=F^{R}(\tau,\sigma,\nu)
    \sum_{a\in Z} q^{2a^{2}}z^{-2a} {1-q^{2a}z^{-1}\over 
    (1+q^{a}z^{-1/2}\zeta^{-1/2})(1+q^{a}z^{-1/2}\zeta^{1/2})}
\end{equation}
The specialized character  turns out to be finite, despite 
the divergence of $F^{R}$:
\begin{equation}
    \chi^{R,IV}_{h=0}=\prod_{1}^{\infty}\left(
    {1+q^{n}\over 
    1-q^{n}}
    \right)^{4}\left[1-16\sum_{a=1}^{\infty}
    a q^{2a^{2}}{1-q^{a}\over 1+q^{a}}-8\sum_{a=1}^{\infty}
    {q^{2a^{2}+a}\over (1+q^{a})^{2}}\right]=1+8q+24q^{2}+\ldots
\end{equation}
We thus see that in the Ramond sector, all 
conformal weights are  of the form $h=0\hbox{ Mod integers}$ or
$h={1\over 8}\hbox{ Mod integers}$. 
Setting $\sigma=1$ gives us $Tr(-1)^{F}$ (a supertrace) and evaluates what is called 
the super character; we check that 
$\chi^{sR,I}_{h=1/8}(\tau)=0$ while $\chi^{sR,IV}_{h=0}(\tau)=1$. 

It is possible to write the specialized character in a more compact 
form using the function \cite{Eguchi}
\begin{equation}
    h_{3}(\tau)={1\over \eta(\tau)\theta_{3}(\tau)}\sum_{a\in Z}
    {q^{a^{2}-1/8}\over 1+q^{a-1/2}}
\end{equation}
One finds then
\begin{equation}
    \chi^{R,IV}_{h=0}(\tau)={\chi_{0}^{2}-\chi_{1/2}^{2}\over 
    \chi_{0}^{2}+\chi_{1/2}^{2}}(\tau)+2\chi_{0}\chi_{1/2}h_{3}(\tau)
\end{equation}

\subsubsection{Neveu-Schwarz sector}

In the NS sector, the fermions have antiperiodic boundary conditions, 
and the $SU(2/1)$ symmetry is broken down to a sub $SU(2)\times U(1)$. 
The lowest energy state in this sector has negative conformal 
dimension, leading to an effective central charge equal to $c_{eff}=3$. 
The corresponding character is 
\begin{equation}
    \chi^{NS,I}_{h=-1/8}(\tau,\sigma,\nu)={1\over \eta(\tau)}
    \left[\chi_{1/2}(\tau,\sigma)\chi_{1/2}(\tau,\nu)+
    \chi_{0}(\tau,\sigma)\chi_{0}(\tau,\nu)\right]
\end{equation}
The specialized characters reads 
\begin{equation}
    \chi^{NS,I}_{h=-1/8}(\tau)=q^{-1/8}\left(1+4q^{1/2}+\ldots\right)
\end{equation}
so half integer gaps appear in this sector. 

The other character is again more complicated
\begin{equation}
    \chi^{NS,IV}_{h=1/4}(\tau,\sigma,\nu)
    =q^{1/4}z^{1/2}F^{NS}(\tau,\sigma,\nu)
    \sum_{a\in Z} q^{2a^{2}+2a}
    z^{2a} {1-zq^{2a+1}\over 
    (1+q^{a+1/2}z^{1/2}\zeta^{-1/2})(1+q^{a+1/2}z^{1/2}\zeta^{1/2})}
\end{equation}
where, 
\begin{equation}
    F^{NS}=\prod_{n=1}^{\infty}
    {(1+z^{1/2}\zeta^{1/2}q^{n-1/2})(1+z^{-1/2}\zeta^{1/2}q^{n-1/2})
    (1+z^{1/2}\zeta^{-1/2}q^{n-1/2})(1+z^{-1/2}\zeta^{-1/2}q^{n-1/2})
    \over (1-q^{n})^{2}(1-zq^{n})(1-z^{-1}q^{n-1})}
\end{equation}
The specialized character is again finite, despite 
the divergence of $F^{NS}$:
\begin{equation}
    \chi^{NS,IV}_{h=1/4}(\tau)=h_{3}(\tau)\left[\chi_{0}(\tau)^{2}+\chi_{1/2}(\tau)^{2}\right]
\end{equation}
and like $\chi^{NS,I}_{h=-1/8}$, exhibits half integer gaps. 

\subsection{The remaining operator content }

It is crucial to realize that  the 
 characters we wrote down all have complicated or unusual 
 modular 
transformations. A possible strategy to proceed is then to 
try to supplement these 
characters with others in order to obtain a finite dimensional 
representation of the modular group, and maybe modular invariants.

\subsubsection{A continuum above $h=1/8$ in the R sector}

\bigskip
\noindent{\bf $\bullet$ Transformation of $\chi_{h=-1/8}^{sNS,I}$}
\bigskip

Consider for instance the Neveu Schwarz characters. 
They correspond to fermions being antiperiodic in both directions. 
Aternatively, this corresponds to an antiperiodic chain along the 
space direction, while one takes a trace along the (imaginary) time 
direction. Imagine instead taking a supertrace along the imaginary 
time direction, or setting $\sigma=1$ in the expression of the 
characters, giving rise again to  a supercharacter. The 
specialized supercharacter for $h=-1/8$ reads
\begin{equation}
\chi^{sNS,I}_{h=-1/8}(\tau)={\chi_{0}^{2}-\chi_{1/2}^{2}\over\eta}(\tau)
\end{equation}
Recall now  the modular transformation of the basic $SU(2)_{1}$ WZW 
characters
\begin{eqnarray}
   \left(\begin{array}{c}\chi_{0}\\
   \chi_{1/2}\end{array}\right)(-1/\tau)
   =\left(\begin{array}{cc}
   {1\over\sqrt{2}} &{1\over\sqrt{2}}\\
   {1\over\sqrt{2}} &{-1\over\sqrt{2}}\end{array}\right)
\left(\begin{array}{c}\chi_{0}\\\chi_{1/2}\end{array}\right)(\tau)
\end{eqnarray}
Meanwhile the $\eta$ function obeys 
$\eta(-1/\tau)=\sqrt{-i\tau}\eta(\tau)$. So the only way to write the 
modular transform of this $SU(2/1)$ (super) character as a sum of powers of 
$q$ is to introduce an integral representation of the $\eta$ factor
\begin{equation}
    {1\over\sqrt{-i\tau}}=\int_{-\infty}^{\infty}d\alpha 
   e^{i\pi\tau\alpha^{2}}
\end{equation}
and thus
\begin{equation}
    \chi^{sNS,I}_{h=-1/8}(-1/\tau)=2\chi_{0}(\tau)\chi_{1/2}(\tau)\times 
    \int_{-\infty}^{\infty} d\alpha
    {q^{\alpha^{2}/2}\over \eta(\tau)}=\chi^{R,I}_{h=1/8}\int d\alpha 
    q^{\alpha^{2}/2}\label{new}
\end{equation}
Now observe that under modular transformation the boundary conditions 
which determine the super NS character turn into boundary conditions 
where the fermions are periodic in the space direction, and one takes 
a trace in the imaginary time direction. Hence the right hand side 
of (\ref{new}) should appear in the generating function of the 
spectrum in the $sl(2/1)$ symmetric sector of our  chain if the 
associated conformal field theory contains the Neveu Schwarz 
characters discussed previously. This means {\bf there should be a 
continuous component starting at the $h=1/8$ value}. 

To find out about this continuous component, we can keep track of 
the $B_{0}$ and $J_{0}^{3}$ numbers by keeping the parameters 
$\sigma,\nu$:
\begin{eqnarray}
\chi^{sNS,I}_{h=-1/8}(\tau,\sigma,\nu)\equiv 
Tr\left[(-1)^{F}q^{L_{0}}z^{J_{0}^{3}}\zeta^{B_{0}}\right] 
={1\over \eta(\tau)}
\left[\chi_{0}(\tau,\sigma)\chi_{0}(\tau,\nu)-\chi_{1/2}(\tau,\sigma)
\chi_{1/2}(\tau,\nu)
\right]
\end{eqnarray}
The modular transformation of general $SU(2)$ characters  reads
\begin{eqnarray}
   \left(\begin{array}{c}\chi_{0}\\
   \chi_{1/2}\end{array}\right)(-1/\tau,\theta/\tau)
   =e^{i\pi \theta^{2}/2\tau}\left(\begin{array}{cc}
   {1\over\sqrt{2}} &{1\over\sqrt{2}}\\
   {1\over\sqrt{2}} &{-1\over\sqrt{2}}\end{array}\right)
\left(\begin{array}{c}\chi_{0}\\\chi_{1/2}\end{array}\right)(\tau 
,\theta )
\end{eqnarray}
and thus 
\begin{equation}
    \chi^{sNS,I}_{h=-1/8}(-1/\tau,\sigma/\tau,\nu/\tau)=
    {1\over\sqrt{-i\tau}\eta(\tau)}e^{i\pi(\sigma^{2}+\nu^{2})/2\tau}
    \left[\chi_{0}(\tau,\sigma)\chi_{1/2}(\tau,\nu)+
    \chi_{1/2}(\tau,\sigma)\chi_{0}(\tau,\nu)\right]\label{newmod}
    \end{equation}
Due to the indefinite metric for the $SU(2/1)$ supergroup, the 
overall factor in modular transformations should however not be
 $e^{i\pi(\sigma^{2}+\nu^{2})/2\tau}$, but rather 
 $e^{i\pi(\sigma^{2}-\nu^{2})/2\tau}$ (since the Casimir is 
 proportional to $j^{2}-b^{2}$). We are then left with an 
 overall factor $e^{i\pi\nu^{2}/\tau}$ whose interpretation is 
 slightly delicate. What we propose is to write
 \begin{equation}
     {1\over\sqrt{-i\tau}}e^{i\pi\nu^{2}/\tau}={1\over\sqrt{-1}}
     \int_{-\infty}^{\infty}db e^{-i\pi\tau b^{2}}e^{2i\pi\nu 
     b}
     \end{equation}
 The phase factor is ambiguous and depends on a choise of cut. As for 
 the integral, is is divergent in the physical situation where 
 $\hbox{Im }\tau>0$. If we  assume  however that this integral is 
 defined by some sort of analytical continuation from the case $\hbox{Im }\tau<0$, 
 we see that we can formally interpret the leftover $\nu^{2}$ exponential in our 
 modular transform (\ref{newmod}) as 
 resulting from an integral over fields with conformal dimension and 
 $B_{0}$ charges:
 \begin{equation}
     h=-{1\over 2}b^{2},~~~B_{0}=b,~~~-\infty<b<\infty
     \end{equation}
Moreover, the right hand side of the modular
transform is, for each value of $b$, of the form
\begin{equation}
{1\over\eta(\tau)} e^{-i\pi \tau b^{2}}e^{2i\pi \nu b}
    \left[\chi_{0}(\tau,\sigma)\chi_{1/2}(\tau,\nu)+
    \chi_{1/2}(\tau,\sigma)\chi_{0}(\tau,\nu)\right]
\end{equation}
which is exactly the character for the affine representation with 
base $[b,1/2]$ (integrable character in class I in the language of 
\cite{Taormina}). It thus seems that the model should contain in 
fact, not only the representation $[0,1/2]$, but the continuum 
$[b,1/2]$ with $b$ real.  This is at odds with the first 
section where we speculated that $[-n+1/2,1/2]$ only apppeared.
See below for a discussion of this issue. 

While the conformal weights associated with 
those representations become arbitrarily large and negative, the 
analysis of the characters and modular transforms 
suggests that the finite size spectrum is given by a sort 
of analytical 
continuation, whose result, by the mechanism discussed 
above, is to make the continuum appear {\sl 
above} the $1/8$, not below. Note also that while the $B_{0}$ numbers 
in the chiral and antichiral sector appear continuous, the lattice analysis 
anyway only 
identifies the combination $L+R$, which is made of 
integers. 
    
\bigskip
\noindent{\bf $\bullet$ Transformation of $\chi_{h=1/4}^{sNS,IV}$}
\bigskip

We can perform the same analysis for the specialized supercharacter 
for $h=1/4$:
\begin{equation}
    \chi^{sNS,IV}_{h=1/4}(\tau)=-2{\chi_{0}\chi_{1/2}\over
    \chi_{0}^{2}+\chi_{1/2}^{2}}+h_{3}\left(\chi_{0}^{2}-\chi_{1/2}^{2}\right)
\end{equation}
The function $h_{3}$ obeys
\begin{equation}
    h_{3}(-1/\tau)=-h_{3}(\tau)+{1\over\eta}\int_{-\infty}^{\infty} 
    d\alpha {q^{\alpha^{2}/2}\over 2\cosh\pi\alpha}
\end{equation}
Thus
\begin{equation}
    \chi_{h=1/4}^{sNS,IV}(-1/\tau)={\chi_{1/2}^{2}-\chi_{0}^{2}\over
    \chi_{1/2}^{2}+\chi_{0}^{2}}(\tau)
    +2\chi_{0}\chi_{1/2}(\tau)\left[-h_{3}(\tau)+
    {1\over \eta(\tau)}\int_{-\infty}^{\infty}d\alpha
    {q^{\alpha^{2}/2}\over 2\cosh\pi\alpha}\right]
    \end{equation}
 where we again obtain a continuous spectrum starting at $h={1\over 
 8}$. This time however, instead of a flat measure over all the 
 reals, one has a measure proportional to $1/\cosh\pi\alpha$. Note 
we can 
 write
 \begin{equation}
     \chi^{sNS,IV}_{h=1/4}(-1/\tau)=-\chi^{R,IV}_{h=0}(\tau)+\chi_{h=1/8}^{R,I}\int d\alpha {q^{\alpha^{2}/2}\over 
     2\cosh\pi\alpha}\label{toberecov}
     \end{equation}

 To find out about the spectrum of $B_{0}$ and $J_{0}^{3}$ values 
 is considerably more complicated, and the tired reader may want to jump to 
 formula (\ref{Nice}). The following calculations  rely on the 
 paper \cite{Semikhatov}; one  starts by writing 
 \begin{equation}
     \chi_{h=1/4}^{sNS,IV}(\tau,\sigma,\nu)=-e^{i\pi 
     \tau/2}e^{i\pi\sigma} F^{sNS}(\tau,\sigma,\nu) \left[
     K_{4}(\tau,{\tau+\sigma\over 2},-{\nu\over 2})-
     K_{4}(\tau,{-\tau-\sigma\over 2},-{\nu\over 2})\right]
     \end{equation}
 where 
 \begin{equation}
     F^{sNS}=\prod_{n=1}^{\infty}
     {(1-z^{1/2}\zeta^{1/2}q^{n-1/2})(1-z^{-1/2}\zeta^{1/2}q^{n-1/2})
     (1-z^{1/2}\zeta^{-1/2}q^{n-1/2})(1-z^{-1/2}\zeta^{-1/2}q^{n-1/2})
     \over (1-q^{n})^{2}(1-zq^{n})(1-z^{-1}q^{n-1})}
 \end{equation}
 It is useful to rewrite this prefactor as 
 \begin{equation}
     F^{sNS}={\theta_{10}(\tau,{\sigma+\nu-\tau+1\over 2})
     \theta_{10}(\tau,{\sigma-\nu-\tau+1\over 2})\over 
     \theta_{11}(\tau,\sigma)q^{-1/8}\eta^{3}(q)}
     \end{equation}
The $K$ are Apell functions 
\begin{equation}
    K_{l}(\tau,\sigma,\nu)=\sum_{m\in Z}{e^{i\pi m^{2}l\tau+2i\pi 
    ml\sigma}\over 1-e^{2i\pi(\sigma+\nu+m\tau)}}
    \end{equation}
 and the $\theta$ are usual theta functions
 \begin{eqnarray}
     \theta_{10}(q,z)=\prod_{m\geq 0}(1+z^{-1}q^{m})
     \prod_{m\geq 1}(1+zq^{m})\prod_{m\geq 1}(1-q^{m})\nonumber\\
     \theta_{11}(q,z)=\prod_{m\geq 0}(1-z^{-1}q^{m})
	  \prod_{m\geq 1}(1-zq^{m})\prod_{m\geq 1}(1-q^{m})
	  \end{eqnarray}
%	 
% Note that $\chi^{sNS}$ is  not invariant under $\sigma\rightarrow -\sigma$ 
 %as $F^{sNS}$ obeys 
 %$F^{sNS}(-\sigma)=-e^{-2i\pi\sigma}F^{sNS}(\sigma)$. I am a bit 
 %puzzled by this. 
	 
	  Modular transformations of the theta functions are as follows:
 \begin{eqnarray}
     \theta_{11}(-1/\tau,\mu/\tau)=-i\sqrt{-i\tau}e^{i\pi\mu+i\pi{1\over 
     \tau} (\mu-1/2)^{2}+{i\pi\over 
     4}\tau}\theta_{11}(\tau,\mu)\nonumber\\
     \theta_{10}(-1/\tau,\mu/\tau)=\sqrt{-i\tau}e^{i\pi 
     {(\mu-1/2)^{2}\over\tau}}\theta_{10}(\tau,\mu+1/2-\tau/2)
     \end{eqnarray}
  As for the Appell functions it is much more complicated. One finds 
  \cite{Semikhatov}
  \begin{eqnarray}
      K_{l}(-1/\tau,\sigma/\tau,\nu/\tau)=e^{i\pi l (\sigma^{2}-\nu^{2})/\tau}
      K_{l}(\tau,\sigma,\nu)\nonumber\\
      +\tau\sum_{a=0}^{l-1}e^{i\pi 
      l(\sigma+a\tau/l)^{2}/\tau}\Phi(l\tau,l\nu-a\tau)\theta_{00}(l\tau,l\sigma+a\tau)
      \end{eqnarray}
  where $\theta_{00}(q,z)=\theta_{10}(q,zq^{-1/2})$ and 
  \begin{equation}
      \Phi(\tau,\mu)=-{i\over 2\sqrt{-i\tau}}-{1\over 2}\int dx 
      e^{-\pi x^{2}}{\sinh(\pi 
      x\sqrt{-i\tau}(1+2\mu/\tau))\over\sinh (\pi x\sqrt{-i\tau})}
      \end{equation}
Putting everything together gives
\begin{eqnarray}
    \chi^{sNS,IV}_{h=1/4}(-1/\tau,\sigma/\tau,\nu/\tau)={1\over \tau}
    e^{-i\pi(\sigma^{2}-\nu^{2})/2\tau}e^{2i\pi\sigma/\tau}e^{-i\pi/\tau}
    \nonumber\\
    {\theta_{10}(\tau,-(\sigma+\nu)/2)\theta_{10}(\tau,-(\sigma-\nu)/2)\over
    \theta_{11}(\tau,-\sigma)\eta^{3}(\tau)}\nonumber\\
    \left[K_{4}(-1/\tau,(\sigma-1)/2\tau,-\nu/2\tau)-
    K_{4}(-1/\tau,(-\sigma-1)/2\tau,-\nu/2\tau)\right]\label{big}
    \end{eqnarray}
Using the general Appell transform formulas gives
\begin{eqnarray}
    K_{4}(-1/\tau,(\sigma-1)/2\tau,-\nu/2\tau)=\tau e^{i\pi 
    (\sigma^{2}-\nu^{2})/\tau}e^{i\pi\tau}e^{-2i\pi\sigma/\tau}
    K_{4}(\tau,(\sigma-1)/2\tau,-\nu/2)
    \nonumber\\
    +\tau\sum_{a=0}^{l-1}e^{i\pi 
    (\sigma-1+a\tau/2)^{2}/\tau}\Phi(4\tau,-2\nu-a\tau)\theta(4\tau,-2+
    (2\sigma+a)\tau)\label{transf}
    \end{eqnarray}
 We recognize the expected global factor 
 $e^{i\pi(\sigma^{2}-\nu^{2})/2\tau}$. Also, observe that 
 \begin{equation}
     {\theta_{10}(q,xy)\theta_{10}(q,xy^{-1})\over \theta_{11}(q,x)}=
     -{\theta_{10}(q,x^{-1}y^{-1})\theta_{10}(q,x^{-1}y)\over\theta_{11}(q,x^{-1})}
     \end{equation}
while 
\begin{equation}
    F^{R}(\tau,\sigma,\nu)={\theta_{10}(\tau,(\sigma+\nu)/2)\theta_{10}(\tau(\sigma-\nu)/2)\over
    \theta_{11}(\tau,\sigma)e^{-i\pi\tau/4}\eta^{3}(\tau)}
    \end{equation}
 We also observe that
 $K_{4}(\tau,\nu,\mu)=K_{4}(\tau,\nu\pm 1,\mu)$. Therefore, the 
 contribution to (\ref{big}) coming from the $K_{4}$ term in the 
 transform (\ref{transf}) gives 
 \begin{equation}
   \hbox{First term}=  -e^{i\pi(\sigma^{2}-\nu^{2})/2\tau}
   \chi^{R,IV}_{h=0}(\tau,\sigma,\nu)
     \end{equation}
  The second term reads, after making the $\theta$ function explicit,
 \begin{equation}
     -F^{R}(\tau,\sigma,\nu)   ie^{i\pi(\sigma^{2}+\nu^{2})/2\tau}
     \sum_{a=1}^{3}(-1)^{a}\Phi(4\tau,-2\nu-a\tau) 
     q^{a^{2}/8}z^{a/2}\sum_{m}q^{2m^{2}+ma}\left(z^{2m}-z^{-2m-a}\right)
     \end{equation}
We recognize the prefactor of the $\Phi$ function as the formal 
extension of the Ramond 
character  formula, initially valid only for $j=1/2,b=0,h=1/8$ to 
$j=a,b=0,h=a^{2}/8$, $\chi^{R,I}_{h=a^{2}/8}$. However, 
setting 
\begin{equation}
X_{a}\equiv    q^{a^{2}/8}z^{a/2}\sum_{m}q^{2m^{2}+ma}
\left(z^{2m}-z^{-2m-a}\right)
\end{equation}
it is easy to show that $X_{a}=X_{a-4}=-X_{-a}$. Therefore, 
$X_{0}=X_{\pm 2}=0$, and $X_{1}=-X_{-3}$. Taking advantage of these 
symmetries leads us to a considerably simplified expression
\begin{eqnarray}
    \chi^{sNS,IV}_{h=1/4}(-1/\tau,\sigma/\tau,\nu/\tau)= -e^{i\pi(\sigma^{2}-\nu^{2})/2\tau}
   \chi^{R,IV}_{h=0}(\tau,\sigma,\nu)\nonumber\\
   +
    \chi^{R,I}_{h=1/8}(\tau,\sigma,\nu)e^{i\pi 
    (\sigma^{2}+\nu^{2})/2\tau}
    \int dx e^{-\pi x^{2}}{\cosh (2\pi \nu 
    x\sqrt{-i\tau}/\tau)\over 2\cosh (\pi x\sqrt{-i\tau})}\label{Nice}
    \end{eqnarray}
 As a check we can set $\nu=0$ in this expression.  Using the invariance of $h_{3}(\tau)+h_{3}(-1/\tau)$ 
 we see that
 \begin{equation}
     {1\over \eta(\tau)}\int d\alpha {q^{\alpha^{2}/2}\over 
     2\cosh\pi\alpha}\equiv f(\tau)
     \end{equation}
 is invariant, and thus 
\begin{equation}
    \int dx {e^{-\pi x^{2}}\over 2\cosh\pi x\sqrt{-i\tau}}=\int 
    d\alpha {q^{\alpha^{2}/2}\over 2\cosh\pi\alpha}
    \end{equation}
 which alows us to recover expression (\ref{toberecov}).

We now go back to the more general expression. Crucial is the fact, as 
mentioned earlier, that the exponential has the $\nu^{2}$ term with  
positive sign instead of a negative one.   We thus write in the 
integral $x=y\sqrt{-i\over\tau}$ and thus formally obtain
\begin{eqnarray}
    \sqrt{-i\over\tau}e^{i\pi\nu^{2}/\tau}\int dy e^{i\pi 
    y^{2}/\tau}{\cosh 2\pi\nu y/\tau\over \cosh i\pi y}\nonumber\\
    =\sqrt{-i\over\tau}\int dy {e^{i\pi (\nu+y)^{2}/\tau}\over\cos\pi y}
    \end{eqnarray}
Using the formal integral representation of the exponential 
we write this as
\begin{eqnarray}
    \int dbdy e^{-i\pi\tau b^{2}}{e^{2i\pi(\nu +y)b}\over \cos\pi 
    y}\nonumber\\
    =\int db e^{2i\pi\nu b}e^{-i\pi\tau b^{2}}{1\over \cos\pi b}
    \end{eqnarray}
Finally, we thus rewrite (\ref{Nice}) as 
\begin{eqnarray}
    \chi^{sNS,IV}_{h=1/4}(-1/\tau,\sigma/\tau,\nu/\tau)= -e^{i\pi(\sigma^{2}-\nu^{2})/2\tau}
   \chi^{R,IV}_{h=0}(\tau,\sigma,\nu)\nonumber\\
   +
    \chi^{R,I}_{h=1/8}(\tau,\sigma,\nu)e^{i\pi 
    (\sigma^{2}-\nu^{2})/2\tau}
    \int db e^{2i\pi\nu b}e^{-i\pi\tau b^{2}}{1\over \cos\pi b}
    \label{Nicei}
    \end{eqnarray}
This corresponds again to representations $[b,1/2]$ but this time 
with a measure that is not flat, but proportional to $1/\cos\pi b$.  
The interpretation is discussed below.
    
\bigskip
\noindent{\bf $\bullet$ Transformation of $\chi_{h=-1/8}^{NS,I}$ and $\chi_{h=1/4}^{NS,IV}$}
\bigskip

    Finally, we can also consider the ordinary NS character. 
We know that the boundary conditions corresponding 
to it character are invariant under the modular transformation. But 
using the expression  
\begin{equation}
    \chi^{NS,I}_{h=-1/8}(\tau)={\chi_{0}^{2}+\chi_{1/2}^{2}\over\eta}(\tau)
\end{equation}
we see that the modular transform has a continuous component starting 
at the ground state of the sector, namely $h=-1/8$:  
\begin{equation}
    \chi^{NS,I}_{h=-1/8}(-1/\tau)={1\over\sqrt{-i\tau}}\chi^{NS,I}_{h=-1/8}(\tau)
    =\left(\chi_{0}^{2}+\chi_{1/2}^2\right)(\tau)
    \int d\alpha {q^{\alpha^{2}/2}\over\eta}
\end{equation}
Similarly
\begin{eqnarray}
    \chi_{h=1/4}^{NS,IV}(-1/\tau)=-\chi_{h=1/4}^{NS,IV}(\tau)
    +\chi^{NS,I}_{h=-1/8}(\tau)\int_{-\infty}^{\infty} d\alpha {q^{\alpha^{2}/2}\over 
    2\cosh\pi\alpha} 
\end{eqnarray}

\bigskip
A final note: in the supersymmetric literature (the model has some 
relations to $N=4$ theories \cite{Eguchi}), the subset with discrete quantum 
numbers is refered to as massless, the one with continuous quantum
numbers as massive. 

\subsection{Interpretation}

Comparing the free field analysis of the WZW model with the analysis 
of the modular transformations of basic characters 
suggests that the
spectrum of arbitrarily large negative dimensions 
admits `regularized' expressions for 
characters, and thus partition function and spectra of lattice 
Hamiltonians. For these,  there is a single ground state, and the negative 
dimensions are `folded' back into a continuous spectrum of  positive 
dimensions. 

The slightly surprising thing is that we find in this way indications 
of a {\sl continuous spectrum} based on $[b,1/2]$ while we initially 
expected a discrete one $[-n+1/2,1/2]$.  There are stronger reasons for 
a discrete spectrum of $b$ than the ones given in the first sections. For 
instance 
the free field representation of the generators selects a 
particular radius for the non compact boson, and vertex operators 
corresponding to charges continuous charges $b$ will in general be non 
local with respect to these currents! Of course it might be in fact that the 
free field representation we have used is good only for a subset of 
fields, and does not allow one to explore the full operator content. 
But it could also be that  the continuous spectrum we read from the 
modular transformations is an artifact from the (rather uncontrolled) 
regularization at work.  A hint in this direction comes from the integral 
$$
\int {q^{\alpha^{2}\over 2}\over 2\cosh\pi\alpha}
$$
which, after continuation to the imaginary axis, has poles precisely 
at the 
values $b=-n+{1\over 2}$.  It seems difficult to conclude on this 
issue without more work on the conformal field theory side.

\section{Analysis of the lattice model}
The  next step is to extract as much information about the lattice 
model as possible and compare it with our proposal for the continuum 
limit. Although we have mostly talked about a quantum spin chain so far, 
integrability allows one to consider it as part of a larger family of 
commuting transfer matrices related to an integrable vertex model. 

\subsection{Definition of the integrable vertex model}
%% \subsection{An integrable vertex model}
 \label{sec:vertexmodel}
 
The vertex model is constructed from $R$-matrices acting on tensor products of
spaces carrying the fundamental representation $3$
of the graded Lie algebra $sl(2/1)$ and its dual $\bar{3}$. The $R$-matrices
satisfying Yang-Baxter equations (YBEs) have been constructed in
\cite{LiFo99,AbRi99,Gade99} (see also \cite{Derk00}) and read
\begin{align}
   &R_{33}(v) = 1 -\frac{2}{v} {\cal P}\ ,\quad
   &R_{\bar{3}\bar{3}}(v) = 1 -\frac{2}{v} {\cal P}
\notag\\
   &R_{3\bar{3}}(v) = 1 +\frac{2}{v} {\cal O}\ ,
   &R_{\bar{3}3}(v) = 1 +\frac{2}{v} {\cal O}\ .
\label{Rdef}
\end{align}
Here ${\cal P}$ and ${\cal O}$ are the graded permutation and monoid
operators of $sl(2/1)$, respectively (note that the action of the monoid
${\cal O}$ on $3\otimes\bar{3}$ differs from that on $\bar{3}\otimes3$).  As a
consequence of ${\cal O}^2={\cal O}$ we have
$R_{\bar{3}3}(-u-1)R_{3\bar{3}}(u-1)=1$.

{}From these $R$-matrices we construct two families of monodromy
matrices (labelled by the spectral parameter $v$). The monodromy
matrices act on the tensor product of a ``quantum space''
$\left(3\otimes\bar{3}\right)^{\otimes L}$ (the Hilbert space of the
superspin chains considered below), and a 
three-dimensional ``matrix space'' carrying a representation $3$
($\bar{3}$)
\begin{align}
   &T_3(v) \equiv 
	 R_{33}^{(1)}(v)R_{3\bar{3}}^{(1)}(v-1-\lambda)
	 R_{33}^{(2)}(v)R_{3\bar{3}}^{(2)}(v-1-\lambda) \cdots
	 R_{33}^{(L)}(v)R_{3\bar{3}}^{(L)}(v-1-\lambda)\ ,
\notag\\
   &T_{\bar{3}}(v) \equiv 
	 R_{\bar{3}3}^{(1)}(v-1+\lambda)R_{\bar{3}\bar{3}}^{(1)}(v)
	 R_{\bar{3}3}^{(2)}(v-1+\lambda)R_{\bar{3}\bar{3}}^{(2)}(v) \cdots
	 R_{\bar{3}3}^{(L)}(v-1+\lambda)R_{\bar{3}\bar{3}}^{(L)}(v)\ .
\end{align}
Here the parameter $\lambda$ can be chosen freely without affecting the
essential properties (symmetries and locality of the interactions) of the
models considered in the following.  Variation of $\lambda$ allows one
to tune the coupling between the two sublattices of $3$ and $\bar{3}$
spins, respectively. Later we will focus on the case of $\lambda=0$
which corresponds to the most symmetric coupling.
As a consequence of the YBE
\begin{equation}
   R_{3\bar{3}}(u-v)\ T_3(u)\ T_{\bar{3}}(v-1-\lambda)
	 =  T_{\bar{3}}(v-1-\lambda)\ T_3(u)\ R_{3\bar{3}}(u-v)
\label{RTT}
\end{equation}
the corresponding transfer matrices
\begin{equation}
 \tau_\alpha(v) = {\mathrm{str}}_0
 \left(T_\alpha(v)\right),\quad \alpha=3,\bar{3},
 \label{transferM}
\end{equation}
commute with each other for arbitrary values of the spectral parameter
$v$. Here ${\mathrm{str}}_0$ denotes the supertrace over the matrix space.

\begin{figure}[ht]
 \begin{center}
 \noindent
 \epsfxsize=0.45\textwidth
 \epsfbox{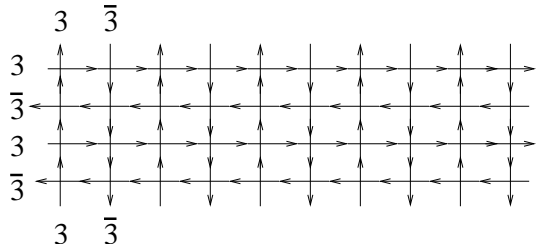}
 \end{center}
 \caption{\label{fig:vertexmodel}
 The vertex model.
 }
 \end{figure}

\begin{figure}[ht]
 \begin{center}
 \noindent
 \epsfxsize=0.15\textwidth
 \epsfbox{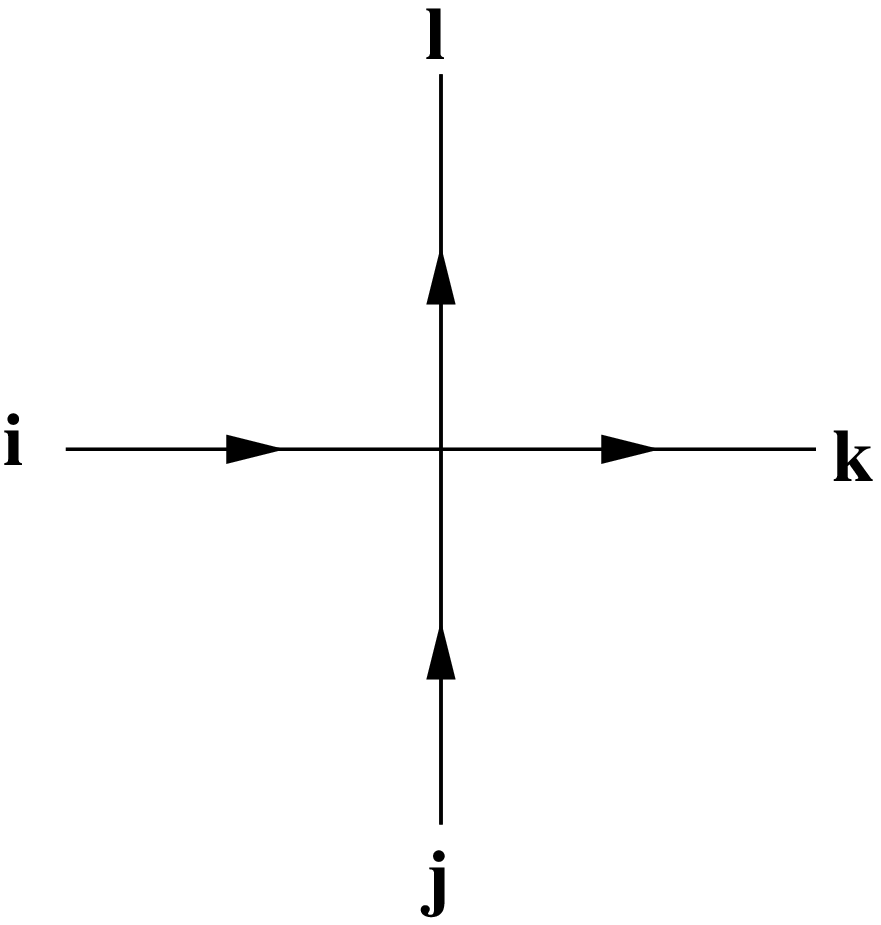}
 \qquad
 \epsfxsize=0.15\textwidth
 \epsfbox{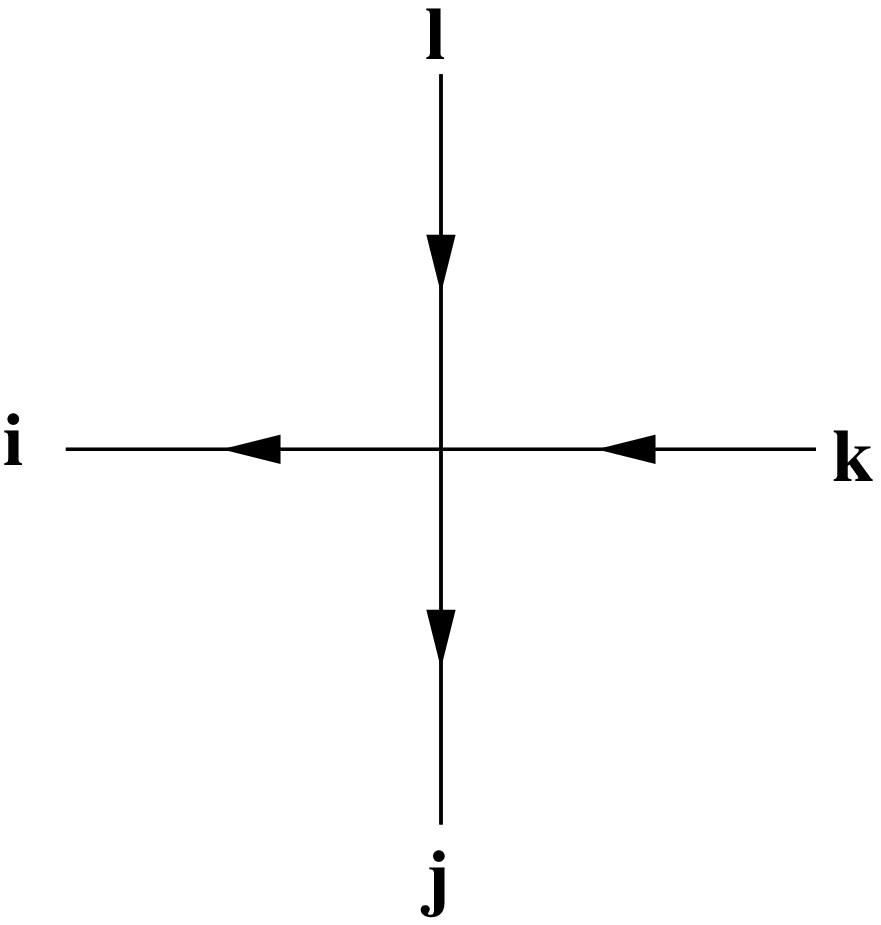}
 \qquad
 \epsfxsize=0.15\textwidth
 \epsfbox{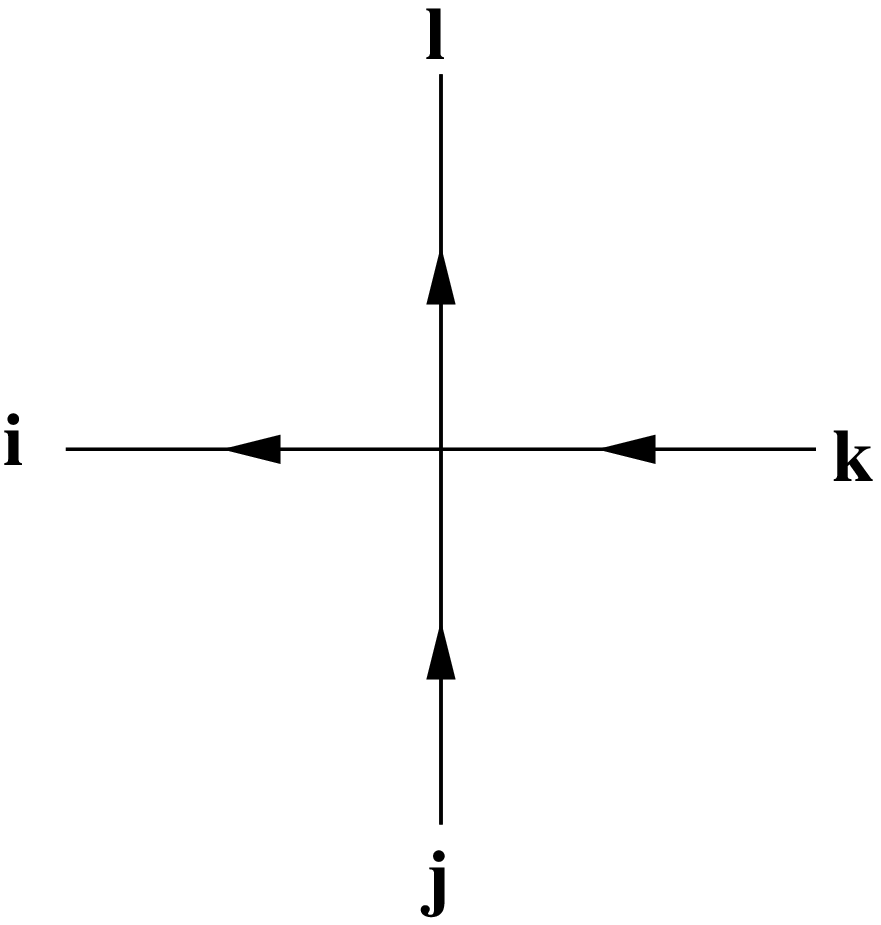}
 \qquad
 \epsfxsize=0.15\textwidth
 \epsfbox{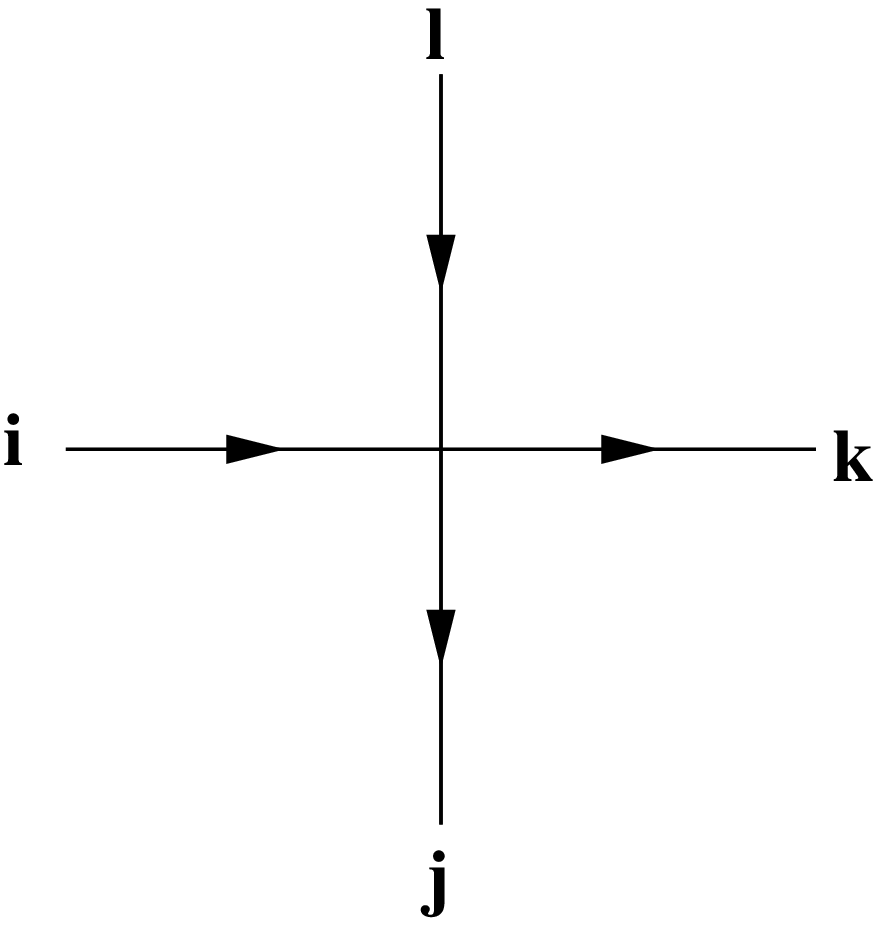}
 \end{center}
 \begin{center}
 $[R_{33}(v)]_{ij}^{kl}$
 \qquad\qquad
 $[R_{\bar{3}\bar{3}}(v)]_{ij}^{kl}$
 \qquad\quad
 $[R_{\bar{3}3}(v-1)]_{ij}^{kl}$
 \qquad\quad
 $[R_{3\bar{3}}(v-1)]_{ij}^{kl}$
 \end{center}
 \caption{\label{fig:vertexweights}
 The vertex weights. Arrows pointing upwards ot to the right (downwards
 or to  the left) indicate that the corresponding link carries the
 fundamental $3$ (the dual $\bar{3}$) representation of $sl(2/1)$.
 }
 \end{figure}

By staggering the transfer matrices \r{transferM} one can construct the
integrable vertex model shown in Fig.\ref{fig:vertexmodel}.
Arrows pointing upwards ot to the right (downwards or to the left)
indicate that the corresponding link carries the fundamental $3$ (the
dual $\bar{3}$) representation of $sl(2/1)$.
%%%%%%%%%%%%%%%%%%%%%%%%%%%%%%%%%%%%%%%%%%%%%%%%%%%%%%%%%%%%%%%%%%%%%%
%%%%%%%%%%%%%%%%%%%%%%%%%%%%%%%%%%%%%%%%%%%%%%%%%%%%%%%%%%%%%%%%%%%%%%
 The vertex weights corresponding to the various arrow configurations
 are shown in Fig.\ref{fig:vertexweights}.
The transfer matrix of this model includes two layers in the vertical
direction and is equal to $\tau(v)=\tau_3(v)\tau_{\bar{3}}(v)$. 

The vertex model constructed in this way was suggested as a description
of the spin quantum Hall effect in Ref. \cite{Gade99}. The notations
of \cite{Gade99} are obtained by setting $v=-2x$ and $\lambda=2u-1$,
where $x$ and $u$ are quantities defined in \cite{Gade99}.

The Hamiltonian of an integrable supersymmetric spin chain with local (nearest
and next-nearest neighbour) interactions only is obtained by taking the
logarithmic derivative of the two-layer transfer matrix at $v=0$\footnote{A
similar Hamiltonian was considered in \cite{LiFo99} from the point of view of
introducing impurities into a t-J model.}
\begin{equation}
   {\cal H} = - \left.\partial_v\ln\left\{
    \frac{\tau_3(v)\tau_{\bar{3}}(v)}{a(v)^{2L}}
    \right\}\right|_{v=0},
 \label{H}
 \end{equation}
where $a(v) = 1-2/v$ \footnote{As usual the overall scale of the
Hamiltonian is at our disposal.}. 
The Hamiltonian \r{H} is not hermitian and its eigenvalues are in general
complex. However, for small systems we find that the ground state and 
low-lying excited states have real eigenvalues. We assume that this
continues to hold true in the thermodynamic limit. Under this assumption
we find that in the thermodynamic limit low lying excited states can
be understood in terms of a conformal field theory and we can extract
the central charge and scaling dimensions from the finite-size
energies for the ground state and low-lying excited states.

The Hamiltonian (\ref{H}) contains a piece proportional to the $sl(2/1)$
invariant product \cite{SNR} $\sum_j\vec{S}_j.\vec{S}_{j+1}$ where $\vec{S}_j$
are $sl(2/1)$ generators on site $j$, and the contraction defines the
invariant form. In addition to this Heisenberg coupling however, our
Hamiltonian (\ref{H}) contains more complicated terms involving three
neighbouring sites. 
 
The eigenvalues and eigenvectors of the transfer matrices were constructed by
Links and Foerster in \cite{LiFo99} by means of he graded quantum inverse
scattering method \cite{SklKul,Kul}. As usual different choices of reference
state lead to different forms for the Bethe ansatz equations
\cite{lai:74,suth:75,Kul,schl:87,esko:92,fk,eks94a,ek94b,PfFr,Frahm99}. In the
following we will analyze the Bethe ansatz equations for the choice of
grading ``$[010]$'' 
% are of the form \cite{LiFo99} 
%\begin{eqnarray}
% \left\{\frac{u_i-1}{u_i+1} \right\}^L &=& \prod_{\beta=1}^M 
%\frac{u_i-\gamma_\beta-1}{u_i-\gamma_\beta+1}\ ,\quad i=1,\ldots,N
%\nonumber\\
% \left\{ \frac{\gamma_\alpha-\lambda-1}{\gamma_\alpha-\lambda+1}
% \right\}^L &=& \prod_{k=1}^N 
%\frac{\gamma_\alpha-u_k-1}{\gamma_\alpha-u_k+1}\ , 
% \quad \alpha=1,\ldots,M\ .  
%\label{bae010} 
%\end{eqnarray}
which read
\begin{align}
   \left( \frac{u_j+i}{u_j-i} \right)^L &= \prod_{\beta=1}^M
   \frac{u_j-\gamma_\beta+i}{u_j-\gamma_\beta-i}\ ,\quad j=1,\ldots,N
 \notag\\
   \left(\frac{\gamma_\alpha+i\lambda+i}{\gamma_\alpha+i\lambda-i}
	  \right)^L &= \prod_{k=1}^N
   \frac{\gamma_\alpha-u_k+i}{\gamma_\alpha-u_k-i}\ ,
	  \quad \alpha=1,\ldots,M\ .
 \label{bae010}
\end{align}
Equations (\ref{bae010}) coincide in fact with the BAE describing the spectrum
of the so-called ``Quantum Transfer Matrix'' of the 1D supersymmetric
$t$--$J$ model % for
%vanishing chemical potential and magnetic field as studied in
%Ref.~\cite{JuKS97} where the inhomogeneity $\lambda$ is related to the ratio
%$\beta/L$ ($L$ being the Trotter number) 
-- see Appendix~\ref{app:JKS}.

Each solution of (\ref{bae010}) parametrizes an eigenstate with quantum
numbers $B=(N-M)/2$ and $J^3=L-(M+N)/2$ (we use the notation $B\equiv 
B_{0}+\bar{B}^{0}$ and $J^{3}=J_{0}^{3}+\bar{J}_{0}^{3}$).  The eigenvalues of the transfer
matrices (\ref{transferM}) are \cite{LiFo99}
\begin{align}
   \Lambda(v) &=
     \left[ a(v) \right]^L \prod_{k=1}^N a(1+iu_k-v)
   + \left[ a(1+\lambda-v) \right]^L \prod_{\beta=1}^M
     a(v-i\gamma_\beta)
 \notag\\
   &- \prod_{k=1}^N a(1+iu_k-v)\prod_{\beta=1}^M a(v-i\gamma_\beta)\ ,
\notag\\ \label{Lam33b}\\
   \bar{\Lambda}(v) &= 
     \left[ a(v) \right]^L \prod_{\beta=1}^M
     a(i\gamma_\beta+1-\lambda-v)
   + \left[ a(1-\lambda-v) \right]^L \prod_{k=1}^N a(v+\lambda-iu_k)
 \notag\\
   &- \prod_{\beta=1}^M a(i\gamma_\beta+1-\lambda-v)
     \prod_{k=1}^N a(v+\lambda-iu_k)\ .
 \notag
\end{align}
The eigenvalues of the Hamiltonian (\ref{H}) are obtained by taking the
appropriate logarithmic derivatives of the eigenvalues of the transfer matrix
\begin{equation}
   E(\{u_k\},\{\gamma_\beta\}) = -\left.\partial_v\ln\left\{
    \frac{\Lambda(v)\bar\Lambda(v)}{a(v)^{2L}}
    \right\}\right|_{v=0}
   = -\sum_{k=1}^N \frac{2}{u_k^2+1} 
   - \sum_{\beta=1}^M \frac{2}{(\gamma_\beta+i\lambda)^2+1}\ .
 \label{energy}
\end{equation}
Similarly the eigenvalues of the momentum operator are obtained are
 \begin{equation}
   P(\{u_k\},\{\gamma_\beta\}) = \frac{i}{2}\ln\left\{
    \frac{\Lambda(v)\bar\Lambda(v)}{a(v)^{2L}}
    \right\}\bigg|_{v=0}=
   i\sum_{k=1}^N \ln\left(\frac{u_k+i}{u_k-i}\right)
   +i\sum_{\beta=1}^M
    \ln\left(\frac{\gamma_\beta+i\lambda+i}{\gamma_\beta+i\lambda-i}\right)\
    . 
 \label{mtm}
 \end{equation}
In the definition of the momentum we have taken into account that the
underlying translational symmetry is by {\sl two} sites of the lattice.

%%%%%%%%%%%%%%%%%%%%%%%%%%%%%%%%%%%%%%%%%%%%%%%%%%%
\subsection{The ground state of the lattice model}
\label{ssec:gs}
%%%%%%%%%%%%%%%%%%%%%%%%%%%%%%%%%%%%%%%%%%%%%%%%%%%
In order to gain some insight in the structure of the spectrum of the
Hamiltonian \r{H} we have diagonalized the transfer matrix and the
Hamiltonian explicitly on small lattices. For $L=1$ and $2$ (i.e.\ 2
and 4 sites) this was done analytically and for $L=3,4,5$ 
(corresponding to 6,8,10 sites respectively) numerically. The results
of this analysis are summarized in Appendix \ref{sec:small}.

Perhaps the most striking result of this analysis is that the space of true
singlets (i.e. states annihilated by all the generators of the
algebra, which are not the image of any other state under the action
of one of these generators) appears to be one dimensional. 
This is not entirely obvious: although the superdimension of the
tensor product is always equal to one, it is a priori conceivable that
there are more singlets, and that the superdimension one results 
from cancellations. But the result can be established rigorously (see
Appendix~\ref{app:algebra}). Note however that one  can always
combine an invariant (i.e. annihilated by all generators) state that
is not the image of any other state with another invariant state, and
still get a state that is not the image of any other state. In that 
respect, although the singlet appears only once in the decomposition 
of the spectrum into representations of $sl(2/1)$, the exact formula 
for this singlet depends on the Hamiltonian. 

Based on our results for small systems (see Appendix \ref{sec:small})
we conjecture the following form for the eigenvalues of the transfer
matrices $\tau_3(v)$, $\tau_{\bar{3}}(v)$ of this singlet state
\begin{equation}
  \Lambda_1(v) = \left\{{(v-2)(v+1-\lambda)\over v(v-1-\lambda)}\right\}^L\
    ,\quad
  \bar{\Lambda}_1(v) = \left\{{(v-2)(v+1+\lambda)\over v(v-1+\lambda)}\right\}^L\ .
\end{equation}
Interestingly, there exists no non-degenerate solution of the BAE
(\ref{bae010}) giving rise to these eigenvalues.  Applying a twist in
the boundary conditions for small systems and studying the evolution
of BA roots as this twist goes to zero we were able to verify,
however, that the singlet is in fact described by the \emph{degenerate}
solution $u_k\equiv \lambda$, $\gamma_\alpha\equiv 0,
k,\alpha=1,\ldots,L$ which clearly
gives $\Lambda_1(v)$ from the general form (\ref{Lam33b}).  This
finding coincides with the observation in \cite{JuKS97} regarding the
distribution of BA-roots for the ground state of the quantum transfer
matrix for the $t$--$J$ model in the thermodynamic limit.  The
eigenvalue of the full transfer matrix corresponding to the singlet
state is
\begin{equation}
 \Lambda(v)=\Lambda_1(v)\bar{\Lambda}_1(v)=
 \left(\frac{(v-2)^2(v+1-\lambda)(v+1+\lambda)}{v^2(v-1-\lambda)(v-1+\lambda)}
 \right)^L .
 \label{Lamsinglet}
\end{equation}
For $\lambda=0$ this state is the ground state of the Hamiltonian with
energy 
\begin{equation}
 E_0=-{4} L\ .
\end{equation}
The energy of this singlet is proportional to $L$ without any finite size
corrections which implies that the central charge of the conformal field
theory arising in the scaling limit of the lattice model is zero
\begin{equation}
  c=0\ .
\end{equation}
%%HF States with vanishing values of the $U(1)$ charge are thus organized in non
%%HF invariant states, invariant states embedded in indecomposable
%%HF representations, and a singlet.

%To prove the statement, observe first that all states with $S^{z}=b=0$ 
%in the product $\left(3\otimes \bar{3}\right)^{L}$ 
%have to be bosonic since they are necessary built out of {\sl even} 
%numbers of states with $S^{z}=\pm 1/2$ (bosons) and states with 
%$S^{z}=0$ (fermions). In the tensor product, we will have in general 
%typical representations which all have vanishing superdimension, and 
%then atypical representations. The irreducible ones come in pairs
%$[\pm j,j]$ (for $j\neq 0$)  with   the same 
%superdimension (a result of the invariance under $b\rightarrow -b$). 
%Suppose that for a given $L$, the singlet $[0,0]$ 
%appears only once. Recall $3\otimes\bar{3}=[0,0]+[0,1]$. It is easy 
%to check by inspection that the product of $[0,1]$ with a typical 
%representation or 
%an irreducible atypical one different from $[0,0]$
%never leads to a true singlet. I think 
%the same is true for the product of $[0,1]$ with an indecomposable 
%block - it can presumably be shown by a limiting procedure, 
%obtaining the indecomposable block as a mix of two irreducible 
%representations in the proper limit. If so, we are left with the products of $[0,0]$, and since this 
%is a singlet, it can give rise to a singlet only once, through the 
%product $[0,0]\otimes [0,0]=[0,0]$. 

We now turn to a discussion of Bethe ansatz results on the low-lying
excitations of the lattice system. We present a rather short discussion here,
and refer to Appendix \ref{latticemodel} for the details.

%%%%%%%%%%%%%%%%%%%%%%%%%%%%%%%%%%%%%%%%%%%%%%%%%%%%%%%%%%%%%%%%%%%%%% Based
\subsection{Excitations}
 \label{sec:strange}
%%%%%%%%%%%%%%%%%%%%%%%%%%%%%%%%%%%%%%%%%%%%%%%%%%%%%%%%%%%%%%%%%%%%%% Based
Based on an analysis of the analytical properties of the bare scattering phase
shifts appearing on the right hand sides of the Bethe ansatz equations
(\ref{bae010}) we are able to classify the possible root configurations, so
called strings, arising in their solution (see Appendix \ref{sec:strings}).
We identify the root configurations giving rise to low energy excited states
by considering small systems.
We find that many low-lying states are built from two types of
complexes involving one spectral parameter $u$ and one spectral
parameter $\gamma$ each (the $n=1$ ``strange string'' configurations,
see (\ref{strings6}) and (\ref{strings7})) 
\begin{equation}
\label{sstr2a}
\begin{aligned}
{\rm type\ -}:\ u^{(-)}&= x-\frac{i}{2}+\epsilon_-\ ,\qquad  
\gamma^{(-)}=x+\frac{i}{2}+\epsilon_-^*\ ,\\
{\rm type\ +}:\ u^{(+)}&= x+\frac{i}{2}+\epsilon_+\ ,\qquad  
\gamma^{(+)}= x-\frac{i}{2}+\epsilon_+^*\ .
\end{aligned}
\end{equation}

A particular class of low-lying excited states is obtained by
combining $N_+$ strange strings of type $+$ with
$N_-$ strange strings of type $-$
\begin{equation}
\label{sstr2}
\begin{aligned}
u_j^{(-)}&= x_j-\frac{i}{2}+\epsilon_{j,-}=
\left(\gamma_j^{(-)}\right)^*\ , \quad\quad j=1,\ldots,N_-\ ,\\
u_k^{(+)}&= y_k+\frac{i}{2}+\epsilon_{k,+}=
\left(\gamma_k^{(+)}\right)^*\ , \quad\quad k=1,\ldots,N_+\ .
\end{aligned}
\end{equation}

Our numerical results indicate that the corrections $\epsilon_\pm$ to the
``ideal'' string solutions are of order $1/L$.  Each of the two
complexes \r{sstr2a} constitutes an acceptable, non self-conjugate,
solution of the Bethe ansatz equations (\ref{bae010}).  Such solutions
have been encountered in other studies of non unitary Hamiltonians,
like the $sl(3)$ integrable spin chain \cite{SW}.

To make contact to the analysis of the WZW model we have to determine
the spectrum of scaling dimensions for the field theory describing the
scaling limit of the lattice model.  Conformal invariance implies that
the dimensions can be extracted from a finite size scaling analysis of
the low-lying energies of the latter.  Before evaluating these
energies in large, finite systems it is very useful to consider an
infinite volume first. In the thermodynamic limit we can neglect the
deviations $\epsilon^\pm$ from the ideal string solutions as they do
not contribute to the leading ${\cal O}(1)$ behaviour of the
energies. Then, using (\ref{sstr2}) in (\ref{bae010}) and multiply the
equations for the components of a strange string we arrive at the
following set of equations involving only the real centres of the
strange strings
\begin{equation}
\begin{aligned}
   &\left(\frac{x_j+\frac{3i}{2}}{x_j-\frac{3i}{2}}\
   \frac{x_j+\frac{i}{2}}{x_j-\frac{i}{2}}\right)^L=-
   \prod_{k=1}^{N_+}\left(\frac{x_j-x_k+2i}{x_j-x_k-2i}\right)
   \prod_{k=1}^{N_-}\left(\frac{x_j-y_k+i}{x_j-y_k-i}\right)^2 ,
   \quad j=1,\ldots, N_+, \\
 &\left(\frac{y_j+\frac{3i}{2}}{y_j-\frac{3i}{2}}\
   \frac{y_j+\frac{i}{2}}{y_j-\frac{i}{2}}\right)^L=-
   \prod_{k=1}^{N_+}\left(\frac{y_j-x_k+i}{y_j-x_k-i}\right)^2
   \prod_{k=1}^{N_-}\left(\frac{y_j-y_k+2i}{y_j-y_k-2i}\right),
   \quad j=1,\ldots, N_-.
\end{aligned}
\label{baesstr}
\end{equation}
The algebraic equations \r{baesstr} can be turned into coupled
integral equations for root densities in the limit $L\to\infty$,
$N_\pm\to\infty$, $\frac{N_\pm}{L}=n_\pm$ fixed
(see Appendix \ref{app:inteqs}) 
\begin{equation}
\label{igl:rhos}
\begin{aligned}
   &\rho_+(x) + \rho^h_+(x) + a_4*\rho_+\bigr|_x
	     + 2a_2*\rho_-\bigr|_x   = a_1(x) + a_3(x)\ ,
 \\
   &\rho_-(y) + \rho^h_-(y) + a_4*\rho_-\bigr|_y
	     + 2a_2*\rho_+\bigr|_y   = a_1(y) + a_3(y)\ .
\end{aligned}
\end{equation}
Here $\rho_{\pm}(x)$ are the root densities of the two types of
strings (\ref{sstr2}), $\rho^h_{\pm}$ are the corresponding densities
of `holes' in these distributions,
\begin{equation}
   a_n(u)=\frac{1}{\pi}\ \frac{2n}{4u^2+{n^2}}\ ,
\end{equation}
and $*$ denotes a convolution, i.e.
\be
a_n*f\bigr|_x=\int_{-\infty}^\infty dx'\ a_n(x-x')\ f(x')\ .
\ee
The energy per site of such states in the thermodynamic limit is
obtained from (\ref{energy})
\begin{equation}
 \label{ess}
 \begin{aligned}
 \lim_{L\to\infty}\frac{1}{L}{E_{N_+,N_-}} &=-2\pi\sum_{\sigma=\pm}\int dx\
     \rho_\sigma(x)\left[a_1(u)+a_3(u)\right].
 \end{aligned}
\end{equation}
The eigenstates of lowest energy within this class are obtained by filling the
states with negative energy.  Such states correspond to distributions with
$\rho_\sigma(x)= 0$ ($\rho^h_\sigma(x)=0)$ for $|x|>A_\sigma$
($|x|<A_\sigma$).  They are characterized by the total density of type-$\pm$
strange strings
\begin{equation}
    \frac{N_\pm}{L} = \int_{-A_\pm}^{A_\pm} dx \rho_\pm(x)\ .
\end{equation}
For the low-lying multiplets identified for small systems in the
Appendix~\ref{ssec:config} we have $N_++N_-\approx L$. Due to the
symmetry of the equations we can restrict our attention to the case
$N_+\ge N_-$.
Within the description in terms of string densities $\rho_\pm$ these states
correspond to the choice of $A_+=\infty$: it is straightforward to see in this
case that ${N_{+}+N_{-}\over L}=1$ independent of $A_-$.

We can now proceed and apply standard Bethe ansatz methods to extract
the finite size energy gaps $\Delta E$ of these low-lying states with
respect to the completely filled Fermi seas of type-$\pm$ strange
strings (corresponding to $A_\pm=\infty$).
Note that to extract the scaling dimensions from this spectrum
according to the finite size scaling predictions of conformal field
theory the gaps have to be measured with respect to the true ground
state discussed in Section \ref{ssec:gs}.  This offset between the
lowest excitation considered in this section and the ground state
energy $E_0$ is not accessible to the analytical methods applied below
and will be determined by numerical solution of the Bethe equations.

%%%%%%%%%%%%%%%%%%%%%%%%%%%%%%%%%%%%%%%
\subsubsection{Excitations with $N_+=N_-$}
%%%%%%%%%%%%%%%%%%%%%%%%%%%%%%%%%%%%%%%

The particular subset of solutions discussed above with $N_+=N_-$
needs to discussed separately.  Let us first consider the
thermodynamic limit.  Within the integral equation approach the lowest
energy state of this type is described by $A_+=A_-\equiv A$ and the
densities of $+$ type and $-$ type strings are found to coincide.
Introducing $\rho_+(x)=\rho_-(x)\equiv \rho(x)$ one is left with a
single integral equation instead of (\ref{igl:rhos}). For states with
$N_+=N_-\approx L/2$ such that
\begin{equation}
   \lim_{L\to\infty} \frac{N_\pm}{L}=\frac{1}{2}
\end{equation}
this integral equation reads
\begin{equation}
     \rho(x)+\int_{-\infty}^\infty dx^\prime\ \left[2\
     a_2(x-x^\prime)+a_4(x-x^\prime)\right]
     \rho(x^\prime)=a_1(x)+a_3(x)\ ,
     \label{rhooct}
\end{equation}
which is solved by Fourier transformation giving
\begin{equation}
   \rho(x) = \frac{1}{2\cosh \pi x}\ .
 \label{rhopm}
\end{equation}
Substituting the solution of the integral equation into (\ref{ess})
we then find that the energy for states with ${N_{\pm}\over L}=\frac{1}{2}$
\be
\lim_{L\to\infty}\frac{E_{N_+,N_-}}{L} = -4\ .
\ee
All states (of the type we are considering) 
with $\frac{N_{+}}{L}=\frac{N_{-}}{L}\approx\frac{1}{2}$ 
are thus degenerate with the
singlet ground state in the thermodynamic limit. 

Let us now turn to solutions of the Bethe ansatz equations in the
finite volume.  Solving these equations for finite systems we observe
that an important role is played by root configurations in which each
strange-string of type $+$ becomes ``degenerate'' with a
strange-string configuration of type $-$.  After suitable relabelling
of the indices this implies $y_j=x_j$ for $j=1,\ldots,N_+$ in
(\ref{sstr2}).
%\begin{equation}
%\begin{aligned}
%u_j^{(-)}&= x_j-\frac{i}{2}+\epsilon_{j}=
%\left(\gamma_j^{(-)}\right)^*\ , \\
%u_j^{(+)}&= x_j+\frac{i}{2}+\epsilon_{j}=
%\left(\gamma_j^{(+)}\right)^*\ .
%\end{aligned}
%\end{equation}
%
In such a configuration the spectral parameters of the two levels of
the Bethe ansatz equations (\ref{bae010}) coincide 
\be
\label{sstr2:deg}
u_j^{(-)}=\gamma_j^{(+)}\ ,\qquad u_j^{(+)}=\gamma_j^{(-)}\ .  
\ee 
We will analyze such solutions to the Bethe ansatz equations next.

%%%%%%%%%%%%%%%%%%%%%%%%%%%%%%%%%%%%%%%%%%%%%%%%%%%%%%%%
\subsection{The Takhtajan-Babujian subset}
%%%%%%%%%%%%%%%%%%%%%%%%%%%%%%%%%%%%%%%%%%%%%%%%%%%%%%%%
The results of our numerical solution indicate that such
``degenerate''solutions are a subset of an even larger class of
solutions to the Bethe ansatz equations (\ref{bae010}) in sectors with
same number of roots on both levels, i.e.\ $N=M$.  These solutions are
obtained by setting all spectral parameters $u_j$ on the first level
equal to the spectral parameters on the second level $\gamma_j$,
i.e. by requiring that 
\be u_j\equiv \gamma_j\ ,\qquad j=1,\ldots N.
\ee 
The roots then have to satisfy
\begin{equation}
    \left({u_{j}+i\over 
       u_{j}-i}\right)^{L}=\prod_{k=1}^{n}{u_{j}-u_{k}+i\over
       u_{j}-u_{k}-i}\ .
\end{equation}
This system of equations is identical with the Bethe equations for the
spin-1 antiferromagnetic Takhtajan-Babujian (TB) chain with $L$ lattice
sites and {\sl antiperiodic} boundary conditions \cite{BabuT,AM}
\begin{equation}
\label{H_TB}
  H^{TB} = \sum_{j=1}^L \left( \mathbf{S}_j.\mathbf{S}_{j+1} 
                  - (\mathbf{S}_j.\mathbf{S}_{j+1})^2 \right)\ .
\end{equation}
Here $\mathbf{S}_j$ are $sl(2)$ spin-1 operators.
The degenerate strange string solutions (\ref{sstr2}) with $N_+=N_-$
and (\ref{sstr2:deg}) discussed above become the 2-strings $u_j^\pm =
u_j \pm i/2$ in the Takhtajan-Babujian-classification.
We note that the normalization of the Hamitonian \r{H_TB} is such
that the gaps in our system will be {\sl twice} the gaps in the TB chain.
Alcaraz and Martins have carried out a finite-size scaling analysis of
the anisotropic TB chain with general twisted boundary conditions
\cite{AM} and identified the operator content of the underlying
conformal field theory. We summarize some of their results in Appendix
\ref{app:spin1}.

Boundary conditions must be handled with extreme care. This is
especially true in theories with super group symmetries: since one
deals with graded tensor products, periodic and antiperiodic boundary
conditions for fermions do not necessarily translate into periodic and
antiperiodic boundary conditions for the spin operators in the chain.

\subsubsection{Periodic boundary conditions}
The sector we are considering here, $N=M$, is simple in that respect: in that
case, since the total spin $J^3$ is integer, we are exploring only bosonic
states. Therefore, periodic boundary conditions for the fermions translate
into antiperiodic boundary conditions for the spins of the
Takhtajan-Babujian chain (for a detailed discussion of boundary
conditions in chains with supergroup symmetries, see \cite{KaufSal}). 

To proceed, we now recall some results about
that chain \cite{AM} (see also Appendix \ref{app:spin1}).  The best
way to encode the part of the spectrum that comes from the
Takhtajan-Babujian chain is to use the generating function of levels -
the conformal partition function \cite{DiFrancesco}. 
Let us introduce the quantities 
 \begin{eqnarray}
     \Delta_{nm}={1\over 8}\left[n+(m+\Phi/\pi)\right]^{2}\nonumber\\
     \bar{\Delta}_{nm}={1\over 8}\left[n-(m+\Phi/\pi)\right]^{2}\ .
\end{eqnarray}
We further introduce the partition functions for the Ising model with 
twisted boundary conditions
 \begin{eqnarray}
     Z^{Ising}(0,0)&=&\left|\chi^{Ising}_{0}\right|^{2}+
     \left|\chi^{Ising}_{1/2}\right|^{2}={1\over 2}
     \left({|\theta_{3}|\over |\eta|}+{|\theta_{4}|\over |\eta|}\right)\nonumber\\
     Z^{Ising}(0,1)&=&Z_{Ising}(1,0)=\left|\chi^{Ising}_{1/16}\right|^{2}=
     {1\over 2}{|\theta_{2}|\over |\eta|}\nonumber\\
     Z^{Ising}(1,1)&=&\chi^{Ising}_{0}\bar{\chi}^{Ising}_{1/2}+
     \bar{\chi}^{Ising}_{0}\chi^{Ising}_{1/2}={1\over 2}
     \left({|\theta_{3}|\over |\eta|}-{|\theta_{4}|\over |\eta|}\right)
\end{eqnarray}
 The generating function  of the levels in the Takhtajan-Babujian chain is then
 \begin{equation}
     Z_{even}^{TB}=\sum_{r=0}^{1}\sum_{s=0}^{1}Z^{Ising}_{r,s}\sum_{
     \begin{array}{c}
     m=r+2Z \\
     n=s+2Z\end{array}}
     q^{\Delta_{nm}-1/24}\bar{q}^{\bar{\Delta}_{nm}-1/24}\label{genfct}
 \end{equation}
and coincides with the partition function of the 
$SU(2)$ WZW model  at level two \cite{DiFrancesco}. 

We notice that, in the equivalence between the Takhtajan-Babujian spectrum and a 
subset of our $sl(2/1)$ integrable spin chain spectrum, 
the Takhtajan-Babujian model has an even number of sites if the $sl(2/1)$ model has 
an even number  $L$ of 3 (and thus $\bar{3}$) representations. 
Expression (\ref{genfct}) is relevant to this case only.  One can 
also consider the case of this number being odd. Then 
the generating function of levels reads 
 \begin{equation}
     Z_{odd}^{TB}=\sum_{r=0}^{1}\sum_{s=0}^{1}Z^{Ising}_{r,s}\sum_{m=r+2Z+1,n=s+2Z+1}
     q^{\Delta_{nm}-1/24}\bar{q}^{\bar{\Delta}_{nm}-1/24}\label{genfctI}
 \end{equation}
%
%(In this case, the 
%corresponding modular invariant is obtained by summing the odd-odd, 
%odd-even and even-odd contributions, and reads
%
% \begin{eqnarray}
%     Z=\sum_{r=0}^{1}\sum_{s=0}^{1}Z^{Ising}_{r,s}
%     \left(\sum_{
%    \begin{array}{c}
%     m=r+1+2Z\\
%     n=s+1+2Z\end{array}}+\sum_{\begin{array}{c}     
%     m=r+1+2Z
%     \\n=s+2Z\end{array}}\right.\nonumber
%     \left.+\sum_{\begin{array}{c}
%     m=r+2Z\\
%     n=s+1+2Z\end{array}}\right)
%     q^{\Delta_{nm}-1/24}\bar{q}^{\bar{\Delta}_{nm}-1/24}\label{genfctI}
% \end{eqnarray}
%
%Adding this to the other invariant, and using the self duality of the 
%Ising model, leads to a decoupled object $Z=Z^{Ising}\times 
%Z^{freeboson}$. In algebraic terms, it corresponds to combining the 
%affine $SU(2)$ symmetry with $N=1$ supersymmetry to form an $N=3$ 
%superconformal algebra \cite{Ginsparg}.)

In the case $\Phi=\pi$ however, expressions
(\ref{genfct}) and (\ref{genfctI}) can be shown to coincide!  

Let us now extract some conformal weights for the $sl(2/1)$ chain 
from these expressions. To do so, one has to 
\begin{itemize}
\item{} {\sl Double} the gaps, because the normalizations of the
  Hamiltonians are such that the energies of the $sl(2/1)$ chain are
  twice the energies of the TB chain.
\item{} Shift the energy differences by a constant $\frac{\pi^2}{2L}$
  (after the rescaling). This shift in necessary, because the
  finite-size spectrum in the TB chain with antiperiodic boundary
  conditions is calculated with respect to the ground state with
  {\sl $L$ even} and {\sl periodic boundary conditions} \cite{AM}. The
energy difference between the true ground state and the TB ground
state for even $L$ and periodic boundary conditions is equal to
$\frac{\pi^2}{2L}$. 
\end{itemize}

For $r,s=0$, we get from (\ref{genfct}) (that is $L$ even)
dimensions in the $c=0$ theory which read
\begin{eqnarray}
    h+\bar{h}&=&{n^{2}\over 2}+{(m+1)^{2}\over 2}-{1\over 4}\nonumber\\
    h-\bar{h}&=&n(m+1);~~~n,m\in~ 2Z
\end{eqnarray}
Hence we have the possibility $n=0$, $m=0,m=-2$, giving rise to 
$h=\bar{h}={1\over 8}$. All  other scalar operators have  
$h=\bar{h}={1\over 8}+\hbox{integer}$. The lowest non scalar  operator
corresponds to $n=2,m=-2$ with $\bar{h}=17/8=1/8+2,{h}=1/8$. 

For $r=1$  and $s=0$ 
\begin{eqnarray}
    h+\bar{h}&=&{n^{2}\over 2}+{(m+1)^{2}\over 2}\nonumber\\
    h-\bar{h}&=&n(m+1);~~~n\in~ 2Z,~m\in~ 2Z+1
\end{eqnarray}
 Here all  the dimensions are integers.  
 For $r=0$  and $s=1$,
 \begin{eqnarray}
     h+\bar{h}&=&{n^{2}\over 2}+{(m+1)^{2}\over 2}\nonumber\\
     h-\bar{h}&=&n(m+1);~~~n\in~ 2Z+1,~m\in~ 2Z
 \end{eqnarray}
with again only integer dimensions. Finally, the case $r=s=1$ leads to
 \begin{eqnarray}
     h+\bar{h}&=&{n^{2}\over 2}+{(m+1)^{2}\over 2}+{3\over 4}\nonumber\\
     h-\bar{h}&=&n(m+1)+1;~~~n\in~ 2Z+1,~m\in~ 2Z+1
 \end{eqnarray}
 The case $n=1,m=-1$ corresponds to $h=9/8,\bar{h}=1/8$. 
 
 In fact, the  generating function of the $sl(2/1)$ dimensions deduced from 
 the TB spectrum can be written 
 after a few manipulations as the sum of two simple contributions
 \begin{equation}
     Z^{TB~subset}_{1/8}=2q^{1/8}
     \prod_{n=1}^{\infty}(1+q^{n})(1+q^{2n})\times(q\rightarrow\bar{q})
     \end{equation}
 and
 \begin{equation}
     Z^{TB~subset}_{0}=
     \prod_{n=0}^{\infty}(1+q^{n+1})(1+q^{2n+1})\times(q\rightarrow\bar{q})
     \end{equation}
Of course these two contributions bear resemblance to the two Ramond 
characters discussed in the first section. However,  it is important
to understand that we cannot deduce from these 
 the degeneracies of the $sl(2/1)$ chain. Rather, all we can deduce 
 form the argument is that the 
 the dimensions appearing in the expansion of these two objects 
 appear also in the $sl(2/1)$ spectrum, with a multiplicity at least 
 equal to the multiplicity it has in this expansion. We denote such a 
 relationship by 
 \begin{eqnarray}
     Z^{TB~subset}_{1/8}&\leq Z^{sl(2/1)~chain}_{1/8}&\nonumber\\
     Z^{TB~subset}_{0}&\leq Z^{sl(2/1)~chain}_{0}\label{info}
  \end{eqnarray}
It can be checked that this is compatible with the spectrum of the 
chain containing at least the modulus square of the integrable 
characters, that is 
 \begin{eqnarray}
    Z^{TB~subset}_{1/8}&\leq &|\chi_{h=1/8}^{R,I}|^{2}\nonumber\\
    Z^{TB~subset}_{0}&\leq &|\chi_{h=0}^{R,IV}|^{2}\label{infoi}
 \end{eqnarray}

 We can make some more detailed guess about which states on right are 
 represented by a term on the left in (\ref{info}), (\ref{infoi}). For
 instance,  $Z^{TBsubset}_{1/8}$ 
 contains two fields with dimension $h=\bar{h}=1/8$, and we know that these 
 fields must have $B=0$. These fields must also have vanishing $SU(2)$ 
 spin, since they are obtained by taking the $n$ number of the Takhtajan-Babujian 
 chain equal to zero. This means that their $SU(2)$ spin in the 
 $sl(2/1)$ chain must also vanish,  and thus, in the continuum limit, 
 they  must be the symmetric and antisymmetric combinations 
 of $J^{3}_{0}=1/2,~\bar{J}^{3}_{0}=-1/2$ and $\bar{J}^{3}_{0}=-1/2,~
 J^{3}_{0}=1/2$. The fields 
 with $h=1/8$ and $J^{3}_{0}=\bar{J}^{3}_{0}=0$ are, in particular,  absent in the 
 subspectrum obtained from the Takhtajan-Babujian chain.  But we know they
 are there 
 in the full spectrum.
 
 Similarly, $Z_{0}^{TB subset}$ contains a single field with $h=\bar{h}=1$;
 once again it must have $B=J^3=0$. On the right hand side, we have $8\times
 8$ fields with such weights, obtained by tensoring the adjoint with
 itself. It is highly likely that the field on the left comes from the part of
 this tensor product where the $J^3_{0}=0,B_{0}=0$ field for right movers is tensored with
 the $\bar{J}^3_{0}=0,\bar{B}_{0}=0$ field for left movers. 

In Tables \ref{table:spin1-odd} and \ref{table:spin1-even} we
summarize the lowest scaling dimensions (that is, $x=h+\bar{h}$, 
irrespective of whether $h=\bar{h}$ or not) in the TB subset and their
$sl(2/1)$ quantum numbers for $L$ even and $L$ odd.
\begin{table}[ht]
\begin{center}
\begin{tabular}{|l|l|l|l|l|}
\hline
& (B,S)=(0,0) & (B,S)=(0,1) & (B,S)=(0,2) & (B,S)=(0,3)\\ \hline\hline
m=-1 & 0 & $\frac{1}{4}$,$\frac{9}{4}$ & $2$ &
$\frac{17}{4}$,$\frac{25}{4}$\\ \hline
m=0 &$\frac{5}{4}$, $\frac{5}{4}$ & $1$ & $\frac{13}{4}$, $\frac{13}{4}$
 & $5$\\
\hline
m=1 & $2$ & $\frac{9}{4}$,$\frac{17}{4}$ & $4$ &
$\frac{25}{4}$,$\frac{33}{4}$\\ 
\hline
\end{tabular}
\caption{Smallest scaling dimensions and $sl(2/1)$ quantum numbers in
  the TB subset for $L$ odd.}  
\label{table:spin1-odd}
\end{center}
\end{table}
\begin{table}[ht]
\begin{center}
\begin{tabular}{|l|l|l|l|l|}
\hline
& (B,S)=(0,0) & (B,S)=(0,1) & (B,S)=(0,2) & (B,S)=(0,3)\\ \hline\hline
m=-1 & 0 & $\frac{5}{4}$, $\frac{5}{4}$ & $2$ & $\frac{21}{4}$, $\frac{21}{4}$
\\ \hline
m=0  & $\frac{1}{4}$,$\frac{9}{4}$ & $1$ & $\frac{9}{4}$,
$\frac{17}{4}$ & $5$\\ \hline
m=1  & 2 & $\frac{13}{4}$, $\frac{13}{4}$ & $4$ & 
$\frac{29}{4}$, $\frac{29}{4}$
\\ \hline
\end{tabular}
\caption{Smallest scaling dimensions and $sl(2/1)$ quantum numbers in
  the TB subset for $L$ even.} 
\label{table:spin1-even}
\end{center}
\end{table}
We note that these scaling dimensions occur with the same
multiplicities as in the odd $L$ case, but the $sl(2/1)$ quantum
numbers are different.

%%%%%%%%%%%%%%%%%%%%%%%%%%%%%%%%%%%%%%%%%%%%%%%%%%%%%%%%%%%%%%%%%%%%%%
\subsubsection{Antiperiodic boundary conditions and the Neveu Schwarz
  sector}  
%%%%%%%%%%%%%%%%%%%%%%%%%%%%%%%%%%%%%%%%%%%%%%%%%%%%%%%%%%%%%%%%%%%%%%

We can also consider the case of antiperiodic boundary conditions 
for the spin chain in the sector $N=M$. This should correspond to 
antiperiodic boundary conditions for the fermions, ie the NS sector. 
Note that the $sl(2/1)$ symmetry is then {\sl broken}. In Appendix
\ref{app:BC_anti} we summarize some results of a numerical solution of
the Bethe ansatz equations with antiperiodic boundary conditions in
the $sl(2/1)$ chain for small $L$.

The same argument we used above in the periodic case shows that the
$sl(2/1)$ chain with antiperiodic boundary conditions has a particular
sector that is related to the spectrum of the Takhtajan-Babujian chain
with periodic boundary conditions, i.e. $\Phi=0$. This time, the
generating function of the $sl(2/1)$ chain levels can be written as
the sum of two contributions 
\begin{equation}
Z^{TB~subset}_{1/4}=q^{1/4}\prod_{1}^{\infty}(1+q^{2n})^{3}\times
(q\rightarrow\bar{q})\ ,
\end{equation}
and 
\begin{equation}
Z^{TB~subset}_{-1/8}=2q^{-1/8}\times (q\rightarrow\bar{q})\ .
\end{equation}
 Now the correspondence between R and NS weights is as follows
 \begin{eqnarray}
     h_{NS}=h_{R}+{1\over 4}-J^{3,R}_{0}\ ,\nonumber\\
     \bar{h}_{NS}=\bar{h}_{R}+{1\over 4}+\bar{J}^{3,R}_{0}\ .
 \end{eqnarray}
 We thus see that the two fields in $Z^{chain}_{1/8}$ with weight 
 $h_{R}=\bar{h}_{R}=1/8$ should get, respectively, 
 $h_{NS}=\bar{h}_{NS}=-1/8$, and $h_{NS}=\bar{h}_{NS}=7/8=-1/8+1$. The 
 field in the $1/8$ multiplet whose dimension becomes 
 $h_{NS}=\bar{h}_{NS}=3/8=-1/8+1/2$ is the one with 
 $J^{3,R}_{0}=\bar{J}^{3,R}_{0}=0$, and it is not present in the Takhtajan-Babujian 
 spectrum. Thus the form of $Z^{chain}_{-1/8}$ is compatible with the 
 $SU(2/1)$ WZW hypothesis, in the NS sector. 
 
Similarly, the singlet in $Z^{TB~subset}_{0}$ has $h=J^{3}_{0}=0$ and
thus acquires, under spectral flow, a weight
$h_{NS}=\bar{h}_{NS}=1/4$. As for the multiplet with
$h_{R}=\bar{h}_{R}=1$, under spectral flow, it acquires 
$h_{NS}=\bar{h}_{NS}=5/4$. The field with $h_{R}=\bar{h}_{R}=1$ and 
$J^{3,R}_{0}=-\bar{J}^{3,R}_{0}=1/2$ 
which acquires under spectral flow $h_{NS}=\bar{h}_{NS}=3/4=1/4+1/2$ 
is absent from the Takhtajan-Babujian part of the spectrum. 
 
 It is also satisfactory to see that, within the Takhtajan-Babujian part of the 
 spectrum, we have 
 \begin{eqnarray}
     h_{NS}={1\over 4}(n+m)^{2}={1\over 4}(n+m+1)^{2}-{1\over 
     4}-{1\over 2}(n+m)\equiv h_{R}+{1\over 4}-J^{3,R}_{0}\ ,\nn
     \bar{h}_{NS}={1\over 4}(n-m)^{2}={1\over 4}(n-m-1)^{2}-{1\over 
     4}+{1\over 2}(n-m)\equiv h_{R}+{1\over 4}+\bar{J}^{3,R}_{0}\ ,
 \end{eqnarray}
and therefore
 \begin{eqnarray}
     J^{3,R}_{0}={1\over 2}(n+m+1)\ ,\nn
     \bar{J}^{3,R}_{0}={1\over 2}(n-m-1)\ .
 \end{eqnarray}
 Observe in particular that $J^{3,R}_{0}+\bar{J}^{3,R}_{0}=n$, the scalar
 $J^3$ quantum number. All these relations are compatible with our 
 WZW conjecture for the continuum limit of the $sl(2/1)$ chain.

%%%%%%%%%%%%%%%%%%%%%%%%%%%%%%%%%%%%%%%%%%%%%%%%%%%%%%%%%%%%%%%%%%%%%%
\subsection{Strange string excitations with $N_+\ne N_-$}
%%%%%%%%%%%%%%%%%%%%%%%%%%%%%%%%%%%%%%%%%%%%%%%%%%%%%%%%%%%%%%%%%%%%%%

The integral equations (\ref{igl:rhos}) for strange string excitations
with $N_+\neq N_-$ can be analyzed by Fourier transformation. This
allows one to express the Fourier transforms of the hole densities
$\tilde{\rho}^h_\pm(\omega)$ in terms of the Fourier transforms of the
root densities $\tilde{\rho}_\pm(\omega)$. Inverting the resulting
matrix equation, we obtain 
\begin{equation}
\label{igl:rhoh}
\begin{aligned}
    \tilde{\rho}_{+}(\omega)&=\frac{1}{2\cosh(\omega/2)}-{1+e^{2|\omega|}\over 
    4\sinh^{2}\omega}\ \tilde{\rho}_{+}^{h}(\omega)+{e^{|\omega|}\over 
    2\sinh^{2}\omega}\ \tilde{\rho}_{-}^{h}(\omega)\ ,\\
    \tilde{\rho}_{-}(\omega) &={1\over 2\cosh(\omega/2)}-{1+e^{2|\omega|}\over 
	4\sinh^{2}\omega}\ \tilde{\rho}_{-}^{h}(\omega)+{e^{|\omega|}\over 
	2\sinh^{2}\omega}\ \tilde{\rho}_{+}^{h}(\omega)\ .
\end{aligned}
\end{equation}
Here we have defined the Fourier transform by $\tilde{f}(\omega)=\int dx
e^{i\omega x}f(x)$ and $\tilde{a}_{n}(\omega)=e^{-n|\omega|/2}$.
The right-hand side of \r{igl:rhoh} is most conveniently written in
matrix form using
\be
\tilde{R}(\omega)={e^{|\omega|}\over 2\sinh^{2}\omega}\left(
\begin{array}{cc}
\cosh(\omega) & -1 \\
-1 & \cosh(\omega) \\
  \end{array} \right) .
\ee
The matrix $\tilde{R}$ is the Fourier transform of the
resolvent $R=(1-K)^{-1}$ of the (matrix) integral operator in our
original equations (\ref{igl:rhos}) with kernel
\begin{equation}
  K(x-y) = - \left(
  \begin{array}{cc}
  a_4(x-y) & 2 a_2(x-y)\\
  2 a_2(x-y) & a_4(x-y)
  \end{array} \right) .
\end{equation}
Starting from (\ref{igl:rhos}) (or, equivalently, (\ref{igl:rhoh}))
and (\ref{ess}) we can analyze the finite size spectrum of these
excitations by standard methods.  Cases similarly to the one
considered here with complete symmetry between the excitations in the
$+$ type and $-$ type sectors have been studied in
Refs.~\cite{SuzJ,DeVega}. There the following general formula for the
finite size spectral gaps $\Delta E$ of low lying excitations over a
filled Fermi sea with several components in a relativistic invariant
model (i.e.\ with a single Fermi velocity) was derived. In our case it
takes the form
\begin{equation}
\label{finitegap}
  {L\over {2\pi v}}\Delta E = 
    \frac{1}{4}(\Delta N)^{T}\tilde{R}^{-1}(0)\Delta N
    +D^{T}\tilde{R}(0)D\ .
\end{equation}
Here $v=\pi$ is the characteristic velocity of the system and is
calculated in Appendix \ref{app:fermiv},
$\Delta N$ is a 2-component vector $\begin{pmatrix}
\Delta N_{+}
\cr \Delta N_{-}\end{pmatrix}$ where $\Delta N_{\sigma}$ is the
change in the number of roots of type $\sigma$ compared to the ground
state. Similarly, $D$ is the 2-component vector
$\begin{pmatrix}D_{+} \cr D_{-}\end{pmatrix}$
where $D_{\sigma}$ is the number of solutions of type $\sigma$
``backscattered'' from the left to the right of the Fermi sea.  

In our case there is a slight additional subtlety here coming from the fact
that the ground state is made of complex solutions (\ref{sstr2}): in such a
situation it is known that -- even in the usual cases such as $sl(2)$ -- on
top of the Gaussian spectrum, deduced from our general formula
(\ref{finitegap}), the dimensions of other operators may appear, as well as
selection rules.  At the same time it has been observed in these cases that
most of the basic {\sl qualitative} conclusions drawn from an analysis based
on the string picture are valid and provide the basis for a quantitative
description when amended for example by results from a numerical computation
of these gaps (see e.g.\ \cite{AM,FrYF90,FrYu90}).

Applying (\ref{finitegap}) blindly to our case requires a bit of care as the
matrix $R$ has divergent elements. This rather unusual feature can be traced
back to the divergence of the Fourier transform of the kernel describing the
scattering between complexes of type $+$ and $-$ for $N_+=N_-$.  We can
regularize these elements by giving $\omega$ a small value
\begin{equation}
    \tilde{R}(\epsilon)\approx {1\over 2\epsilon^{2}} \left(\begin{array}{cc}
    1+\epsilon+2\epsilon^{2}/3&-1-\epsilon-\epsilon^{2}/6\\
    -1-\epsilon-\epsilon^{2}/6&1+\epsilon+2\epsilon^{2}/3\end{array}\right),~
    \tilde{R}(0)^{-1}=\left(\begin{array}{cc}
    2&2\\
    2&2
    \end{array}
    \right)
\end{equation}
{}From this it follows that
\begin{equation}
     {L\over {2\pi v}}\Delta E =
     {1\over 2}(\Delta N_{+}+\Delta N_{-})^{2}+0\times (\Delta N_{+}-\Delta 
     N_{-})^{2}+{1\over 8} (D_{+}+D_{-})^2+\infty\times 
     (D_{+}-D_{-})^{2}
\label{gaps}
\end{equation}
%While setting $\rho_{+}=\rho_{-}$ gives a well behaved kernel in
%(\ref{rhooct}), each of the kernels present here exhibits a divergence as
%$k\rightarrow 0$. 
A divergence of the kernels similar to the one leading to (\ref{gaps})
here has been encoutered in the study of the $osp(2/2)$ chain (that
is, based on the four dimensional typical representation $[0,1/2]$)
and the study of the antiferromagnetic Potts model.  In both cases,
the divergence of the kernels (in Fourier space) at $\omega=0$ has
been interpreted \cite{JacobsenSaleur} as a manifestation of a
continuous spectrum -- which is equivalent to the infinite degeneracy
of the finite size gaps between states with $N_++N_-={\rm const.}$ in
(\ref{gaps}).

We emphasize that this result is obtained for ${\cal O}(1)$ deviations
in $N_\pm$ compared to the state with equal numbers of the $+$ and $-$
type strings (\ref{rhooct}). In other words,
the result holds for {\sl finite} values of $\Delta N$, $D$.

In fact, the state with the largest possible asymmetry between these
distributions, i.e.\ $N_+=L$, $N_-=0$, can be analyzed in the
thermodynamic limit giving
\begin{equation}
   \rho_+(x) =
   \frac{1}{\sqrt{2}}\ \frac{\cosh\frac{\pi}{2}x}{\cosh\pi x}\ ,\quad
   \rho_-(x) \equiv 0
\end{equation}
for the densities. The energy of this state is
$E_{N_+,0}=-2L(1+\ln2)$ and the state is clearly highly
excited. 
In summary, we expect on the basis of the above analysis that there is
a {\sl macroscopic degeneracy} of the finite size spectrum identified
in the Takhtajan-Babujian subset.

As we have previously mentioned, these results ought to be taken {\sl cum
grano salis} as they have been obtained within the framework of the
string hypothesis. The latter neglects the finite size deviations
$\epsilon_\pm$ in (\ref{sstr2}), which also contribute to order ${\cal
O}(L^{-1})$ to the gaps. Furthermore, in order to extract conformal
dimensions from the finite size spectrum we need to determine the
difference between the ground state energy $E_0$ (see Sect.\
\ref{ssec:gs}) to that of the state $\Delta N_\pm=0=D_\pm$ to 
order $1/L$ which is beyond the scope of the analytical methods
applied here.
In light of these complications and in order to verify the observation
from (\ref{gaps}) that each gap is infinitely degenerated (or
equivalently that the spectrum has a {\sl continuous} component added
to the discrete part identified via the Takhtajan-Babujian mapping), we
have carried out an extensive numerical analysis of the Bethe
ansatz equations (\ref{bae010}) for large, finite $L$. Due to the
presence of strong logarithmic corrections (i.e.\ 
corrections of order $1/(L\ln L)$ to (\ref{finitegap}) we had to study
systems with $L$ up to 5.000, i.e.\ 10.000 lattice sites.

\begin{figure}[ht]
 \begin{center}
 \noindent
 \epsfxsize=0.7\textwidth
 \epsfbox{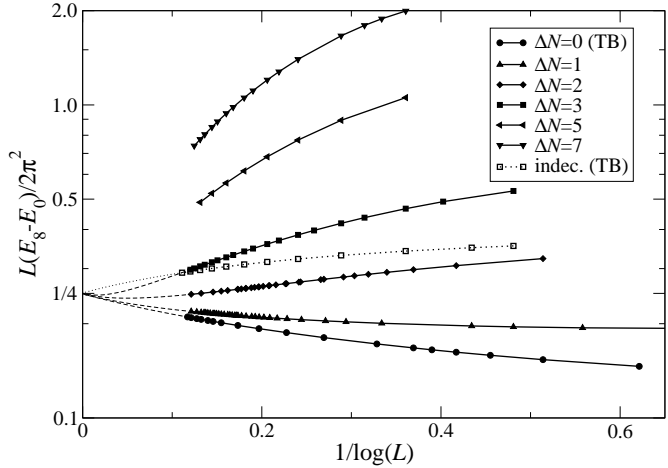}
 \end{center}
 \caption{\label{fig:scaling}
 Corrections to scaling of the energies of the low-lying states studied in
   Section~\ref{sec:ox}.  The dashed lines indicate the rational
   function extrapolation $L\to\infty$.
 }
\end{figure}

We summarize the results of this finite-size scaling analysis for
several low-lying $[0,1]$ octet states ($N_++N_-=L-1$) as well as
the octet and the indecomposable $[0,-1/2,1/2,0]$ representations
($N_+=N_-=L/2$) in the Takhtajan-Babujian sector in Appendix
\ref{sec:ox}. 
The scaling of their energies expected from conformal invariance is
\begin{equation}
\frac{L}{2\pi v} \left[E(N_+,N_-)-E_0\right] \xrightarrow[L\to\infty]{}
h+\bar{h}=x\ ,
\label{fsscaling}
\end{equation}
where $x$ scaling dimension.
We plot the left-hand side of equation \r{fsscaling} as a function of
the lattice length for several different excited states in
Fig.~\ref{fig:scaling}.

\begin{figure}[ht]
 \begin{center}
 \noindent
 \epsfxsize=0.7\textwidth
 \epsfbox{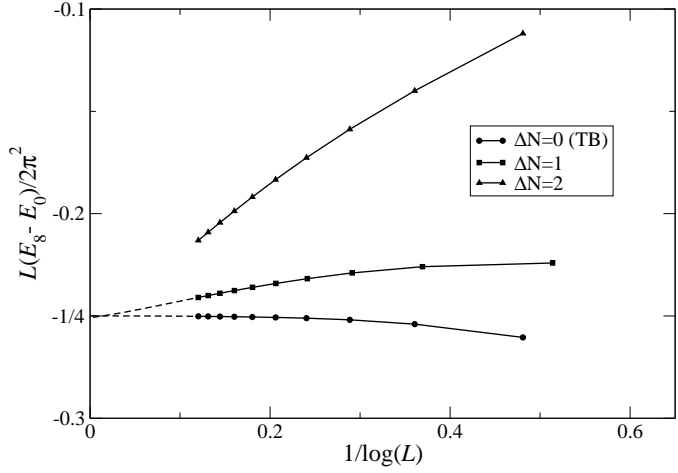}
 \end{center}
 \caption{\label{fig:twist}
 Finite-size scaling of low-lying energy levels for the system  with
 antiperiodic boundary conditions. $E_0$ is the ground state in the
 sector with periodic boundary conditions (see Section \ref{ssec:gs}).
 }
\end{figure}

Within rational function extrapolation of our numerical data we find that
the energies of the states considered all become degenerate to order
$1/L$ in the thermodynamic limit. The scaling dimension obtained from
this extrapolation of \r{fsscaling} is 
\be
x=\frac{1}{4}\ .
\ee
As has been discussed above, we have to extend the analysis of the
finite size gaps to the case of antiperiodic boundary conditions for a
proper identification fo the continuum limit.  For this we have
studied states build from the strings (\ref{sstr2}) in some detail
which again are the relevant configurations for the lowest
excitations.  The details of the numerical analysis are given in the
appendix, here we just present the main results which give some
evidence that again there are are many states degenerate with the
ground state (in the Takhtajan-Babujian sector) in the thermodynamic ($L\to\infty$)
limit.  All of these states have $N_++N_-=L$, but different values of
$\Delta N=N_+-N_-$.  Their scaling dimensions extrapolate to $x=-1/4$
relative to the ground state of the chain with periodic boundary
conditions.  The corrections to this scaling for some states are shown
in Fig.~\ref{fig:twist}.
 
\subsection{Summary}

\begin{figure}[ht]
 \begin{center}
 \noindent
 \epsfxsize=0.2\textwidth
 \epsfbox{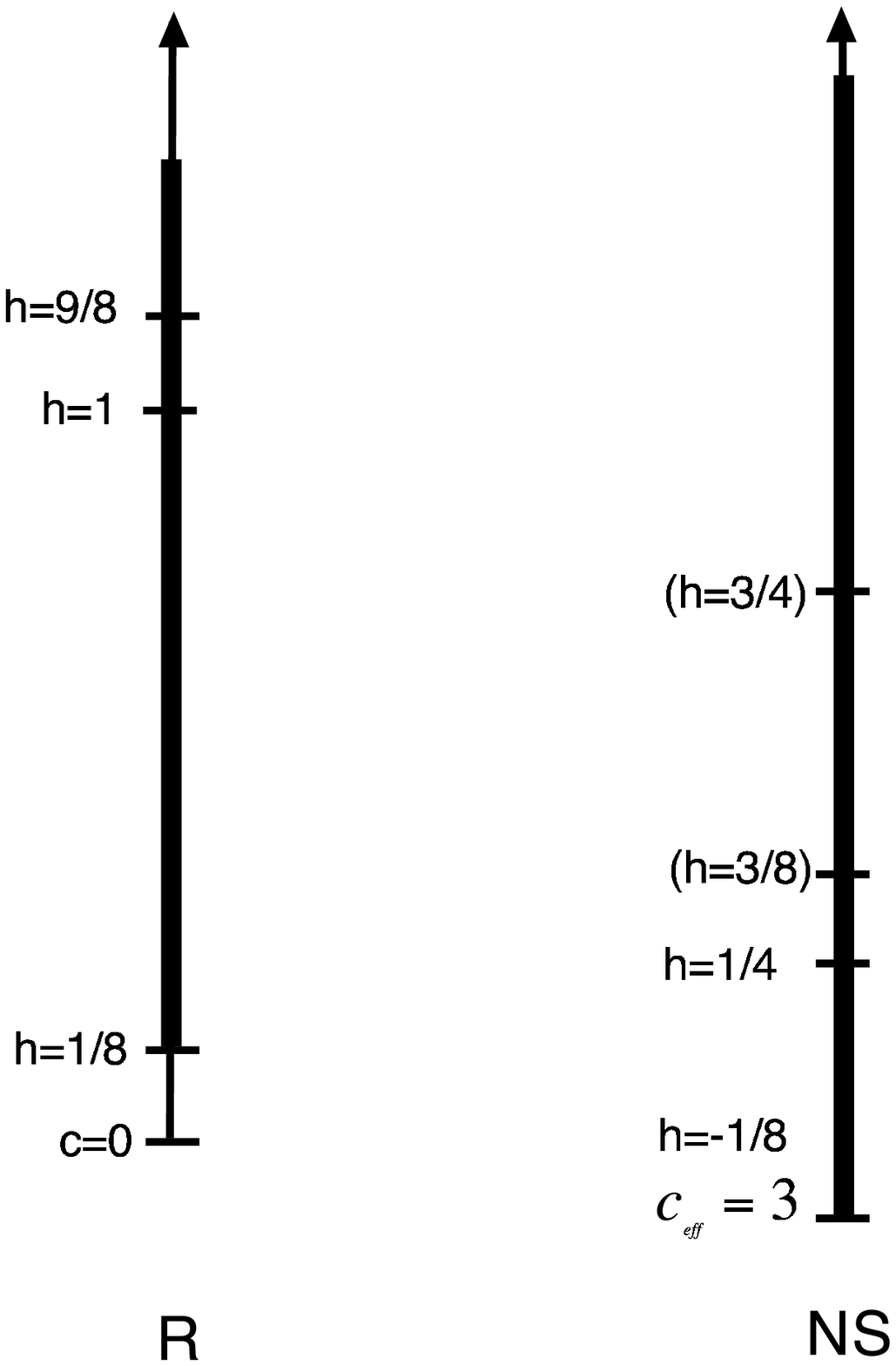}
 \end{center}
 \caption{\label{fig:spectra}
The spectra from the lattice model in the R and NS sector, in 
agreement with the WZW prediction. The thick 
lines represent a continuum of critical exponents (extending to 
infinity). Values in 
parenthesis were expected but beyond our reach. 
 }
\end{figure}

As a summary, we represent on  figure  \ref{fig:spectra} the lowest gaps we 
have observed in the R and NS sectors. Values in parenthes is are 
expected from the study of integrable characters of the WZW model, but 
we have not been able to investigate them numerically.

 \section{Higher level}
 
 It is interesting to consider the same problem with larger 
 representations - more precisely, the fully supersymmetrized 
 atypical representations of type $[\pm j,j]$ ( $s=2j$ boxes in the 
 Young diagram representation). In this 
 case, the Bethe equations allow for a sector reproducing twice the 
 gaps of the spin $s$ Takhtajan-Babujian chain. For the latter, the gaps are obtained 
 once again by combining the gaps from a free boson and disorder 
 operators from a $Z_{2s}$ parafermionic theory. The gaps from the 
 free boson sector read now
 \begin{eqnarray}
     \Delta_{nm}={1\over 
     8s}\left[n+(m+s\Phi/\pi)\right]^{2}\nonumber\\
     \bar{\Delta}_{nm}={1\over 
	  8s}\left[n-(m+s\Phi/\pi)\right]^{2}
 \end{eqnarray}
 The $sl(2/1)$ symmetric case (R sector) corresponds to $\Phi=\pi$. The leading 
 gaps with respect to the $c=0$ theory read then
 \begin{eqnarray}
     h+\bar{h}&=&{(m+s)^{2}\over 2s}-{s\over 2(s+1)}+{r(2s-r)\over 
     2s(s+1)}\nonumber\\
     h&=&\bar{h}
 \end{eqnarray}
 with the constraint that $m=r\hbox{ mod }2s$. The choice $m=r-2s$ 
 gives then
  \begin{equation}
      h=\bar{h}={(s-r)^{2}\over 4(s+1)},~~r=0,\ldots,s-1
  \end{equation}
 This obviously reproduces the conformal weights in the $SU(2/1)$ WZW 
 model for representations $[0,j]$ with $j=0,1/2,\ldots,{s-1\over 2}$ 
 at level $s$. 
 
 Meanwhile in the Neveu Schwarz sector, $c_{eff}={6s\over s+1}$, and 
 for the weights we get the other formula
 \begin{equation}
     h=\bar{h}={m^{2}\over 4(s+1)}+{(m-s+r)(m+s-r)\over 4s(s+1)}
 \end{equation}
 The case $m=r-2s $ now gives 
 \begin{equation}
 h=\bar{h}={(2s-r+1)^{2}\over 4(s+1)}-{1\over 4}
 \end{equation}
This can be written as 
\begin{equation}
    h=\bar{h}={(s-r)^{2}\over 4(s+1)}+{s\over 4}+{s-r\over 2}
\end{equation}
matching the formula for the spectral flow in the $SU(2/1)$ level $s$ 
WZW theory. Hence central charges and TB subset match the WZW 
prediction in this case as well, and it is likely that a more 
thorough study would find similar agreement for the continuum parts of 
the spectrum too.

 %It is particularly interesting to discuss in this light the hull 
 %operator which is a scalar with $h=\bar{h}=5/8$. The only way to get 
 %this field in the WZW model would be to have $j^{2}-b^{2}=5/4$, for 
 %instance with $j=3/2,b=\pm 1$. But the field with $j=3/2$ does not 
 %seem to be allowed since it is not in the $SU(2)_{1}$ model, so we do 
 %not expect to see the hull operator.
 
 %Notice however that one may get an operator for which 
 %$h+\bar{h}={5\over 4}$ with $h=1/8,\bar{h}=9/8$, and spin unity. 

\section{Conclusions and speculations}

To conclude, from the information at our disposal using the Takhtajan-Babujian part 
of the spectrum, 
analytical study of the Bethe ansatz, and 
extensive numerical calculations, we have obtained consistency with the hypothesis 
that the continuum limit is the $SU(2/1)$ level one WZW model. This 
consistency includes the 
set of all dimensions in the R sector, and only a subset of 
dimensions in the   NS sector. It is highly indicative that our 
conjecture is correct that the effective central charge in the NS 
sector is $c=3$. As for the part of the NS spectrum whose weights 
are of the form $h_{NS}=-{1\over 8}+{1\over 2}+\hbox{ integers}$ 
$h_{NS}={1\over 4}+{1\over 2}+\hbox{ integers}$, we have argued they 
are situated outside of the Takhtajan-Babujian part of the spectrum, and thus we 
have nothing to say about them based on this chain of arguments.  It 
would of course be a crucial further test to extract these exponents 
from the Bethe ansatz solution  (this might be difficult as there are 
sitting inside the continuum). We note that it would 
 be extremely useful to compare more thoroughly than we have done 
 the group content of 
 the lattice and field theory models: this however may be a very non 
 trivial exercise, in particular because fine structures of the 
 representations (such as indecomposability) may not evolve 
 straightforwardly in the continuum limit process \cite{Woynarovich}.

One of the main difficulties we have encountered is the the lack of 
understanding of the WZW theory itself. In particular, the free field 
representation as well as algebraic arguments point to the existence 
of a spectrum of dimensions unbounded from below, and maybe continuous. 
While this is somewhat expected in view of the unboundedness of the 
action in the field theory, it is not totally  clear how this should 
be related to the spectrum and partition function of  
a  lattice model. We have suggested that some sort of analytical 
continuation of the characters is at work, folding back the spectrum 
of arbitrarily large negative dimensions into a continuum of positive 
dimensions. This is strongly supported by calculations involving 
modular transformations of characters, and gives rise to predictions 
in remarkable agreement with lattice calculations. A point that 
remains unclear is whether this continuation is a formal operation, or 
has the meaning of changing the target space to a Riemannian symmetric 
superspace as argued in \cite{Bocquet}. 
Clearly more 
work is needed to clarify these questions, which we believe are crucial 
to our understanding of CFT description of the transition between 
plateaux.
  
 Finally, we come back to our original question concerning the 
 universality of  $3\otimes 
 \bar{3}\otimes 3\otimes\bar{3}\ldots$ $sl(2/1)$ spin chains.
 It is easy to check that the analysis in \cite{ReadSaleur}
 for the ``Heisenberg'' model (with $\vec{S}.\vec{S}$ nearest 
 neighbour 
 coupling only) in the R 
 sector, gives a totally different spectrum. First, multiplicities are all 
 finite. Then, the following scalar operators
 \begin{equation}
     h=\bar{h}={(3P+1)^{2}-1\over 24},~~~h=\bar{h}= {4M^{2}-1\over 24}
 \end{equation}
 appear, as well 
 as many other non scalar operators whose chiral weights generically 
 read (modulo integers)
 \begin{equation}
     h={(3P/N+2M)^{2}-1\over 24},~~N|M
 \end{equation}
 Except for the lowest weights, there is no indication that any of 
 these weights appear in the spectrum of the integrable $sl(2/1)$ 
 chain. Moreover, imposing antiperiodic boundary conditions for the 
 fermions (ie the NS sector) gives results which are totally 
 different from the ones of the integrable spin chain; in particular, 
 instead of being three, the effective central charge is an irrational 
 number $c=1+{9\over\pi^{2}}\ln^{2}{3+\sqrt{5}\over 2}\approx 1.84$!
 In fact, pretty much the only features common to the two systems are 
 the vanishing of the central charge in the R sector, and the lowest 
 gap $h=\bar{h}={1\over 8}$ - extremely ubiquitous features of 
 superalgebra field theories. 
 
 Since the integrable model requires fine tuning of nearest neighbour 
 and next to nearest neighbour couplings, we expect that it describes 
 the least stable of the two fixed points, as evidenced by the higher 
 value of its effective central charge in the NS sector.

 It seems desirable to investigate the complete phase diagram of the 
 model, a task that could be made easier by using the loop 
 representation. We hope to get back to this question later. A property 
 that could prove useful in this analysis is that the ground state of the Heisenberg 
 chain, like the ground state of the integrable spin chain, is a true 
 singlet, with trivial value of the energy. It does not seem to be 
 the same state however.
 
 \vskip1cm
 \noindent{\bf Acknowledgments:} 
We are grateful to G. Goetz, V. Gurarie, A.~Kl\"umper, J. Links,
A.W.W. Ludwig, M. Martins, N. Read, V. Schomerus, M.~Shiroishi, 
A. Taormina, J. Teschner and
A.M. Tsvelik for helpful discussions. HS is especially thankful to J. 
Germoni for his wonderful  crash courses on super algebra 
representation theory.  FHLE thanks the Department of
Mathematics at the University of Brisbane, where part of this work was
carried out, for hospitality. This work was supported in part by
the Deutsche Forschungsgemeinschaft under grants Fr~737/2 and Fr~737/3
(HF), the DOE (HS), the EPSRC under grant GR/R83712/01 (FE), the
 Humboldt Foundation (HS), and the network EUCLID (TMR, network contact 
 HPRN-CT-2002-00325).
     
%\begin{appendix}
\appendix
\section{Some algebraic considerations}
\label{app:algebra}   
   Numerical study indicates that there is a unique singlet in 
   the product $(3\otimes \bar{3})^{\otimes L}$:
   for instance, the tensor product 
     $(3\otimes\bar{3})^{\otimes 2}$ decomposes as 
   \begin{equation}
       (3\otimes \bar{3})^{2}=[0,0]+4[0,1]+[0,2]+[1/2,3/2]+[-1/2,3/2]+
       [0,-1/2,1/2,0]
   \end{equation}
One has to be careful in defining what one means by singlet in this 
problem, as 
there are invariant 
states (that is, states annihilated by all the generators of the 
algebra) within some of the indecomposable representations too. We will 
call `true singlet' a representation isomorphic to the trivial 
representation, and which is not a quotient of a larger 
indecomposable representation. When decomposing 
$(3\otimes\bar{3})^{\otimes L}$, it appears as 
an invariant state which does not lie in the image of the raising or 
lowering generators of the algebra.   

It is known from 
   representation theory of $sl(2/1)$ \cite{Marcu} that in the tensor product 
   we are interested in, 
   there can only appear typical representations, simple (irreducible) 
   atypical of the type $[\pm j,j]$ and non simple (indecomposable) atypical, 
  representations. The latter 
   have vanishing superdimension, as do the typical 
   representations. It is obvious that the superdimension of our 
   tensor product is {\sl one} but one could imagine that this {\sl one}
   does not come from the existence of a single $[0,0]$ representation. 
   For instance, it could arise as the result of a cancellation 
   between {\sl several} $[0,0]$ reps (each of superdimension one, as 
   it is easy to see that all $[0,0]$ states have to be bosonic in our 
   system) and several simple atypical representations with 
   negative superdimension. It is easy to exclude this possibility if 
   one recognizes that the typical representations as well as the 
   indecomposable atypical appearing in our tensor 
   product are {\sl projective}, while simple (irreducible) atypical are 
   not (recall here \cite{Benson} that a representation is 
   projective if it cannot appear as a quotient of a bigger, 
   indecomposable block). 
   In particular, $[0,0]$ is not projective. A well known result of 
   algebra \cite{Martin, Benson, CurtisReiner}
   says that
   \begin{equation}
      \hbox{projective } \otimes \hbox{ anything} =\hbox{ projective}
      \end{equation}
   But consider $3\otimes \bar{3}=[0,0]+[0,1]$. The adjoint appearing 
   here is typical and thus projective. Hence, tensor products 
   $[0,1]^{\otimes L}$ decompose onto projectives only, that is 
   typical representations and non simple atypical. No simple atypical 
   ever can epppear in these tensor products.  
   Thus, $[0,0]$ can appear only once, as the result of the tensor 
   product of $[0,0]$ with itself, which is our initial claim. 
   
   Note that the result could also be proven by using the fact that 
   the tensor product of any indecomposable representation (either 
   irreducible like a simple typical, or reducible like a non simple 
   atypical) of vanishing superdimension decomposes as a sum of 
   representations each with vanishing superdimension themselves 
   \cite{Mathieu}.

   Notice that a similar result is known to take place in the 
    $SU(2)_{q}$ product of an even number of 
    fundamental (two dimensional) representations 
    when $q^{3}=-1$, so the spin $j=1$ representation has vanishing 
    q-dimension. In this case, there is a unique  singlet
    representation, which coincides with the space $\hbox{Ker 
    }S^{+}/\hbox{Im }S^{+}$ \cite{PasquierSaleur,ChariPressley}.
 
 Of course, if we get back for instance 
 to $(3\otimes \bar{3})^{\otimes 2}$, there are two invariant (that is, annihilated by all the 
 generators in the algebra) vectors, one in $[0,0]$ and one in the 
 indecomposable. The vector in $[0,0]$ is a `true singlet', which 
 means simply that it
 is not in the image of the generators. Obviously, a generic 
 combination of the two invariant vectors will still have this property. 
 This means that the way $[0,0]$ will be expressed in terms of the 
 basis vectors in $3$ and $\bar{3}$ will in fact depend on the 
 Hamiltonian (one 
 could imagine selecting the `good true singlet' by requiring that it 
 be orthonormal to the image of the generators in the subspace of 
 vanishing $U(1)$ charge. However  this will not work, as the 
 invariant state in the indecomposable has zero norm square). 
 Remarkable expressions for such states have been obtained recently 
 (albeit in a different language) in the context of the `Razumov 
 Stroganov conjecture' \cite{Razumov}. We leave this point to 
 further study.

%%%%%%%%%%%%%%%%%%%%%%%%%%%%%%%%%%%%%%%%%%%%%%%%%%%%%%%%%%%%%%%%%%%
\section{Details of the analysis of the lattice model}
\label{latticemodel}
%%%%%%%%%%%%%%%%%%%%%%%%%%%%%%%%%%%%%%%%%%%%%%%%%%%%%%%%%%%%%%%%%%%

 %%%%%%%%%%%%%%%%%%%%%%%%%%%%%
 \subsection{String hypothesis}
 \label{sec:strings}
 %%%%%%%%%%%%%%%%%%%%%%%%%%%%%
 The standard way to classify the solutions of (\ref{bae010}) is to
 consider configurations of spectral parameters that sit on poles of
 the bare scattering phase shifts (the right-hand-sides in
 (\ref{bae010})). A simple calculation yields the following types of
 ``strings''
 \begin{itemize}
 \item[{\bf (1)}] \underline{``Reals'':}

 unpaired, purely real spectral parameters $u_j$ and
 $\gamma_\beta$. 
 \item[{\bf (2)}] \underline{``Wide strings'':}

 ``type-1'': composits containing $n-1$ $\gamma$'s and $n$ $u$'s ($n>1$)
 \begin{eqnarray}
   u_{\alpha,k}^{(n,1)} &=& u_\alpha^{(n,1)} +i \left(n+1-2k\right),\quad
	   k=1\ldots n\nonumber\\
	   \gamma_{\alpha,j}^{(n,1)} &=& u_\alpha^{(n,1)} + i
	   \left(n-2j\right),\quad
	   j=1\ldots n-1\ ,\quad u_\alpha^{(n,1)}\in {\rm I\!R}
 \label{strings1}
 \end{eqnarray}

 ``type-2'': composits containing $n$ $\gamma$'s and $n-1$ $u$'s ($n>1$)
 \begin{eqnarray}
   \gamma_{\alpha,k}^{(n,2)} &=&  \gamma_\alpha^{(n,2)} 
	 +i \left(n+1-2k\right),\quad
	   k=1\ldots n\nonumber\\
	   u_{\alpha,j}^{(n,2)} &=&  \gamma_\alpha^{(n,2)} + i
	   \left(n-2j\right),\quad
	   j=1\ldots n-1\ ,\quad \gamma_\alpha^{(n,2)}\in {\rm I\!R}
 \label{strings2}
 \end{eqnarray}

 \begin{figure}[ht]
 \begin{center}
 \noindent
 \epsfxsize=0.4\textwidth
 \epsfbox{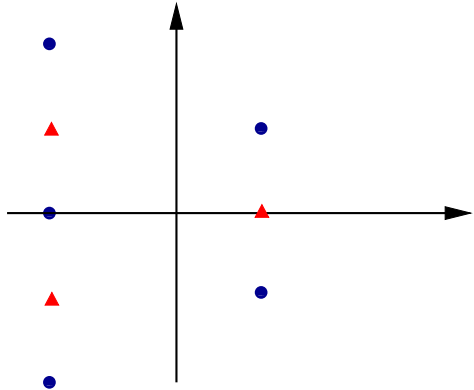}
 \end{center}
 \caption{\label{fig:widestr}
 Wide strings of lengths three and two respectively. The
 circles/triangles denote the positions of the $u$'s/$\gamma$'s
 involved in the string.
 }
 \end{figure}

 \item[{\bf (3)}] \underline{``Strange strings'':}

 composits containing $n$ $\gamma$'s and $n$ $u$'s ($n\geq 1$)
 \begin{eqnarray}
   u_{\alpha,k}^{(n,-)}&=& u_\alpha^{(n)} + i(n+1-2k-\frac{1}{2}),\quad
   k=1,\ldots,n\ ,\nonumber\\
   \gamma_{\alpha,k}^{(n,-)}&=& u_\alpha^{(n)} - i(n+1-2k-\frac{1}{2})
 \ ,\quad u_\alpha\in {\rm I\!R},
 \label{strings6}
 \end{eqnarray}
 or
 \begin{eqnarray}
   u_{\alpha,k}^{(n,+)}&=& u_\alpha^{(n)} - i(n+1-2k-\frac{1}{2}),\quad
   k=1,\ldots,n\ ,\nonumber\\
   \gamma_{\alpha,k}^{(n,+)}&=& u_\alpha^{(n)} + i(n+1-2k-\frac{1}{2})
 \ ,\quad u_\alpha\in {\rm I\!R}.
 \label{strings7}
 \end{eqnarray}
 We note that solutions of this type of solution is quite different
 from usual string solutions in that the set of roots on one level
 of the Bethe equations is not invariant under complex conjugation. 
 The other solutons discussed above (reals and wide strings) are
 invariant under this operation. This is somewhat similar to what was
 found recently in \cite{SW}. It is important to note that although
 ``strange strings'' are not invariant under complex conjugation, the
 corresponding energy (\ref{energy}) in the case $\lambda=0$ is still
 real. 

 \begin{figure}[ht]
 \begin{center}
 \noindent
 \epsfxsize=0.4\textwidth
 (a)\epsfbox{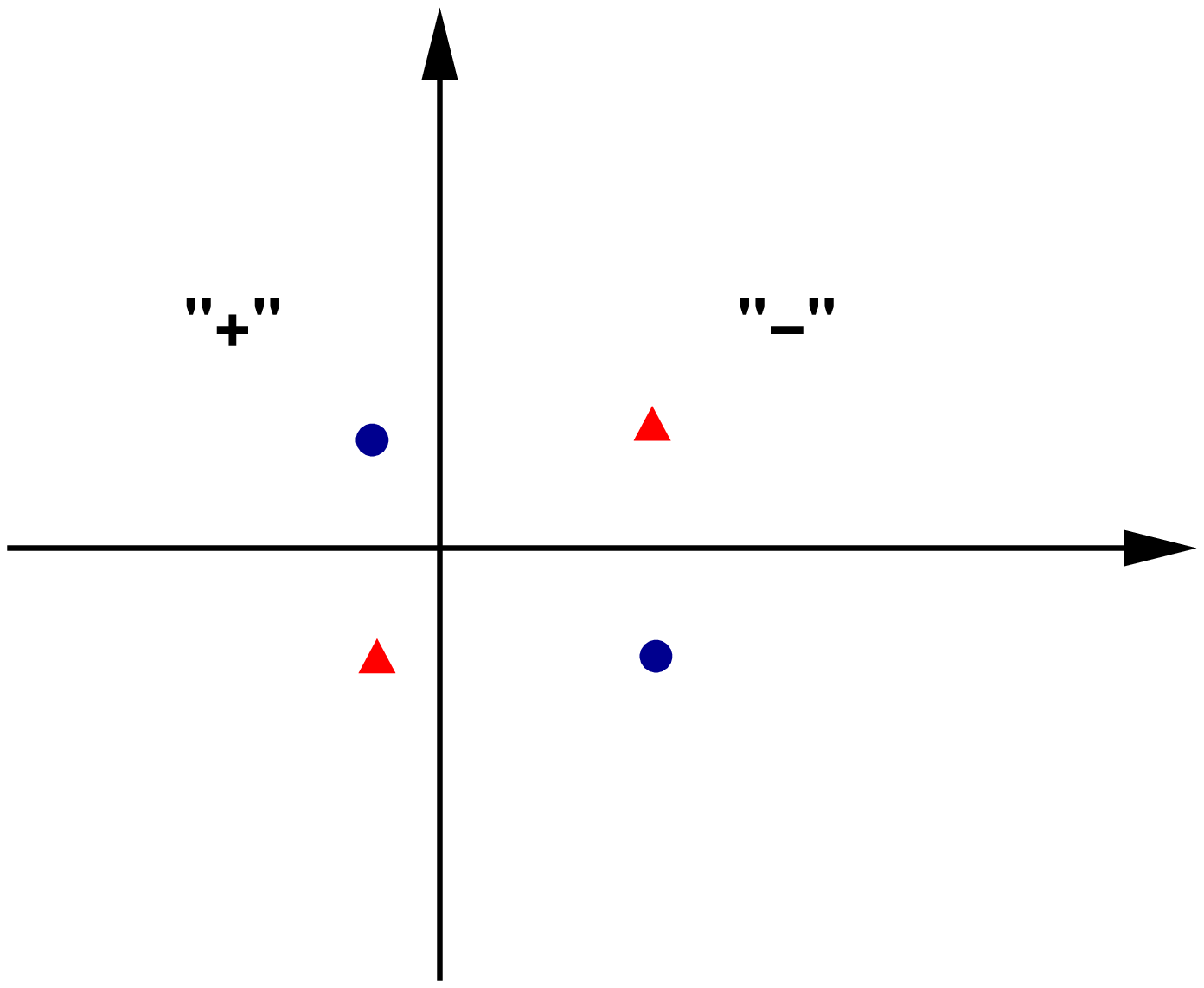}
 \epsfxsize=0.4\textwidth
 (b)\epsfbox{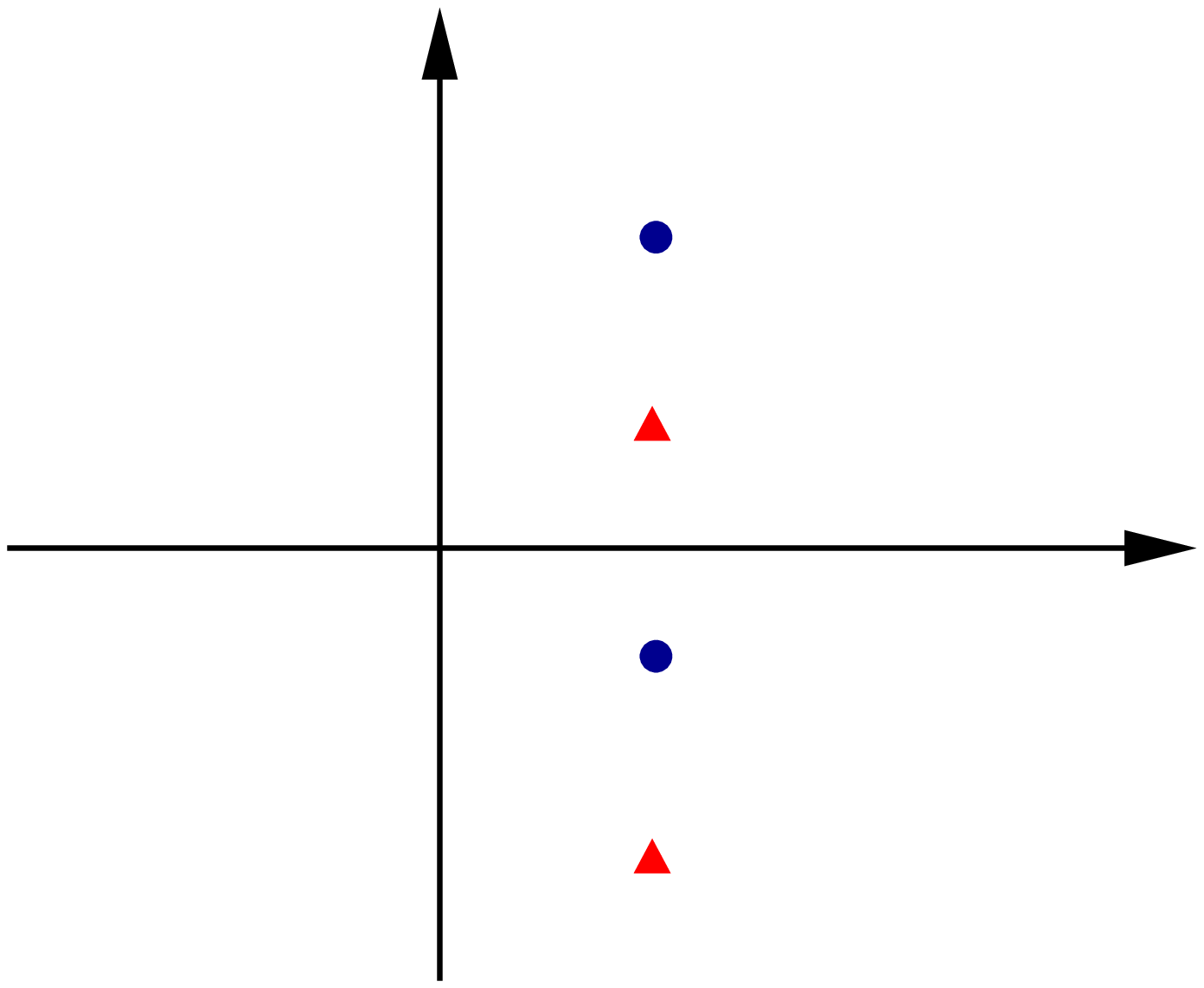}
 \end{center}
 \caption{\label{fig:sstr}
 (a) The two types of $n=1$ strange strings;
 (b) A ``+'' type strange string with $n=2$.
 }
 \end{figure}

 \item[{\bf (4)}] \underline{``Narrow strings'':}

 composits containing $n$ $\gamma$'s and $n$ $u$'s ($n>1$)
 \begin{eqnarray}
   u_{\alpha,k}^{(n,n)} &=&  u_\alpha^{(n,n)} 
	 + \frac{i}{2}\left(n+1-2k\right),\quad
	   k=1\ldots n\nonumber\\
	   \gamma_{\alpha,j}^{(n,n)} &=&  u_\alpha^{(n,n)} +
	   \frac{i}{2}\left(n+1-2j\right),\quad
	   j=1\ldots n\ ,\quad u_\alpha^{(n,n)}\in {\rm I\!R}
 \label{strings3}
 \end{eqnarray}
 Narrow strings may be thought of as special cases of strange strings
 or wide strings in the following sense. 
 \begin{itemize}
 \item{}
 Combining a ``+''-strange string of length $n$ and centre $u^{(n)}$
 with a a ``-''-strange string of length $n$ and centre $u^{(n)}$ we
 obtain a narrow string of even length $2n$. This is shown for $n=1$
 in Fig.\ref{fig:narrowstr}(a). 
 \item{} 
 Combining a type-1 wide string of length n and centre $u^{(n,1)}$
 with a type-2 wide string of length n and centre $u^{(n,1)}$
 we obtain a narrow string of length $2n-1$. This is shown for the
 case $n=2$ in Fig. \ref{fig:narrowstr}(b). 
 \end{itemize}

 \begin{figure}[ht]
 \begin{center}
 \noindent
 (a)
 \epsfxsize=0.4\textwidth
 \epsfbox{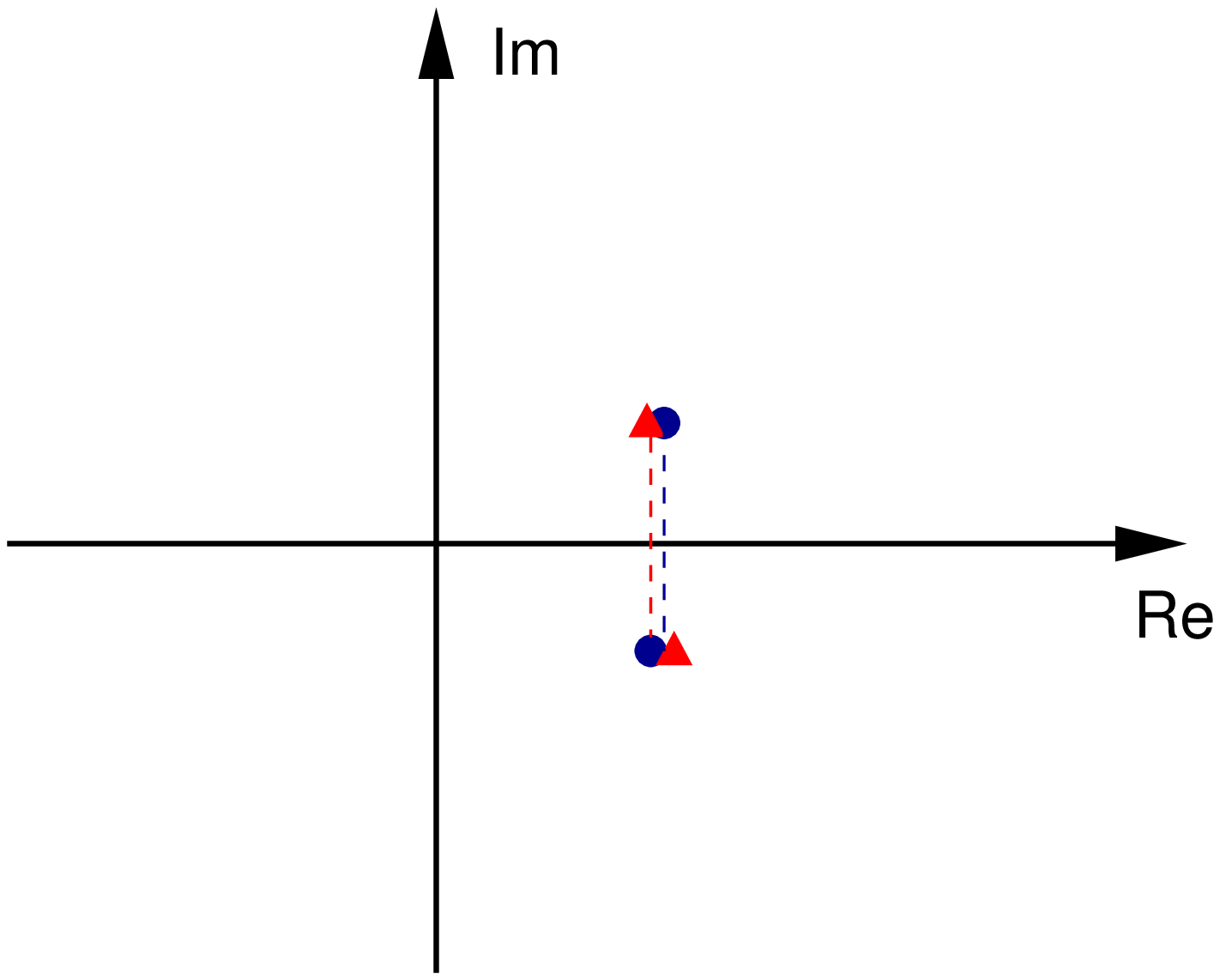}
 (b)
 \epsfxsize=0.4\textwidth
 \epsfbox{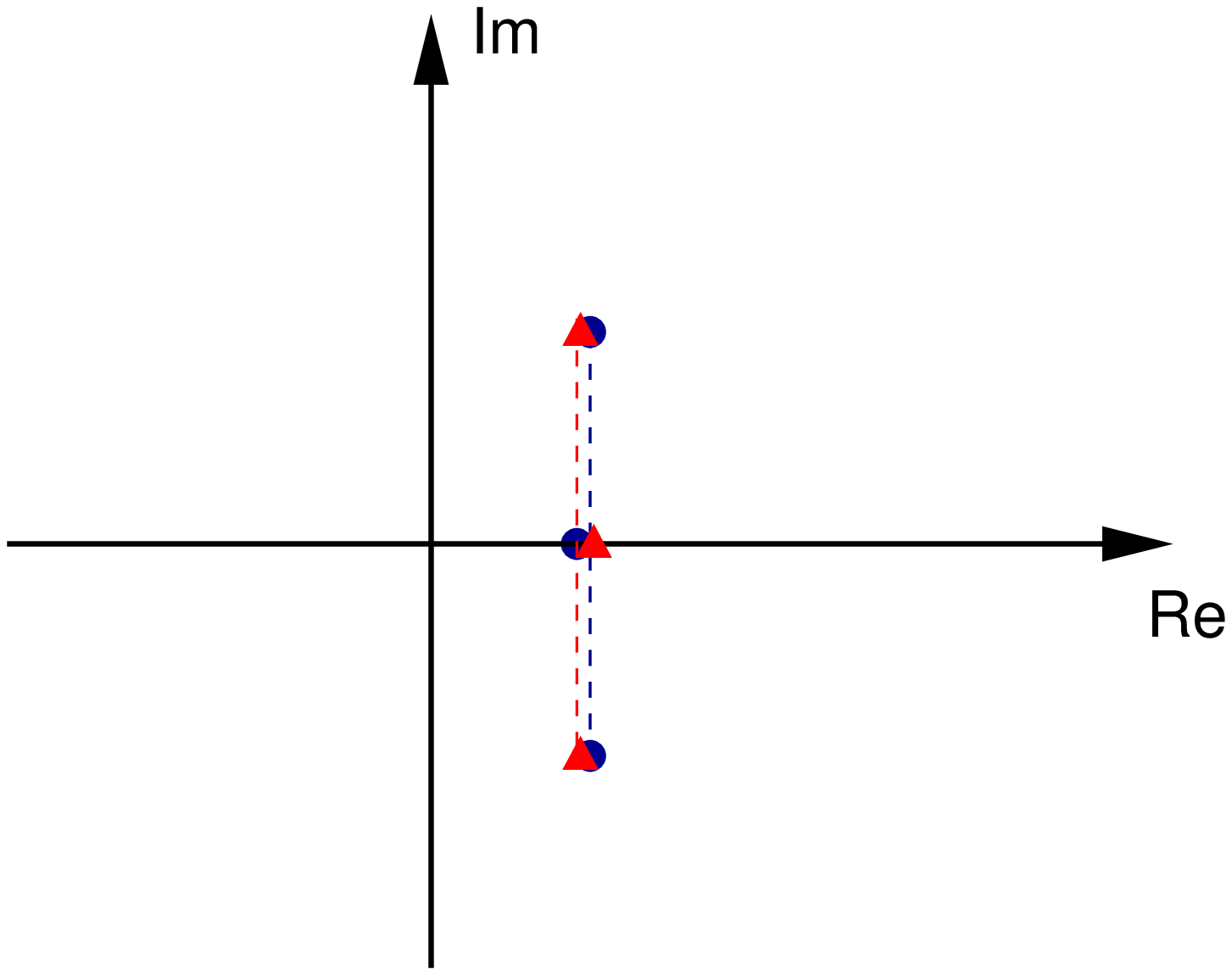}
 \end{center}
 \caption{\label{fig:narrowstr}
 (a) Combining a pair of ``+'' and ``-'' strange strings of length $1$
 gives a narrow string of length $2$;
 (b) Combining a type-1 wide string of length 2 with a type-2 wide
 string of length 2 gives a narrow string of length $3$;
 }
 \end{figure}
 It is clear from our discussion that narrow strings are not
 ``fundamental'' but are merely degenerate cases of strange string
 solutions of the Bethe ansatz equations.
 \end{itemize}

 We have verified explicitly by numerical solution of the Bethe ansatz
 equation (\ref{bae010}), mainly for the case $\lambda=0$, that
 solutions of the above types exist at least for small string-lengths
 $n$. On the other hand it is entirely possible that the above
 classification is not complete. 

 However, it is important to note that if 
 \begin{itemize}
 \item[(a)] our classification is complete and
 \item[(b)] in the thermodynamic limit solutions of the
 Bethe ansatz equations (\ref{bae010}) with the same values of $J^3$
 per site and $B$ per site as the ground state tend towards ideal
 string solutions (as is often the case in Bethe ansatz solvable
 models), 
 \end{itemize}
 then all energies are real for $\lambda=0$. Given the string
 hypothesis (\ref{strings1})--(\ref{strings7}) the usual way of
 proceeding in a Bethe ansatz solvable model is to construct the
 thermodynamic Bethe ansatz (TBA) equations at finite temperature and
 then use them to determine the distribution of roots 
 in the zero temperature ground state. This applicability of this
 procedure to our case is questionable as it is not clear whether the
 minimization of the free energy functional is a meaningful procedure
 for the case of a non-hermitian Hamiltonian. Disregarding this issue
 one finds that the standard ``integer method'' of classifying
 solutions of the Bethe ansatz equations does not seem to work.
In particular, we find that in some cases the numbers of solutions of
a particular type for fixed  $L$, $N$ and $M$ are found to differ
from what the integer method  predicts. We don't have a good
 explanation for this fact at present and leave this issue for future
studies. Irrespective of this, the Bethe
 ansatz does not provide a complete set of states in any case. This
 follows from the peculiar representation theory of superalgebras. It
 was shown in \cite{LiFo99} that for the model at hand all eigenstates
 obtained from the Bethe ansatz are highest weight states of $sl(2/1)$.
 For integrable models exhibiting a classical Lie-algebra symmetry a
 complete set of states can be obtained from these highest-weight
 states by acting with all possible combinations of lowering operators
 \cite{completeness}. In the present case there are states that belong
 to reducible but indecomposable respresentations, for which the
 highest weight state is not a cyclic one \cite{Marcu}.  As a
 consequence, acting on the Bethe ansatz states with lowering operators
 is not sufficient to obtain a complete set of states.  This
 complication is absent in other integrable, supersymmetric models, for
 which complete sets of states have been obtained by the aforementioned
 procedure \cite{fk,eks,BeFr95,bef,dopes}.

 Given these complications we will follow a different, less ambitious 
 approach: by
 studying finite systems we will try to identify the ground state and
 some low-lying excited states of the Hamiltonian (\ref{H}) in terms of
 distributions of roots of the Bethe ansatz equations (\ref{bae010}).

%%%%%%%%%%%%%%%%%%%%%%%%%%%%%%%
\subsection{Small Finite Systems}
\label{sec:small}
%%%%%%%%%%%%%%%%%%%%%%%%%%%%%%%
%
For $L=1,2$ the transfer matrix $\tau_3(v)$ can be diagonalized directly.
\subsubsection{$L=1$}

The tensor product $3\otimes \bar{3}$ is fully reducible and decomposes into
a singlet $[0]$ and the typical octet $[b,j]=[0,1]$.
\begin{equation}
    3\otimes \bar{3}= [0,0]+[0,1]
\end{equation}
The corresponding eigenvalues of $\tau_3(v)$ are readily obtained
\begin{equation}
 \label{spec1}
\begin{aligned}
   \Lambda_8(v) &= 1 - \frac{2}{v} + \frac{2}{v-1-\lambda}
\\
   \Lambda_1(v) &= 1 -\frac{\lambda-1}{\lambda+1}\,\frac{2}{v}
+\frac{\lambda-1}{\lambda+1}\,\frac{2}{v-1-\lambda}\ .
\end{aligned}
\end{equation}
Similarly we find for the eigenvalues of $\tau_{\bar{3}}(v)$
\begin{equation}
\label{spec1b}
\begin{aligned}
   \bar\Lambda_8(v) &= 1 - \frac{2}{v} + \frac{2}{v-1+\lambda}
\\
   \bar\Lambda_1(v) &= 1 -\frac{\lambda+1}{\lambda-1}\,\frac{2}{v}
+\frac{\lambda+1}{\lambda-1}\,\frac{2}{v-1+\lambda}\ .
\end{aligned}
\end{equation}
Obviously, $\Lambda_8(v)$, $\bar\Lambda_8(v)$ are the eigenvalues on the
pseudo vacuum ($M=0=N$) while $\Lambda_1(v)$, $\bar\Lambda_1(v)$ can be
obtained from (\ref{Lam33b}) in the sector with $(M,N)=(1,1)$ for $u_1 =
\lambda$, $\gamma_1=0$ which clearly solve the BAE (\ref{bae010}).  We
note that the singlet eigenvalue satisfies the functional equation
 \begin{equation}
   \Lambda_1(v)\bar\Lambda_1(v-1-\lambda)
	 = \left(\frac{(v+1-\lambda)(v-3-\lambda)}
		      {(v-1-\lambda)^2}\right)^{L}\ .
 \label{invrel}
 \end{equation}
 Furthermore, for $\lambda=0$, we find $\Lambda_1(v)|_{\lambda=0} =
 (v-2)(v+1)/v(v-1) = \bar\Lambda_1(v)|_{\lambda=0}$. The eigenvalues of
 the Hamiltonian are
 \begin{equation}
 E_1=\frac{4}{\lambda^2-1}\ ,\qquad E_8=\frac{2}{\lambda^2-1}.
 \end{equation}

\subsubsection{$L=2$}
The tensor product of representations appearing in the $L=2$ (4 site) system
can be decomposed as
\begin{equation}
\label{decom2}
    (3\otimes \bar{3})^{2}=[0,0]+4[0,1]+[0,2]+[1/2,3/2]+[-1/2,3/2]+
    [0,-1/2,1/2,0]
\end{equation}
Here, the atypical representation $[0,-1/2,1/2,0]$ is an eight dimensional
reducible but indecomposable representation which is the semi direct sum of
$[1/2,1/2]$, $[-1/2,1/2]$, and two $[0,0]$ representations
\cite{Marcu}) (see section 2 of the main text).  Note that this block is the same
indecomposable block as the one which appears in the product $[0,1/2]\otimes
[0,1/2]$. 

Diagonalization of the transfer matrix yields the eigenvalues of the 4-site
system.  Below we present their values together with the labels of the sectors
$(M,N)$ in which they appear in the BA and the corresponding
irreducible representation of the algebra $sl(2/1)$ in the
decomposition (\ref{decom2}): 

\begin{table}[ht]
 \begin{center}
 \begin{tabular}{|l|l|l|l|}
 \hline
  $(M,N)$  & $[B,S]$ & eigenvalue of $\tau_3(u), \bar\tau_3(u)$
 & Bethe roots\\ \hline\hline
  (0,0) & $\left[0,2\right]$ &  $\Lambda_{16}(v) = 1 - \frac{4}{v} + \frac{4}{v^2}
	   +\frac{4}{v-1-\lambda} + \frac{4}{(v-1-\lambda)^2}$ & \\
 & & $\bar\Lambda_{16}(v) = 1-\frac{4}{v}+\frac{4}{v^2} 
	 +\frac{4}{v-1+\lambda}+\frac{4}{(v-1+\lambda)^2}$
 & vacuum\\ \hline
  (0,1) &$\left[+\frac{1}{2},\frac{3}{2}\right]$
  & $\Lambda_{12}^{(+)}(v) =  1 - \frac{4}{v} - \frac{4}{v^2}
	 + \frac{4}{v-1-\lambda} + \frac{4}{(v-1-\lambda)^2}$ & \\
 & & $\bar\Lambda_{12}^{(+)}(v) =
	 1-\frac{4}{v} +\frac{4}{v^2} 
	 +\frac{4}{v-1+\lambda} -\frac{4}{(v-1+\lambda)^2}$
 & $u_1=0$ \\ \hline
  (1,0) & $\left[-\frac{1}{2},\frac{3}{2}\right]$
  & $\Lambda_{12}^{(-)}(v) = 1 - \frac{4}{v} + \frac{4}{v^2}
	 + \frac{4}{v-1-\lambda} - \frac{4}{(v-1-\lambda)^2}$ & \\
 & & $ \bar\Lambda_{12}^{(-)}(v) = 
	 1-\frac{4}{v} -\frac{4}{v^2} 
	 +\frac{4}{v-1+\lambda} +\frac{4}{(v-1+\lambda)^2}$
 & $\gamma_1=-i\lambda$ \\   \hline
  (1,1) & $\left[0,1\right]^2$ & $\Lambda_8^{(1,2)}(\l,v)$ 
 & $u_1=-i[\lambda\pm\sqrt{\lambda^2-1}]$\\
  & & $\bar\Lambda_8^{(1,2)}(\l,v)$  & $\gamma_1=u_1+i\lambda$\\ \hline
  (1,1) & $\left[0,1\right]^2$ & $\Lambda_8^{(3,4)}(\l,v)$
 & $u_1=-i[\frac{\lambda\pm\sqrt{\lambda^2+3}}{3}]$\\
  & & $\bar\Lambda_8^{(3,4)}(\l,v)$  & $\gamma_1=-i\lambda - u_1$\\ \hline
  (1,2) & $\left[0,-\frac{1}{2},\frac{1}{2},0\right]$ &
    $\Lambda_x(v) =  1 - \frac{4}{v} - \frac{4}{v^2}
	 + \frac{4}{v-1-\lambda} - \frac{4}{(v-1-\lambda)^2}$
 & $u_{1,2}=-i[\lambda\pm\sqrt{1+\lambda^2}]$ \\
 & &  $\bar\Lambda_x(v) = 1-\frac{4}{v}-\frac{4}{v^2}
	 +\frac{4}{v-1+\lambda}-\frac{4}{(v-1+\lambda)^2}$
 & $\gamma_1=-i\lambda$
 \\ \hline
  (2,1) & $\left[0,-\frac{1}{2},\frac{1}{2},0\right]$ &
    $\Lambda_x(v) =  1 - \frac{4}{v} - \frac{4}{v^2}
	 + \frac{4}{v-1-\lambda} - \frac{4}{(v-1-\lambda)^2}$
 & $u_1=0$ \\
 & &  $\bar\Lambda_x(v) = 1-\frac{4}{v}-\frac{4}{v^2}
	 +\frac{4}{v-1+\lambda}-\frac{4}{(v-1+\lambda)^2}$
 & $\gamma_{1,2}=\mp i\sqrt{1+\lambda^2}$\\ \hline
  (2,2) & $\left[0,-\frac{1}{2},\frac{1}{2},0\right]$ &
    $\Lambda_x(v) =  1 - \frac{4}{v} - \frac{4}{v^2}
	 + \frac{4}{v-1-\lambda} - \frac{4}{(v-1-\lambda)^2}$
 & $\gamma_1=-i\lambda+\frac{1}{\gamma_2+i\lambda}$\\
 & &  $\bar\Lambda_x(v) = 1-\frac{4}{v}-\frac{4}{v^2}
	 +\frac{4}{v-1+\lambda}-\frac{4}{(v-1+\lambda)^2}$
 & $u_{1,2}=\frac{1\pm\sqrt{1-(\frac{1}{\gamma_1}
 +\frac{1}{\gamma_2})^2}}{\frac{1}{\gamma_1}+\frac{1}{\gamma_2}}$\\ \hline
  (2,2) & $\left[0\right]$ &
    $\Lambda_1(v) = \left( \frac{(v-2)(v+1-\lambda)}
			 {v(v-1-\lambda)} \right)^{2}$& \\
 & &  $\bar\Lambda_1(v) =
	 \left(\frac{(v-2)(v+1+\lambda)}{v(v-1+\lambda)}\right)^2$ & \\
 \hline
\end{tabular}
\caption{Spectrum of the transfer matrices for $L=2$.}
\end{center}
\end{table}

 %\begin{eqnarray}
 % (0,0) & \left[0,2\right] &
 % \Lambda_{16}(v) = 1 - \frac{4}{v} + \frac{4}{v^2}
 %	  +\frac{4}{v-1-\lambda} + \frac{4}{(v-1-\lambda)^2}
 %\nonumber\\
 % (0,1) &\left[+\frac{1}{2},\frac{3}{2}\right]
 % & \Lambda_{12}^{(+)}(v) =  1 - \frac{4}{v} - \frac{4}{v^2}
 %	+ \frac{4}{v-1-\lambda} + \frac{4}{(v-1-\lambda)^2}
 %\nonumber\\
 % (1,0) & \left[-\frac{1}{2},\frac{3}{2}\right]
 % & \Lambda_{12}^{(-)}(v) = 1 - \frac{4}{v} + \frac{4}{v^2}
 %	+ \frac{4}{v-1-\lambda} - \frac{4}{(v-1-\lambda)^2}
 %\nonumber\\
 % (1,1) & \left[0,1\right]^4 & \Lambda_8^{(a)},\, a=1\ldots4
 %\nonumber\\
 % (1,2),(2,1),(2,2) & \left[0,-\frac{1}{2},\frac{1}{2},0\right] &
 %   \Lambda_x(v) =  1 - \frac{4}{v} - \frac{4}{v^2}
 %	+ \frac{4}{v-1-\lambda} - \frac{4}{(v-1-\lambda)^2}
 %\nonumber\\
 % (2,2) & \left[0\right] &
 %   \Lambda_1(v) = \left\{ \frac{(v-2)(v+1-\lambda)}
 %			{v(v-1-\lambda)} \right\}^{2}\
 %\nonumber
 %\end{eqnarray}

 Here the eigenvalues corresponding to the four (typical) octets
 $[0,1]$ are
 \begin{eqnarray}
   \Lambda_8^{(1,2)}(\l,v) &=&
     1 - \frac{\lambda}{1+\lambda}\,\frac{4}{v}
       \pm \sqrt{\frac{\lambda-1}{\lambda+1}}\,\frac{4}{v^2}
       + \frac{\lambda}{1+\lambda}\,\frac{4}{v-1-\lambda}
       \mp \sqrt{\frac{\lambda-1}{\lambda+1}}\,\frac{4}{(v-1-\lambda)^2}\nonumber\\
  \bar\Lambda_8^{(1,2)}(\l,v) &=&
    1-\frac{\lambda}{\lambda-1}\,\frac{4}{v}
     \mp\sqrt{\frac{\lambda+1}{\lambda-1}}\,\frac{4}{v^2}
     +\frac{\lambda}{\lambda-1}\,\frac{4}{v-1+\lambda}
     \pm\sqrt{\frac{\lambda+1}{\lambda-1}}\,\frac{4}{(v-1+\lambda)^2}
 \nonumber\\
   \Lambda_8^{(3,4)}(\l,v) &=&
     1 -\frac{\lambda^2+\lambda+4\pm2\sqrt{\lambda^2+3}}{(1+\lambda)^2}\,
		 \frac{4}{v}
       -\frac{2\pm\sqrt{\lambda^2+3}}{1+\lambda}
		 \frac{4}{v^2}\nonumber\\
 &&      +\frac{\lambda^2+\lambda+4\pm2\sqrt{\lambda^2+3}}{(1+\lambda)^2}\,
		   \frac{4}{v-1-\lambda}
       -\frac{2\pm\sqrt{\lambda^2+3}}{1+\lambda}\,
		   \frac{4}{(v-1-\lambda)^2}\\
  \bar\Lambda_8^{(3,4)}(\l,v) &=& \Lambda_8^{(3,4)}(-\l,v)\ .
 %   1-\frac{\lambda^2-\lambda+4\pm2\sqrt{\lambda^2+3}}{(\lambda-1)^2}\,
 %	\frac{4}{v}
 %   +\frac{2\pm\sqrt{\lambda^2+3}}{\lambda-1}\,\frac{4}{v^2}\\
 %   &&+\frac{\lambda^2-\lambda+4\pm2\sqrt{\lambda^2+3}}{(\lambda-1)^2}\,
 %	\frac{4}{v-1+\lambda}
 %   +\frac{2\pm\sqrt{\lambda^2+3}}{\lambda-1}\,\frac{4}{(v-1+\lambda)^2}
 \nonumber
 \end{eqnarray}

 %{}From the residues of these expressions at $v=0$ and $v=1+\lambda$ we
 %can determine the corresponding roots of the BAE (\ref{bae010}):
 %\begin{eqnarray}
 %  \Lambda_{12}^{(+)} && u_1=0
 %\nonumber\\
 %  \Lambda_{12}^{(-)} && \gamma_1=\lambda
 %\nonumber\\
 %  \Lambda_{8}^{(1,2)} && u_1=\lambda\pm\sqrt{\lambda^2-1},\
 %		         \gamma_1=u_1-\lambda
 %\nonumber\\
 %  \Lambda_{8}^{(3,4)} &&
 %  u_1=\frac{1}{3}\left(\lambda\pm\sqrt{\lambda^2+3}\right),\
 %  \gamma_1=\lambda-u_1
 %\nonumber\\
 %  \Lambda_x && (2,1):\quad u_1 = 0,\
 %			   \gamma_{1,2}=\pm\sqrt{1+\lambda^2}
 %\nonumber\\
 %	    && (1,2):\quad u_{1,2}=\lambda\pm\sqrt{1+\lambda^2},\
 %			   \gamma_1=\lambda
 %\nonumber\\
 %	    && (2,2):\quad \gamma_1 = \lambda-\frac{1}{\gamma_2-\lambda},\
 %			   u_{1,2} = \frac{1}{\gamma_1+\gamma_2}\
 %				\left\{\gamma_1\gamma_2\pm
 %				 \sqrt{(\gamma_1\gamma_2)^2
 %				      +(\gamma_1+\gamma_2)^2}\right\}
 %\nonumber
 %\end{eqnarray}

 Let us consider the $[0,-\frac{1}{2},\frac{1}{2},0]$ multiplet in more
 detail. 
 %The structure of states in this multiplet is shown in
 %Fig.\ref{fig:indec}. 
%\begin{figure}[ht]
 %\begin{center}
 %\noindent
 %\epsfxsize=0.45\textwidth
 %(a)\epsfbox{weights.eps}
 %\epsfxsize=0.45\textwidth
 %(b)\epsfbox{indec.eps}
 %\end{center}
 %\caption{\label{fig:indec}
 %(a) Root diagram of $gl(2|1)$.
 %(b) The 8 states forming the $[0,-\frac{1}{2},\frac{1}{2},0]$
 %representation of $gl(2|1)$.
 %}
 %\end{figure}
 There are altogether three highest-weight states, i.e. states
 annihilated by the $sl(2/1)$ raising generators
 Their quantum numbers are $B=\pm\frac{1}{2}$, $J^{3}=\frac{1}{2}$
 and $B=J^{3}=0$ respectively. From the table above we see that the Bethe
 ansatz yields three states in this multiplet, which by \cite{LiFo99}
 are highest weight states. The Bethe ansatz states with $(M,N)=(1,2)$
and $(M,N)=(2,1)$ are readily identified as $B=J^{3}=\frac{1}{2}$,
 $B=-J^{3}=-\frac{1}{2}$ highest-weight states. The solution to the Bethe
 ansatz equations with $(M,N)=(2,2)$, leading to the eigenvalue
 $\Lambda_x$, is very peculiar to say the least: it contains a free
 parameter! For any value of $\gamma_2$ the roots given above satisfy
 the BAE (\ref{bae010}) \emph{and} give the eigenvalue $\Lambda_x(v)$
 of the transfer matrix!
 We identify this last highest-weight state with the invariant singlet
 state of $[0,-\frac{1}{2},\frac{1}{2},0]$. 
 The second singlet state in the multiplet is a cyclic state, i.e. all
 states in the representation can be obtained from it by acting with
 symmetry generators. However, it appears that it is not possible to
 obtain this state from the Bethe ansatz! In other words, the Bethe
 ansatz is manifestly incomplete due to the peculiar representation
 theory of superalgebras.

 For the eigenvalue $\Lambda_1(v)$ corresponding to the unique
 \emph{regular} singlet in the tensor product there exists no
 non-degenerate solution of the BAE (\ref{bae010}).  Applying a twist
 in the boundary conditions and studying the evolution of BA roots as
 this twist goes to zero we were able to verify that the singlet is in
 fact described by the \emph{singular} solution $u_k\equiv \lambda$,
 $\gamma_\alpha\equiv 0$ which clearly gives $\Lambda_1(v)$ from the
 general form (\ref{Lam33b}).  This finding coincides with the
 observation in \cite{JuKS97} about the distribution of BA-roots for
 the ground state of the quantum transfer matrix for the $t$--$J$ model
 in the thermodynamic limit. For finite $L$ the authors of
 \cite{JuKS97} find this degeneracy to be lifted which is a
consequence of the fact that their system is subject to a phase shift
when compared to ours (see Appendix \ref{app:JKS}).  Note that this
phase shift also breaks the supersymmetry of the system. 
Furthermore it is not obvious that
the two states actually are mapped onto one another when the phase
shift is changed adiabatically: the distribution of roots of the
function ${\cal D}$ in \cite{JuKS97} seems to be very different.

 We note further that $\Lambda_1(v)$ is just the square of the
 eigenvalue obtained for the singlet in the 2-site system
 (\ref{spec1}).  Hence, $\Lambda_1(v)$ is again the solution of the
 functional equation (\ref{invrel}).
 This observation lead us to formulate the following conjecture for the 
 ground state eigenvalue of the transfer matrix
 \begin{equation}
   \Lambda_1(v) =
     \left\{ \frac{(v-2)(v+1-\lambda)}{v(v-1-\lambda)} \right\}^{L}\ .
 \label{conj}
 \end{equation}
 For the $L=2$ system the eigenvalues of the Hamiltonian are (\ref{energy}):
 \begin{eqnarray}
 E_0 &=& \frac{8}{\lambda^2-1}\ ,\quad
 E_x = -4\ ,\quad
 E_8^{(1,2)} = -2,\quad E_{12}^{(\pm)}= -2\ ,\quad
 E_{16} = 0\ ,\nonumber\\
 E_8^{(3,4)} &=&\frac{-2\l^8-8\l^6+56\l^4+40\l^2+42
 \pm\sqrt{3+\l^2}(-8\l^6+28\l^4+20\l^2+24)}
 {(\l^2\pm 1)^3(2\pm\sqrt{3+\l^2})^2}\ .
 \nonumber
 \end{eqnarray}
 For $\l=0$ we find that $\left.E_8^{(3,4)}\right|_{\lambda=0}=-6$,
 i.e.\ the eigenvalues are equally spaced integers.
 We note that the conjecture (\ref{conj}) implies that the ground state
 $E_0=4L/(\lambda^2-1)$ is strictly extensive without any finite size
 corrections.  For a conformally invariant system this implies that the
 central charge of the underlying Virasoro algebra is $c=0$ --- as
 expected for the class of models considered here.

 %%%%%%%%%%%%%%%%%%%%%%%%%%%%%%%%%%%%%%%%%%%%%%%%%%%%%%%%%%%%%%%%%%%%%%
 \subsubsection{L=3,4,5}
 %%%%%%%%%%%%%%%%%%%%%%%%%%%%%%%%%%%%%%%%%%%%%%%%%%%%%%%%%%%%%%%%%%%%%%
 By numerically diagonalizing the transfer matrix for system sizes up to
 $L=5$ we find that in general there is exactly one eigenstate with an
 eigenvalue given by 
 \begin{equation}
 \Lambda(v)=\Lambda_1(v)\bar{\Lambda}_1(v)=
 \left(\frac{(v-2)^2(v+1-\lambda)(v+1+\lambda)}{v^2(v-1-\lambda)(v-1+\lambda)}
 \right)^L .
 \label{singlet}
 \end{equation}
 This is in agreement with the conjecture (\ref{conj}). For $\lambda=0$
 this state is the ground state of the Hamiltonian with energy
 \begin{equation}
 E_0=-4L\ .
 \end{equation}
 From now on we set $\lambda=0$. The first few low-lying multiplets for
 $L=3$ are shown in Table \ref{l=3}. We see that the first excited
 state is an octet with energy $-11.1231$. Then there are two
 degenerate octets with energy $E=-9.60555$. Finally there are two
 degenerate $[0,-\frac{1}{2},\frac{1}{2},0]$ multiplets. We only list
 one Bethe ansatz solution for each of these representations. Like in the
 two-site case $L=2$, the invariant singlets of
 $[0,-\frac{1}{2},\frac{1}{2},0]$ multiplets correspond to
 one-parameter families of solutions to the Bethe ansatz equations.
 %
 %%%%%%%%%%%%%%%%%%%%%%%%%%%%%%%%%%%%%%%%%%%%%%%%%%%%%%%%%%%%%%%%%%%%%%
 \begin{table}
 \begin{center}
 \begin{tabular}{|l|l|l|r|}
 \hline
 $[B,S]$  & $E$ 	  &{\rm BA\ roots}  & $P/2\pi$\\ \hline\hline
 $[0,1]$  & -11.1231   &   $u_1=\gamma_1= 0.530i$,\qquad
 $u_2=\gamma_2=-0.530i$ & 0 \\ \hline
 $[0,1]$	 & -9.60555   &   $u_1=\gamma_1^*=-0.254-0.608i$,\qquad 
			   $u_2=\gamma_2^*=0.254-0.608i$& 0\\ \hline
 $[0,1]$	 & -9.60555   &   $u_1=\gamma_1^*=-0.254+0.608i$,\qquad
			   $u_2=\gamma_2^*=0.254+0.608i$& 0 \\ \hline
 $[0,\frac{1}{2},-\frac{1}{2},0]$ & -9.00000 &
 $\gamma_1=-\gamma_2^*=- 0.144  -0.382 i$,\qquad $u_1=-1.427i$& 0\\ 
 & & $u_2=-u_3^*=- 0.163+0.713i$& \\ \hline
 $[0,-\frac{1}{2},\frac{1}{2},0]$ & -9.00000 & 
 $\gamma_1=-\gamma_2^*=- 0.144  +0.382 i$ ,\qquad $u_1=1.427i$&0\\
 & & $u_2=-u_3^*=- 0.163-0.713i$& \\ \hline
 $[0,1]$  & -6.5   &   $u_1=\gamma_1^*= -0.238-0.547i$,\qquad
 $u_2=\gamma_2^*= 1.104+0.175i$ & $-1/3$\\  \hline
 $[0,1]$  & -6.5   &   $u_1=\gamma_1^*= 0.238+0.547i$,\qquad
 $u_2=\gamma_2^*= -1.104-0.175i$& $1/3$\\  \hline
 $[0,1]$  & -6.5   &   $u_1=\gamma_1^*= -0.238+0.547i$,\qquad
		       $u_2=\gamma_2^*= 1.104-0.175i$&  $-1/3$\\ \hline
 $[0,1]$  & -6.5   &   $u_1=\gamma_1^*= 0.238-0.547i$,\qquad
		       $u_2=\gamma_2^*= -1.104+0.175i$&  $1/3$\\  \hline
 $[0,1]$  & -6.23607&  $u_1=\gamma_1=-0.535- 0.309i$,\qquad
		       $u_2=\gamma_2=-0.535 + 0.309i$& $-1/3$ \\ \hline
 $[0,1]$  & -6.23607&  $u_1=\gamma_1=0.535- 0.309i$,\qquad
		       $u_2=\gamma_2=0.535 + 0.309i$& $1/3$\\ \hline
 \end{tabular}
 \caption{Low-lying multiplets for $L=3$.}
 \label{l=3}
 \end{center}
 \end{table}

 In Table \ref{l=4} we list the low-lying $sl(2/1)$
 multiplets for $L=4$. The first excited states are two octets
 with energy $E=-15.0484$. Next there is a single
 $[0,-\frac{1}{2},\frac{1}{2},0]$ multiplet with energy
 $E=-14.1677$. Clearly the degeneracies for $L=4$ are different from
 $L=3$. This suggests that there might be a difference in structure of
 the low-lying excitation spectra for odd and even $L$.

 %%%%%%%%%%%%%%%%%%%%%%%%%%%%%%%%%%%%%%%%%%%%%%%%%%%%%%%%%%%%%%%%%%%%%%
 \begin{table}
 \begin{center}
 \end{center}
 \begin{center}
 \begin{tabular}{|l|l|l|}
 \hline
 $[B,S]$  & $E$ 	  &{\rm BA\ roots}\\ \hline\hline
 $[0,1]$  & -15.0484   &  $u_1=\gamma_1^*= 0.227 +0.554i$,\qquad
			$u_2=\gamma_2^*= -0.227 +0.554i$\\ 
	  &            &  $u_3=\gamma_3^*= -0.50948i$\\ \hline
 $[0,1]$  & -15.0484   &  $u_1=\gamma_1^*= 0.227 -0.554i$,\qquad
			$u_2=\gamma_2^*= -0.227 -0.554i$\\
	  &            &  $u_3=\gamma_3^*=0.509i$\\ \hline
 $[0,-\frac{1}{2},\frac{1}{2},0]$
	  & -14.1677   &   $u_1=\gamma_1=-0.051+0.521i$,\quad 
			   $u_2=\gamma_2=0.871+0.700i$\\
	  &            &   $u_3=\gamma_3=-0.051-0.521i$,\quad 
			   $u_4=\gamma_4=0.871-0.700i$\\ \hline
 $[0,1]$	 & -13.0749   &   $u_1=\gamma_1^*=-0.426-0.654 i$,\qquad
			 $u_2=\gamma_2^*= 0.426-0.654 i$\\
	  &            &   $u_3=\gamma_3^*= -0.596i$\\ \hline
 $[0,1]$	 & -13.0749   &   $u_1=\gamma_1^*=-0.426+0.654 i$,\qquad
			 $u_2=\gamma_2^*= 0.426+0.654 i$\\
	  &            &   $u_3=\gamma_3^*= 0.596i$\\ \hline
 \end{tabular}
 \caption{Low-lying multiplets for $L=4$.}
 \label{l=4}
 \end{center}
 \end{table}

 Table \ref{l=5} summarizes the lowlying states for $L=5$.
 The first excited multiplet is an octet, then there are two degenerate
 octets, followed by two degenerate $[0,-\frac{1}{2},\frac{1}{2},0]$
 multiplets. The structure of degeneracies coincides with the $L=3$
 case.
 
 %%%%%%%%%%%%%%%%%%%%%%%%%%%%%%%%%%%%%%%%%%%%%%%%%%%%%%%%%%%%%%%%%%%%%%
 \begin{table}
 \begin{center}
 \begin{tabular}{|l|l|l|}
 \hline
 $[B,S]$  & $E$ 	  &{\rm BA\ roots}\\ \hline\hline
 $[0,1]$  & -19.4230   &   $u_1=\gamma_1= +0.200+0.523i$,\qquad
			 $u_2=\gamma_2= +0.200-0.523i$\\
	  &            &   $u_3=\gamma_3= -0.200+0.523i$,\qquad
			 $u_4=\gamma_4= -0.200-0.523i$\\ \hline
 $[0,1]$	 & -18.66651  &   $u_1=\gamma_1^*=+0.390-0.593i$,\qquad
			   $u_3=\gamma_3^*=-0.559i$\\ 
 {}       &  {}        &   $u_2=\gamma_2^*=-0.390-0.593i$,\qquad
			   $u_4=\gamma_4^*=+0.503i$\\\hline
 $[0,1]$	 & -18.66651  &   $u_1=\gamma_1^*=-0.390+0.593i$,\qquad
			   $u_3=\gamma_3^*=+0.559i$\\
 {}       &  {}        &	  $u_2=\gamma_2^*=+0.390+0.593i$,\qquad
			   $u_4=\gamma_4^*= -0.503i$\\\hline
 $[0,-\frac{1}{2},\frac{1}{2},0]$ & -18.28188 & (A)\\
 {} & {} & $u_1=\gamma_1^*=-1.105+0.671i$,\qquad
 $u_4=\gamma_4^*=-0.835-0.743i$\\
 {} & {} & $u_2=\gamma_2^*=-0.113-0.543i$,\qquad
 $u_5=\gamma_5^*= 0.02364+0.50539i$\\
 {} & {} & $u_3=\gamma_3^*= 0.263-0.551i$\\
 \hline
 $[0,-\frac{1}{2},\frac{1}{2},0]$ & -18.28188  &
 (A) with $u_j\longrightarrow -u_j$, 
 $\gamma_\alpha\longrightarrow -\gamma_\alpha$\\ \hline
 \end{tabular}
 \caption{Low-lying multiplets for $L=5$.}
 \label{l=5}
 \end{center}
 \end{table}

 %%%%%%%%%%%%%%%%%%%%%%%%%%%%%%%%%%%%%%%%%%%%%%%%%%%%%%%%%%%%%%%%%%%%%%
 \subsubsection{Root configurations for a class of low-lying states}
 \label{ssec:config}
 %%%%%%%%%%%%%%%%%%%%%%%%%%%%%%%%%%%%%%%%%%%%%%%%%%%%%%%%%%%%%%%%%%%%%%
 Our results for small systems indicate that the lowest-lying
 excitations of the model are in the octet sector, i.e. belong to
 $[B,S]=[0,1]$, or form indecomposable $[0,-\frac{1}{2},\frac{1}{2},0]$
 multiplets.  Within the Bethe ansatz for a system of length $L$ the
 octets are parametrized by $L-1$ roots $u_j$ and $\gamma_\alpha$ while
 the indecomposables are found for example in the sector $B=0=J^3$ with
 $L$ roots on both levels.
 Within our classification of solutions by means of the string
 hypothesis from Section~\ref{sec:strings} the corresponding roots form
 certain sequences of strange strings of length~$n=1$ (or their
 degenerate realization as ``$2-2$'' narrow strings of length
 (\ref{strings3})).  We denote the number of these ``$+$'' and ``$-$''
 type strange strings by $N_+$ and $N_-$ respectively.  We constrain
 ourselves to the case $N_+\geq N_-$; states with $N_->N_+$ are
 obtained by exchanging $u$'s and $\gamma$'s and are by construction
 degenerate.  We denote the energy levels of the states by
 $E_{N_+,N_-}$.
\vskip .25cm
 %%%%%%%%%%%%%%%%%%%%%%%%%%
 \noindent\underline{\bf $4.4.1$ Even $L$:}
 %%%%%%%%%%%%%%%%%%%%%%%%%%
 \begin{itemize}
 \item{} A low-energy state with $B=0=J^3$ belonging to a
   $[0,-\frac{1}{2},\frac{1}{2},0]$ multiplet is obtained by choosing
   $N_+=N_-=\frac{L}{2}$ and considering a realization of this
   configuration as a collection of ``$2-2$'' narrow strings.  This
   implies that after a suitable relabelling of roots we have
   $u_j=\gamma_j$, for $j=1,\ldots,L$.  Using this degeneracy
   the Bethe ansatz equations (\ref{bae010}) can be rewritten as
   \begin{eqnarray}
     \left\{ \frac{u_j+i}{u_j-i} \right\}^L &=& \prod_{k=1}^N
     \frac{u_j-u_k+i}{u_j-u_k-i}\ ,\quad j=1,\ldots,N\ ,
     \label{bae2}
   \end{eqnarray}
   with $N=L$.  These equations are identical to the Bethe ansatz
   equations of the SU(2) symmetric Takhtajan-Babujian spin-1 chain
   \cite{BabuT} for a lattice of $L$ sites and antiperiodic boundary
   conditions.  The operator content of an anisotropic version of this
   model (with twisted rather than antiperiodic boundary conditions)
   was determined by Alcaraz and Martins \cite{AM}. We show in
   Appendix~\ref{app:spin1} how our results relate to theirs.
 \item{} The lowest lying excitation has $N_+=\frac{L}{2}$ and
   $N_-=\frac{L}{2}-1$ and is the highest weight state of $[0,1]$
   multiplet.  The distribution of strings is symmetric around the
   imaginary axis, i.e.\
   \begin{equation}
     \{u_j|j=1,\ldots L-1\}\equiv \{-u_j^*|j=1,\ldots L-1\},
   \end{equation}
   and similarly for $\{\gamma_j|j=1,\ldots L-1\}$. The root
   distribution for this state for $L=10$ is shown in
   Fig.\ref{fig:sstr10}(a).
 %%%%%%%%%%%%%%%%%%%%%%%%%%%%%%%%%%%%%%%%%%%%%%%%%%%%%%%%%%%%%%%%%%%%%%
 \begin{figure}[ht]
 \begin{center}
 \noindent
 \epsfxsize=0.85\textwidth
 \epsfbox{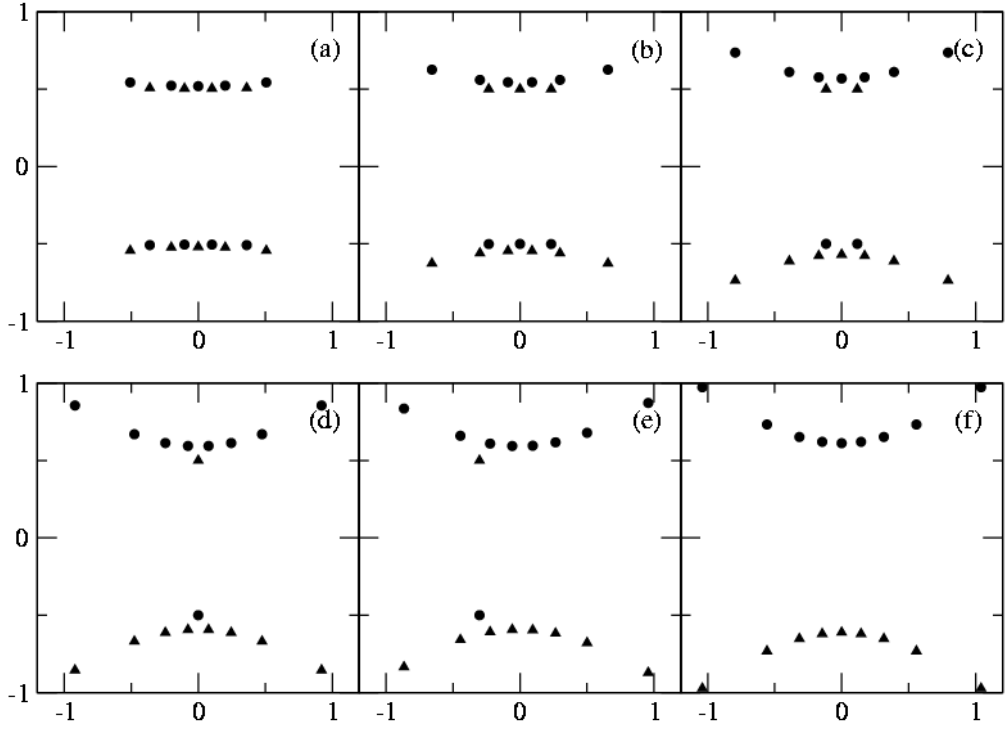}
 \end{center}
 \caption{\label{fig:sstr10} Strange string solutions of the Bethe
    ansatz equations (\ref{bae010}) in the octet sector for a system of
    size $L=10$ with different $\Delta N=N_+-N_-$: circles (triangles)
    denote $u_j$ ($\gamma_\alpha$). (a) $\Delta N=1$, (b) $\Delta N=3$,
    (c) $\Delta N=5$ , (d) and (e) $\Delta N=7$, (f) $N_+=9$, $N_-=0$.}
 \end{figure}
 %%%%%%%%%%%%%%%%%%%%%%%%%%%%%%%%%%%%%%%%%%%%%%%%%%%%%%%%%%%%%%%%%%%%%%
 \item{} Other $[0,1]$ octet states are obtained by taking
   $N_+=\frac{L}{2}+n$ and $N_-=\frac{L}{2}-1-n$, $n=1,2,\ldots$,
   i.e. by replacing $n$ ``$-$'' type strange strings by ``$+$'' type strange
   strings. The corresponding root distributions are shown in
   Fig.\ref{fig:sstr10}(b)-(f). We note that the lowest energy state
   for given $N_\pm$ corresponds to a root distribution that is
   symmetric around the imaginary axis. Higher energy states correspond
   to asymmetric distributions. An example for $L=10$ is shown in
   Fig. \ref{fig:sstr10}(e).
 \end{itemize}
 %%%%%%%%%%%%%%%%%%%%%%%%%%
 \noindent\underline{\bf $4.4.2$ Odd $L$:}
 %%%%%%%%%%%%%%%%%%%%%%%%%%
 \begin{itemize}
 \item{} The lowest lying state has $N_+=N_-=\frac{L-1}{2}$ and is the
   highest weight state of a $[0,1]$ multiplet.  The sets
   $\{u_j|j=1,\ldots L-1\}$ and $\{\gamma_j|j=1,\ldots L-1\}$ are
   closed under complex conjugation.  Hence, after a suitable
   relabelling of roots we have $\gamma_j=u_j$, for $j=1,\ldots L-1$.
   Similarly as for the $B=0=J^3$ state for even $L$ this state can be
   considered as a collection of ``$2-2$'' narrow strings and the Bethe
   ansatz equations can be cast in the form (\ref{bae2}) with $N=L-1$.
 \item{} A sequence of low energy $[0,1]$ octet highest weight states
   is obtained by taking $N_+=\frac{L-1}{2}+n$, $N_-=\frac{L-1}{2}-n$,
   $n=1,2,\ldots$.
 \end{itemize}
 In the following section we carry out a finite-size scaling analysis
 for the states we have just described.
 
 %%%%%%%%%%%%%%%%%%%%%%%%%%%%%%%%%%%%%%%%%%%%%%%%%%%%%%%%%%%%%%%%%%%%%%
 \subsection{Low-lying excitations in the thermodynamic limit}
 \label{app:inteqs}
 %%%%%%%%%%%%%%%%%%%%%%%%%%%%%%%%%%%%%%%%%%%%%%%%%%%%%%%%%%%%%%%%%%%%%%

In this Appendix we derive the coupled integral equations \r{igl:rhos}
from the algebraic equations \r{baesstr}. We start by taking  the
logarithm of \r{baesstr}
\begin{eqnarray}
  \theta(x_j)+\theta(\frac{x_j}{3})=\frac{2\pi I_j}{L}
    +\frac{1}{L}\sum_{k=1}^{N_+} \theta(\frac{x_j-x_k}{4})
    +\frac{2}{L}\sum_{k=1}^{N_-} \theta(\frac{x_j-y_k}{2})\ ,\quad
    j=1,\ldots, N_+,\nonumber\\
\label{logbaes}\\
  \theta(y_j)+\theta(\frac{y_j}{3})=\frac{2\pi J_j}{L}
    +\frac{2}{L}\sum_{k=1}^{N_+} \theta(\frac{y_j-x_k}{2})
    +\frac{1}{L}\sum_{k=1}^{N_-} \theta(\frac{y_j-y_k}{4})\ ,\quad
    j=1,\ldots, N_-,\nonumber
\end{eqnarray}
where $\theta(x)=2\ \arctan(2x)$ and $I_j$ ($J_j$) are integer or half integer
numbers characterizing the state uniquely.  Following Yang and Yang we
introduce counting functions $z_\pm(v)$ by
\begin{eqnarray}
  z_+(x) = \theta(x)+\theta(\frac{x}{3})
    -\frac{1}{L}\sum_{k=1}^{N_+} \theta(\frac{x-x_k}{4})
    -\frac{2}{L}\sum_{k=1}^{N_-} \theta(\frac{x-y_k}{2})\ ,
    \nonumber\\
\label{cfs}\\
  z_-(x) = \theta(y)+\theta(\frac{y}{3})
    -\frac{1}{L}\sum_{k=1}^{N_-} \theta(\frac{y-y_k}{4})
    -\frac{2}{L}\sum_{k=1}^{N_+} \theta(\frac{y-x_k}{2})\ ,
    \nonumber
\end{eqnarray}
By definition, the counting function evaluated at a root of equations
(\ref{logbaes}) yields $2\pi$ times
the corresponding integer numbers divided by $L$,
e.g. $z_+(x_j)=2\pi I_j/L$.
Taking the derivative of (\ref{cfs}) and then $L\to\infty$ we obtain
the integral equations \r{igl:rhos} for the densities of roots
$\rho_\pm$ and holes $\rho^h_\pm$ by using that
\be
2\pi[\rho_{\pm}^h(x)+\rho_\pm(x)]=\frac{dz_\pm(x)}{dx}.
\ee
To proceed further, we restrict ourselves to states with symmetric
distributions of the integers, e.g.
\begin{equation}
  I_j = \frac{1}{2}(-N_+-1) + j,\quad j=1,\ldots,N_+
\end{equation}
and similarly for the $J_k$. For these distributions we have
\be
\rho^h_\pm(x)=0,\ \text{for}\ |x|<A_\pm\ , 
\ee
where $A_\pm$ are fixed by the conditions $z_\pm(A_\pm)=N_\pm/L$. The
integral equations for the root densities may then be simplified to
\begin{eqnarray}
  \rho_+(x) + \int_{-A_+}^{A_+} dx' a_4(x-x') \rho_+(x')
            + 2\int_{-A_-}^{A_-} dy' a_2(x-y') \rho_-(y')
  = a_1(x) + a_3(x)\ ,
\nonumber\\
  \rho_-(y) + \int_{-A_-}^{A_-} dy' a_4(y-y') \rho_-(y')
            + 2\int_{-A_+}^{A_+} dy' a_2(y-x') \rho_+(x')
  = a_1(y) + a_3(y)\ ,
\label{igl:rhos_full}
\end{eqnarray}
where
\begin{equation*}
  a_n(u)=\frac{1}{\pi}\ \frac{2n}{4u^2+{n^2}}\ .
\end{equation*}
The energy per site of such states in the thermodynamic limit is
obtained from (\ref{energy})
\begin{eqnarray}
\lim_{L\to\infty}\frac{1}{L}{E_{N_+,N_-}}
&=&-2\pi \lim_{L\to\infty}
  \left\{ \frac{1}{L}\sum_{k=1}^{N_+} a_1(x_k)+a_3(x_k)
   + \frac{1}{L}\sum_{k=1}^{N_-} a_1(y_k)+a_3(y_k)\right\} \ ,\nonumber\\
&=&-2\pi\sum_{\sigma=\pm}\int_{-A_\sigma}^{A_\sigma} dx\
    \rho_\sigma(x)\left[a_1(u)+a_3(u)\right].
\label{ess2}
\end{eqnarray}

 %%%%%%%%%%%%%%%%%%%%%%%%%%%%%%%%%%%%%%%%%%%%%%%%%%%%%%%%%%%%%%%%%%%%%%
 \subsection{Low-lying excitations for large, finite lattice sizes}
 \label{sec:ox}
 %%%%%%%%%%%%%%%%%%%%%%%%%%%%%%%%%%%%%%%%%%%%%%%%%%%%%%%%%%%%%%%%%%%%%%
 
 %%%%%%%%%%%%%%%%%%%%%%%%%%%%%%%%%%%%%%%%%%%%%%%%%%%%%%%%%%%%%%%%%%%%%
 \subsubsection{$[0,-\frac{1}{2},\frac{1}{2},0]$ states in the ``sector''
   $\{\gamma_j\}\equiv   \{u_j\}$ for even $L$}
 \label{ssec:indec}
 %%%%%%%%%%%%%%%%%%%%%%%%%%%%%%%%%%%%%%%%%%%%%%%%%%%%%%%%%%%%%%%%%%%%%
 The considerations in the previous section are straightforwardly
 applied to the low-energy states with $[B,J^3]=[0,0]$ in the
 indecomposable $[0,-\frac{1}{2},\frac{1}{2},0]$ multiplets for even
 $L$. For these we have to consider solutions satisfying
 \begin{equation}
   \gamma_j=u_j, \quad j=1,\ldots,L\ ,
 \end{equation}
 which consist of degenerate type-$\pm$ strange strings. In the
 thermodynamic limit the corresponding root distributions functions are
 given by (\ref{rhopm}) and hence we may use the counting function
\begin{equation}
  z(x) = 2 \arctan\left(e^{\pi x}\right) -
  \frac{\pi}{2}\ .
\label{countingF}
\end{equation}
 to obtain initial values (\ref{sstr2}) for a
 numerical solution of the Bethe ansatz equations (\ref{bae010}): using
 \begin{equation}
   x_j=y_j=\frac{1}{\pi} \ln
   \tan\pi\left(\frac{I_j}{L}+\frac{1}{4}\right)\ 
 \label{sstart}
 \end{equation}
 with
 \begin{equation}
   \quad I_j=-\frac{1}{4}(L+2)+j,\ j=1,\ldots,\frac{L}{2}
 \end{equation}
 as starting values a small number of iterations is sufficient to
 obtain a numerical solution of the Bethe ansatz equations for lattices
 with more than 10.000 sites or $L\approx5000$. We present our results
 on the finite size energies of an indecomposable in
 Table~\ref{table:indec}.  The scaling behaviour as a function of
 $1/\log(L)$ is shown in Figure~\ref{fig:scaling}.
 \begin{table}
 \begin{center}
 \begin{tabular}{|l|l|l|}
 \hline
  $L$  & $E_{\frac{L}{2},\frac{L}{2}}$ &
  $L(E_{\frac{L}{2},\frac{L}{2}}-E_0)$\\ \hline\hline 
   8 &-31.1255373481719104  & 6.9957012146247166 \\
  10 &-39.3093986078541491  & 6.9060139214585092 \\
 % 16 &-63.5789875395715427 & 6.7361993668553168 \\
 % 32 &-127.79604714728058 & 6.5264912870215994 \\
 % 64 &-255.90067438894053 & 6.3568391078060813 \\
  128 &-511.95141708922984 & 6.2186125785810873 \\
  256 &-1023.9761543904651 & 6.1044760409276932 \\
  512 &-2047.9882636363875 & 6.0090181695995852 \\
  1024 &-4095.9942108167784&  5.9281236189417541 \\
  2048 &-8191.9971392992984&  5.8587150368839502 \\
  4096 &-16383.998584343625&  5.7985285085172748 \\   
 \hline\hline
 $\infty$&	                        & 4.95(2)\ (\rm extrapolated)\\
 \hline
 \end{tabular}
 \caption{Finite size results for the energy of the low-lying
   zero-momentum indecomposable consisting of two-strings with
   $\{u_j\}=\{\gamma_j\}$.}
 \label{table:indec}
 \end{center}
 \end{table}

 %%%%%%%%%%%%%%%%%%%%%%%%%%%%%%%%%%%%%%%%%%%%%%%%%%%%%%%%%%%%%%%%%%%%%
 \subsubsection{Octet states in the ``sector''
   $\{\gamma_j\}\equiv   \{u_j\}$ for odd $L$}
 \label{ssec:oct}
 %%%%%%%%%%%%%%%%%%%%%%%%%%%%%%%%%%%%%%%%%%%%%%%%%%%%%%%%%%%%%%%%%%%%%
 The same approach can be used to obtain the finite size energies of
 the lowest octet states of the odd $L$ systems: these states are
 characterized by $(L-1)/2$ degenerate strange strings of each type $+$
 and $-$.  For the numerical solution of the Bethe ansatz equations we
 use again (\ref{sstart}) but with
 \begin{equation}
    I_j=-\frac{1}{4}(L+1)+j,\ j=1,\ldots,\frac{L-1}{2}\ .
 \end{equation}
 Table \ref{table1} summarizes results for the finite-size energies of
 the lowest-lying state in the octet sector. The momentum of this state
 is zero.  For the extrapolation to $L=\infty$ we assume the
 corrections of $L\Delta E$ to vanish as a function of $1/\log(L)$.
 \begin{table}
 \begin{center}
 \begin{tabular}{|l|l|l|}
 \hline
  $L$  & $E_{\frac{L-1}{2},\frac{L-1}{2}}$ & 
 $L(E_{\frac{L-1}{2},\frac{L-1}{2}}-E_0)$\\
  \hline\hline 
    5  & -19.4230243905225      &  2.88487804738772\\
    7  & -27.5669616952430      &  3.03126813329929\\
 %   9  & -35.6522463838436       & 3.12978254540763\\
 %  11  & -43.7088950969117       & 3.20215393397134\\
 %  13  & -51.7493519959825       & 3.25842405222814\\
 %  15  & -59.7797395948863       & 3.30390607670538\\
 %  21  & -83.8380041499709       & 3.40191285061022\\
 %  41  & -163.913040583785       & 3.56533606483228\\
 %  81  & -323.954341891202       & 3.69830681262141\\
  161  & -643.976352749708       & 3.80720729705104\\
  321  & -1283.98786068577       & 3.89671986852727\\
  641  & -2563.99379470064       & 3.97759688996348\\
 % 961  & -3843.99582753401       & 4.01548842714195\\
 1281  & -5123.99684487686       & 4.04171273971125\\
 %1921  & -7683.99787811611       & 4.07613893683309\\
 2561  & -10243.9983994774       & 4.09893837823756\\
 %3841  & -15363.9989250151       & 4.12901680261996\\
 5121  & -20483.9991898007       & 4.14903019789338\\
 \hline\hline
 $\infty$&	                        & 4.916(4)\ (\rm extrapolated)\\
 \hline
 \end{tabular}
 \caption{Finite size results for the energy of the low-lying
   zero-momentum octet consisting of two-strings with
   $\{u_j\}=\{\gamma_j\}$.  Rational function extrapolation has been
   used to obtain the $L\to\infty$ value.}
 \label{table1}
 \end{center}
 \end{table}
 The scaling of the energies together with the result of the rational
 function extrapolation is shown in Figure~\ref{fig:scaling}.

 %%%%%%%%%%%%%%%%%%%%%%%%%%%%%%%%%%%%%%%%%%%%%%%%%%%%%%%%%%%%%%%%%%%%%%
 \subsubsection{More octet states built from strange strings}
 \label{sec:strange2}
 %%%%%%%%%%%%%%%%%%%%%%%%%%%%%%%%%%%%%%%%%%%%%%%%%%%%%%%%%%%%%%%%%%%%%%
 Octet states from non-degenerate strange string solutions exist for
 both even and odd system lengths and are described by Bethe ansatz
 equations with $2(L-1)$ rapidities $u_j$ and $\gamma_\alpha$ arranged
 in $N_+$ ($N_-$) strange strings of type $+$ ($-$) with $N_++N_-=L-1$.
 We have solved the Bethe ansatz equations (\ref{bae010}) for such
 states numerically for ($\Delta N=N_+-N_-$):
 \begin{itemize}
 \item $L$ even: $\Delta N=1,3,5,7$
 \item $L$ odd: $\Delta N=2$ ($\Delta N=0$ corresponds to the state
   studied in Section~\ref{ssec:oct})
 \end{itemize}
 Starting values for the numerical solution of the Bethe ansatz
 equations for are obtained from counting functions similar to
 (\ref{countingF}) with proper symmetric choice of the quantum numbers
 $I_j$, $J_k$.
 The numerical finite size spectrum of these states is given in
 Table~\ref{tablex}. The energy differences to the ground state as a
 function of $1/\log L$ are plotted in Fig.\ref{fig:scaling}. 
 \begin{table}
 \begin{center}
 \begin{tabular}{|l|l|l|}
 \hline
  $L$  & $E_{\frac{L}{2},\frac{L}{2}-1}$ & $L(
 E_{\frac{L}{2},\frac{L}{2}-1}-E_0)$\\ \hline\hline
 % 30  &-119.86659653902181 &  4.0021038293456002 \\
 % 60 &  -239.93222038813883 &   4.0667767116701725 \\
  120&   -479.96564417058130 &   4.1226995302440628 \\
  240&   -959.98261219919220 &   4.1730721938711213 \\
  480&   -1919.9912162303681 &   4.2162094233208336 \\
  960&   -3839.9955678047386 &   4.2549074509588536 \\
  1920 & -7679.9977658894386 &   4.2894922778941691 \\
  3840 & -15359.998874847095 &   4.3205871549434960 \\
 \hline\hline
 $\infty$&                               & 4.91(2)\ (\rm extrapolated)\\
 \hline\hline
  $L$  & $E_{\frac{L+1}{2},\frac{L-3}{2}}$ & $L(
 E_{\frac{L+1}{2},\frac{L-3}{2}}-E_0)$\\ \hline\hline
 % 31  &-123.82046200666238 &   5.5656777934661648 \\
 % 61  &-243.91200258298872 &   5.3678424376880969 \\
  121 &-483.95681608386576 &   5.2252538522429859 \\
  241 &-9.6397875255777501 &   5.1206335762225308 \\
  481 &-1923.9895168083938 &   5.0424151625950344 \\
  961 &-3843.9948148688195 &   4.9829110644573120 \\
  1921 &-7683.9974299991591&   4.9369716154324124 \\
  3841 &-15363.998724015133&   4.9010578736833850  \\
 \hline\hline
 $\infty$&                               & 4.937(3) (\rm extrapolated)\\
 \hline\hline
  $L$  & $E_{\frac{L}{2}+1,\frac{L}{2}-2}$ & $L(
 E_{\frac{L}{2}+1,\frac{L}{2}-2}-E_0)$\\ \hline\hline
 % 32 & -127.74183473886741&   8.2612883562428578 \\
 % 64 & -255.88145737770967&   7.5867278265814093 \\
  128 & -511.94458966164291&   7.0925233097077580 \\
  256 & -1023.9737491964416&   6.7202057109388988 \\
  512 & -2047.9874366076488&   6.4324568838346750 \\
  1024&  -4095.9939402783793&   6.2051549395546317 \\
  2048&  -8191.9970594424680&   6.0222618254764871\\      
  4096&  -16383.998566214848&   5.8727839789301150\\ 
 \hline\hline
 $\infty$&           & 4.94(3) (\rm extrapolated)\\
 \hline
 \end{tabular}

 \caption{Finite size energies for the low-lying octets consisting of
   ``strange strings'' for $L$ even and odd, respectively.}
 \label{tablex}
 \end{center}
 \end{table}

 Based on our results we conjecture that for any fixed $\Delta N$
 \begin{equation}
   \lim_{L\to\infty}L\left(
   E_{\frac{1}{2}(L-1+\Delta N),
      \frac{1}{2}(L-1-\Delta N)}-E_0\right)
 \approx 4.93\ ,
 \end{equation}
 i.e.\ we expect a macroscopic degeneracy of states in the finite size
 spectrum to order $1/L$ of the super spin chain!
 
%%%%%%%%%%%%%%%%%%%%%%%%%%%%%%%%%%%%%%%%%%%%%%%%%%%%%%%%%%%%%%%%%%%%%%%%%%
\subsection{The Fermi velocity}
\label{app:fermiv}
%%%%%%%%%%%%%%%%%%%%%%%%%%%%%%%%%%%%%%%%%%%%%%%%%%%%%%%%%%%%%%%%%%%%%%%%%%
 In order to determine scaling dimensions  we need to
 know the value of an important parameter, 
 the Fermi velocity $v_{F}$. Indeed, the Hamiltonian of our system is 
 determined up to a scale only, and from conformal field 
 theory we only know, for instance, that 
 the ground state energy $E_0$ of the Hamiltonian on a lattice of
 $L$ sites scales for $L\to\infty$ like
 \begin{equation}
 \frac{E_0}{L}=e_\infty-\frac{\pi c v_{F}}{6L^2}+o(L^{-2}),
 \label{central}
 \end{equation}
 where $e_\infty$ is the ground state energy density and $c$ is the
 central charge of the CFT. The 
 determination of $v_{F}$ is straightforward is many systems solvable 
 by Bethe ansatz, but not so here. 
 We suggest 
 that (we used this value in the main text to deal directly with a 
 ``properly normalized'' Hamiltonian)
 \begin{equation}
   v=\pi.
 \label{vfermi}
 \end{equation}
 Let us briefly list the evidence supporting (\ref{vfermi}).
 \begin{itemize}
 \item[1.] We can determine the dispersion of excited states over the
 lowest octet discussed above, that are described by the integer
 method (not all states are, as pointed out above). It follows from
 (\ref{energy}) and (\ref{mtm}) that the bare energy and bare momentum
 of a 2-2 narrow string are given by 
 \begin{eqnarray}
 \epsilon_0(u)&=&-4\pi[a_1(u)+a_3(u)],\\
 p_0(u)&=&2[\theta(u)+\theta(u/3)],
 \end{eqnarray}
 where $u$ denotes the real centre of the string. Given the integral
 equation (\ref{rhooct}) for the root density of the octet state, we may define
 a dressed energy $\epsilon(u)$ and a dressed momentum $p(u)$ as usual by
 \bea
 \epsilon(u)&=&\epsilon_0(u)-\int_{-\infty}^\infty du^\prime\ \left[2\
 a_2(u-u^\prime)+a_4(u-u^\prime)\right] \epsilon(u^\prime),\\
 p(u)&=&p_0(u)-\int_{-\infty}^\infty du^\prime\ \left[2\
 a_2(u-u^\prime)+a_4(u-u^\prime)\right] p(u^\prime).
 \label{epsoct}
 \eea
 These equations are easily solved by Fourier transform
 \be
 \epsilon(u)=-\frac{2\pi}{\cosh(\pi u)}\ ,\qquad
 p(u)=2\ {\rm arctan}(\sinh(\pi u)).
 \ee
 We may eliminate the spectral parameter and obtain an expression of
 the dressed energy in terms of the dressed momentum directly
 \be
 \epsilon(p)=-2\pi\cos(p/2).
 \ee
 The contribution of a 2-2 string with centre $u$ to the ground state
 energy and momentum is then given by $\epsilon(u)$ and $p(u)$
 respectively. Making a ``hole'' excitation by removing a 2-2 string
 with centre $u$ from the ground state would change the energy and
 momentum by 
 \be
 E_h=-\epsilon(u)\ ,\qquad P_h=-p(u)\ ,
 \ee
 and hence
 \be
 E_h(P_h)=2\pi\cos(P_h/2).
 \ee
 Linearizing this dispersion around the ``Fermi points'' $P_h=\pm
 k_F=\pm\pi$ gives the Fermi velocity
 \be
 v=\frac{dE_h(P_h)}{dP_h}\bigg|_{P_h=-\pi}=\pi.
 \ee
 \item[2.] Following \cite{marnie} we can determine $v$ from functional
 equations for the largest eigenvalue $\Lambda(u)$ of the transfer
 matrix by comparing them to crossing equations for relativistic
 S-matrices. From (\ref{Lamsinglet}) with $\lambda=0$ we obtain 
 \begin{equation}
 \Lambda(u)=\Lambda(1-u)\ ,
 \end{equation}
 which again implies that $v=\pi$.
 \item[3.] All of our finite-size spectra for small $L=5,7,9,11$
 systems are consistent with $v=\pi$ as are the extrapolated values for
 very large systems. 
 \end{itemize}
 Now the scaling dimensions and conformal spins of the excitations
 considered above can be determined from the finite size spectra of
 the Hamitonian and the 
 momentum operator
 \begin{eqnarray}
 x=h+\bar{h}=&=&\frac{L}{2\pi^2}(E_\alpha+4L)\ ,\nonumber\\
 s=h-\bar{h}&=&\frac{L}{2\pi}P\ .
 \end{eqnarray}

%%%%%%%%%%%%%%%%%%%%%%%%%%%%%%%%%%%%%%%%%%%%%%%%%%%%%%%
\subsection{``Antiperiodic'' Boundary Conditions}
\label{app:BC_anti}
%%%%%%%%%%%%%%%%%%%%%%%%%%%%%%%%%%%%%%%%%%%%%%%%%%%%%%%
As discussed in the main text, ``antiperiodic'' boundary conditions 
play a crucial role in the identification of the continuum limit 
(even though they might not play any role in the physical network 
model, for instance). We have studied in some detail the sector
$M=N$. The Hamiltonian is defined by taking the trace
in \r{transferM} instead of the supertrace; observe that 
the boundary conditions break the $sl(2/1)$ symmetry, but respect
the bosonic spin-$su(2)$. The Bethe ansatz equations become

\begin{eqnarray}
  \left[ \frac{u_j+i}{u_j-i} \right]^L &=& -\prod_{\beta=1}^N
  \frac{u_j-\gamma_\beta+i}{u_j-\gamma_\beta-i}\ ,\quad j=1,\ldots,N
\nonumber\\
  \left[ \frac{\gamma_\alpha+i}{\gamma_\alpha-i}
  \right]^L &=& -\prod_{k=1}^N
  \frac{\gamma_\alpha-u_k+i}{\gamma_\alpha-u_k-i}\ ,
  \quad \alpha=1,\ldots,N\ .
\label{bae010anti}
\end{eqnarray}
The sector $\{u_j\}\equiv\{\gamma_\alpha\}$ now reduces to the
Takhtajan-Babujian chain with periodic boundary conditions. By
diagonalizing the transfer matrix for small systems, we find that the
ground state lies in the TB sector. For example for $L=3,4,5,6$ we
have ($\gamma_j\equiv u_j$) 
 \begin{center}
\begin{tabular}{|l|l|l|l|}
\hline
L=3 & L=4 & L=5& L=6\\ \hline\hline
$E=-13.5335$ &
$E=-17.4031$ &
$E=-20.8890$ &
$E=-24.8774$\\ \hline
$u_1= 0.3510 - 0.6258i$ &
$u_1=  0.2958  - 0.5513i$&
$u_1=  0.7564 + 0.8124i$&
$u_1=   0.4385 + 0.5483i$\\
$u_2= -u_1^*$ &
$u_2=  u_1^*$&
$u_2=  -u_1^*$&
$u_2=  u_1^*$\\
$u_3=  0.5219i$ &
$u_3= -0.2958  - 0.5513i$&
$u_3=  0.1694 + 0.5980i$&
$u_3=  -0.4385 + 0.5483i$\\
&$u_4= u_3^*$&
$u_4=  -u_3^*$&
$u_4=  u_3^*$\\
& & $u_5=   -0.5013i$&
$u_5=  -0.5207i$\\
& & &$u_6=  u_5^*$\\ \hline
\end{tabular}
\end{center}
The lowest-lying excited states above the TB ground state are
formed by solutions of the Bethe ansatz equations \r{bae010anti}
involving strange strings. In Fig.\ref{fig:twist} we show the
finite-size scaling of the TB ground state as well as two of these
excited states. The numerical results indicate that both states become
degenerate with the TB ground state in the $L\to\infty$ limit.
The scaling dimensions extrapolate to $-\frac{1}{4}$ relative to the
ground state with periodic boundary conditions. We conjecture that
there again is an infinite degeneracy in the thermodynamic limit.

%%%%%%%%%%%%%%%%%%%%%%%%%%%%%%%%%%%%%%%%%%%%%%%%%%%%%%%%%%
\section{Relation to the spin-1 Takhtajan-Babujian model}
\label{app:spin1}
%%%%%%%%%%%%%%%%%%%%%%%%%%%%%%%%%%%%%%%%%%%%%%%%%%%%%%%%%%
 Alcaraz and Martins \cite{AM} carried out a finite-size scaling
 analysis of the anisotropic spin-1 Takhtajan-Babujian (TB) chain
 based on the following equations

 \begin{eqnarray}
   \left\{ \frac{\sinh(\gamma[u_j+i])}{\sinh(\gamma[u_j-i])} \right\}^L
 &=& -\exp(i\Phi) \prod_{k=1}^{L-n}
 \frac{\sinh(\gamma[u_j-u_k+i])}{\sinh(\gamma[u_j-u_k-i])}\ ,\quad
 j=1,\ldots,L-n. 
 \label{baetb}
 \end{eqnarray}

 \begin{eqnarray}
 {\cal E}&=& \frac{(\sin2\gamma)^2}{2}\sum_{j=1}^{L-n}\frac{1}{\cos 2\gamma
 -\cosh 2\gamma u_j}\ ,\nonumber\\
 {\cal P}&=&\sum_{j=1}^{L-n} 2\arctan[\tanh(\gamma u_j)\cot(\gamma)]\ .
 \label{etb}
 \end{eqnarray}

The twist angle was taken to vary in the interval $0\leq\Phi <\pi$.
Equations (\ref{baetb}),(\ref{etb}) reduce to the equations of
interest here if we take the limits $\Phi\to\pi$,
$\gamma\to 0$ and rescale $\cal E$ by a factor of $4$ and $\cal P$ by
a factor of $2$. The finite size spectrum in the TB chain in the
limit $\Phi\to\pi$, $\gamma\to 0$ is of the form \cite{AM}
 \bea
{\cal E}^\alpha_{M,M^\prime}&=&E_0(\Phi=0,L)+\frac{\pi^2}{L}
\left(x_\alpha+M+M^\prime\right)+o(L^{-1})\ ,\ M,M^\prime=0,1,2,...\nn
\label{TBfse}
{\cal P}^\alpha_{M,M^\prime}&=&\frac{2\pi}{L}
\left(s_\alpha+M-M^\prime\right)+o(L^{-1})\ .
\label{TBfss}
\eea
Here $E_0(\Phi=0,L)$ is the energy of the ground state for zero
twist-angle $\Phi=0$ and {\sl even} $L$. It is equal to
\be
E_0(\Phi=0,L)=-L-\frac{\pi^2}{8L}+o(L^{-1}).
\ee
After the rescaling we then obtain the finite-size energies and
momenta in the $sl(2/1)$ chain
 \bea
 E^\alpha_{M,M^\prime}&=&-4L+\frac{2\pi^2}{L}
 \left(2x_\alpha-\frac{1}{4}+2M+2M^\prime\right)+o(L^{-1})\ ,\
 M,M^\prime=0,1,2,...\nn 
 \label{es=1}
 P^\alpha_{M,M^\prime}&=&\frac{2\pi}{L}
 \left(2s_\alpha+2M-2M^\prime\right)+o(L^{-1})\ .
 \label{es=1b}
 \eea
Here we have used that the Fermi velocity in the $sl(2/1)$ chain is
$v=\pi$ \r{vfermi}. 
This shows that the scaling dimensions in the TB model and the
$sl(2/1)$ chain are related by
\begin{itemize}
\item{} A doubling of the scaling dimensions in the $sl(2/1)$ chain as
  compared to the TB model, combined with a shift by $-\frac{1}{4}$.
\item{} A doubling of the conformal spins.
  sectors). 
\end{itemize}
 
%%%%%%%%%%%%%%%%%%%%%%%%%%%%%%%%%%%%%%%%%%%%%%%%%%%%%%%%%%%%%%%%%%%%%%
\section{Relation to the quantum transfer matrix of the supersymmetric
$t$--$J$ model}
\label{app:JKS}
%%%%%%%%%%%%%%%%%%%%%%%%%%%%%%%%%%%%%%%%%%%%%%%%%%%%%%%%%%%%%%%%%%%%%%
The vertex model considered in this paper also appears in the quantum
transfer matrix (QTM) approach to the thermodynamics of the integrable
supersymmetric $t$--$J$ chain: shifting the rapidities used in
Ref.~\cite{JuKS97} (JKS) as
\begin{equation}
   v_k \to \frac{1}{2}\gamma_k+iu,\quad
   w_j \to \frac{1}{2}u_j +iu+\frac{i}{2},
\end{equation}
we obtain from (JKS.18) the BAE in our notation (\ref{bae010}) with
twisted boundary conditions
\begin{align}
   \left( \frac{u_j+i}{u_j-i} \right)^{N/2} &= {\rm e}^{i\phi}
   \prod_{\beta}
   \frac{u_j-\gamma_\beta+i}{u_j-\gamma_\beta-i}\ ,
 \notag\\
   \left(\frac{\gamma_\alpha+i\lambda+i}{\gamma_\alpha+i\lambda-i}
	  \right)^{N/2} &= {\rm e}^{-i\phi} \prod_{k}
   \frac{\gamma_\alpha-u_k+i}{\gamma_\alpha-u_k-i}\ ,
 \label{baet}
\end{align}
when we set the free parameter $\lambda$ in our model to $\lambda=4u+1$. The
twist entering (\ref{baet}) is related to the chemical potential $\mu$ of the
$t$--$J$ model as $\exp(i\phi) = - \exp(-\beta\mu)$ ($\beta$ is the inverse
temperature).
Note that the QTM deals with the thermodynamic limit of the quantum chain, the
system size $N=2L$ appearing in (\ref{baet}) is the so-called ``Trotter
number'' appearing in the expression of the partition function by means of the
QTM.  The temperature dependence in the QTM approach enters via $u=-4\beta/N$.

The eigenvalues of the QTM (JKS.15/6) can be identified with the
eigenvalue (\ref{Lam33b}) of the transfer matrix carrying the
representation $\bar3$ in the matrix space
\begin{equation}
   \Lambda_{JKS}(v) 
     = -\left(v^2+u^2\right)^{L}\ \bar{\Lambda}(2i(v+iu))
\end{equation}
provided that we choose the twist properly, i.e.\ $\exp(\beta\mu)=-1$.

%%%%%%%%%%%%%%%%%%%%%%%%%%%%%%%%%%%%%%%%%%%%%%%%%%%%%%%%%%%%%%%%%%%%%%

%\end{appendix}


\begin{thebibliography}{99}
  
    \bibitem{Pruisken}  D. 
    E. Khmel'nitskii, JETP Lett 38 (1983) 454; H. Levine, S. 
    B. Libby and A. M. M. Pruisken, Nucl. Phys. B240 (1984) 30,49,71;
    A. M. M. Pruisken, Phys. Rev. Lett. 61 (1988) 
    1297; A. M. M. Pruisken, Nucl. Phys. B235 (1984) 277.
    \bibitem{Altland} A. Altland and M. Zirnbauer, Phys. Rev. B55 
    (1997) 1142.
    \bibitem{SeMF99}
    T. Senthil, J.B. Marston and M.P.A. Fisher,
     Phys. Rev. B{\bf 60}, 4245 (1999), cond-mat/9902062.
     \bibitem{Janssen} M. Janssen, M. Metzler and M. Zirnbauer, Phys. 
    Rev. B59 (1999) 15836.
    \bibitem{Evers} F. Evers, A. Mildenberger and A. D. Mirlin, Phys. 
    Rev. B64 (2001) 241303.
    \bibitem{AffleckHaldane} I. Affleck and D. Haldane, Phys. Rev. 
    B36 (1987) 5291.
    \bibitem{ZamoZamo} A. Zamolodchikov and Al. Zamolodchikov, Nucl. 
    Phys. B379 (1992) 602.
    \bibitem{Efetov} K. B. Efetov, Adv. Phys. 32 (1983) 53.
    \bibitem{Weidenmuller} H. A. Weidenm\"uller, Nucl. Phys. B290 
    (1987) 87.
    \bibitem{Zirnbauer} M. Zirnbauer, J. Math. Phys. 37 (1996) 4986.
    \bibitem{Zirnbaueri} M. Zirnbauer, ``Conformal field theory of 
    the integer quantum Hall plateau transition'', hep-th/9905054.
    \bibitem{Berkovits} N. Berkovits, C. Vafa and E. Witten, 
    JHEP 9903 (1999) 018.
    \bibitem{Bershadsky} M. Bershadsky, S. Zhukov and A. Vaintrob, 
    Nucl. Phys. B559 (1999) 205.
    \bibitem{Superspinchains} M. Zirnbauer, Ann. der Physik 3 (1994) 
    513; D. H. Lee, Phys. Rev. B50 (1994) 10788; J. Kondev and J. B. 
    Marston, Nucl. Phys. B497 (1997) 639. 
    \bibitem{Gruzberg}
    I.A. Gruzberg, A.W.W. Ludwig and N. Read,
     Phys. Rev. Lett. {\bf 82}, 4524 (1999), cond-mat/9902063.
     \bibitem{ReadSaleur}  N. Read and H. Saleur, Nucl. Phys. {\bf B613}, 409 (2001).
\bibitem{Gade} R. Gade, J. Phys. A31 (1998) 4909. 
\bibitem{Affleck} I. Affleck, in ``Fields, Strings and Critical 
Phenomena'', Les Houches 1988
\bibitem{Afflecki} I. Affleck, Nucl. Phys. B305 (1988) 582.
\bibitem{Goddard} P. Goddard, D. Olive and G. Waterson, Comm. Math. 
       Phys. 112 (1987) 591.
\bibitem{Andreas} A. W.W. Ludwig, ``A free field representation of 
the $OSP(2/2)$ current algebra at level $k=-2$ and Dirac fermions in 
a random $SU(2)$ gauge potential'', cond-mat/0012189
 \bibitem{Bhaseen} D. Bernard,  ``Conformal field theory applied to 
 2D disordered systems'', hep-th/9509137;
 M.J. Bhaseen, J.S. Caux, I. Kogan and A. M. Tsvelik, Nucl. Phys. 
 B618 (2001) 465.
 \bibitem{ReadSaleuri} N. Read and H. Saleur, to appear.
 \bibitem{Marcu} M. Marcu, J. Math. Phys. 21 (1980) 1277;1284.
 \bibitem{Leites} D. Leites, Acad. Nauk. CCCP (1983) 764 (in english).
 \bibitem{Germoni} J. Germoni, J. of Algebra 209 (1998) 367. 
 \bibitem{Germonithesis} J. Germoni, ``Repr\'esentations 
 ind\'ecomposables des superalg\`ebres de Lie sp\'eciales 
 lin\'eaires'', Publications de l'Institut de Recherche Math\'ematique 
 Avanc\'ee, Universit\'e de   Strasbourg.
 \bibitem{Benson} D. J. Benson, ``Representations and cohomology'', 
 Cambridge Studies in Adv. Math. 30 
 \bibitem{kausch} H. G. Kausch, Nucl. Phys. B583 (2000) 513; M. 
       Gaberdiel and H. G. Kausch, Nucl. Phys. B538 (1999) 631.
 \bibitem{Chamon} C. de C. Chamon, C. Mudry and X.G. Wen, Phys. Rev. 
{\bf  B53}  (1996) R7638.
 \bibitem{JacobsenSaleur} J. Jacobsen and H. Saleur, to appear. 
 \bibitem{Lesage} F. Lesage, P. Mathieu, J. Rasmussen and H. Saleur, 
  Nucl. Phys. {\bf B647} (2002) 363. 
  \bibitem{Bocquet} M. Bocquet, D. Serban and M. Zirnbauer, Nucl. Phys. 
  B578 (2000) 628.
 \bibitem{Taormina} P. Bowcock, M. Hayes and A. Taormina, Nucl. Phys. 
	 B510 (1998) 739.
\bibitem{yellowpages} P. di Francesco, P. Mathieu, Senechal, 
``Conformal field theory'', Springer. 
\bibitem{Eguchi}T. Eguchi and A. Taormina, Phys. Lett. B200 (1988) 315;
       Phys. Lett. B210 (1988) 125.
       \bibitem{Semikhatov} A.M. Semikhatov, A. Taormina and I. Yu 
	 Tipunin, math.QA/0311314.
	 \bibitem{LiFo99}
	 J. Links and A. Foerster,
	 \newblock J. Phys. {\bf A 32}, 147 (1999), cond-mat/9806129.

	 \bibitem{AbRi99}
	 J. Abad and M. R{\'\i}os,
	 \newblock J. Phys. {\bf A 32}, 3535 (1999), cond-mat/9806106.

	 \bibitem{Gade99}
	 R.M. Gade,
	 \newblock J. Phys. {\bf A 32}, 7071 (1999).

	 \bibitem{Derk00}
	 S. Derkachov, D. Karakhanyan and R. Kirschner,
	 Nucl. Phys. {\bf B583}, 691 (2000).
	 \bibitem{SNR}
	 M. Scheunert, W. Nahm and V. Rittenberg, J. Math. Phys. {\bf 18}, 146
	 (1977).

	 \bibitem{SklKul}
		 P.~P. Kulish and E.~K. Sklyanin, J. Sov. Math. {\bf 19}, 1596 (1982).

	 \bibitem{Kul}
	 P.~P. Kulish, J. Sov. Math. {\bf 35} {2648} (1985).

	
	 \bibitem{lai:74}
	 C.K. Lai, J. Math. Phys. {\bf 15}, 1675 (1974);

	 \bibitem{suth:75}
	 B. Sutherland, Phys. Rev. B {\bf 12},  3795  (1975).                               
	 \bibitem{schl:87}
	 P. Schlottmann, Phys. Rev. B {\bf 36},  5177  (1987).

	 \bibitem{esko:92}
	 F.~H.~L. Essler and V.~E. Korepin, Phys. Rev. B {\bf 46},  9147
	 (1992).
	   
	 \bibitem{fk}
	 A. Foerster and M. Karowski, Nucl. Phys. {\bf B396}, 611 (1993).

	 \bibitem{eks94a}
	 F.H.L. Essler, V.E. Korepin, and K. Schoutens,
	 Int. Jour. Mod. Phys. {\bf B8}, 3205 (1994).

	 \bibitem{ek94b}
	 F.H.L. Essler and V.E. Korepin, 
	 Int. Jour. Mod. Phys. {\bf B8}, 3243 (1994).

	 \bibitem{PfFr}
	   M.~P. Pfannm\"uller and H. Frahm,  Nucl. Phys. \textbf{479} [FS], 575 (1996);\\
	   M.~P. Pfannm\"uller and H. Frahm, J. Phys. A\textbf{30}, L543 (1997);\\
	   H. Frahm and M.~P. Pfannm\"uller, Zap. Nauch. Semin. POMI
	   \textbf{251}, 94 (1998).

	 \bibitem{Frahm99}
	 H. Frahm, Nucl. Phys. \textbf{559} [FS], 613--636 (1999).
	 
	 \bibitem{JuKS97}
	 G. J{\"u}ttner, A. Kl{\"u}mper and J. Suzuki,
	 \newblock Nucl. Phys. {\bf B 487}, 650 (1997), cond-mat/9611058.

	 \bibitem{SW}
	 H. Saleur  and B. Wehefritz-Kaufmann,
	 Phys. Lett. {\bf B481}, 419 (2000).
 \bibitem{BabuT}
	 L. Takhtajan, Phys. Lett. {\bf A87}, 479 (1982),\\
	 H. Babujian, Nucl. Phys. {\bf B215}, 317 (1983).

 \bibitem{AM}
	 F.C. Alcaraz and M.J. Martins, J. Phys. {\bf A23}, 1439 (1990),
	 {\sl ibid} {\bf A22}, 1829 (1989).	
	 
	 \bibitem{KaufSal} H. Saleur and B. Wehefritz-Kaufmann, Nucl. Phys. 
	    B663 (2003) 443.    
	   
	  \bibitem{DiFrancesco} P. di Francesco, H. Saleur and J.B. Zuber, 
    Nucl. Phys. B300 (1988) 393.
   
    \bibitem{SuzJ}
J. Suzuki, J. Phys.{\bf  A21}, L1175 (1988).

    \bibitem{DeVega} H. de Vega and E. Lopes, Nucl. Phys. B362 (1991) 
       261.


\bibitem{FrYF90}
H.~Frahm and N.-C. Yu, J. Phys. A \textbf{23} (1990) 2115.

\bibitem{FrYu90}
H.~Frahm, N.-C. Yu and M.~Fowler, Nucl. Phys. B \textbf{336} (1990)
  396.

\bibitem{eks:xxx}
F.H.L. Essler, V.E. Korepin and K. Schoutens, 
J. Phys. {\bf A25}, 4115 (1992).

 
\bibitem{Woynarovich} F. Woynarovich,  ``On the symmetry of 
 excitations in $SU(2)$ Bethe ansatz systems'', cond-mat/9812415.



 \bibitem{Martin} P. Martin, ``Potts models'', World Scientific
  
\bibitem{CurtisReiner} C.W. Curtis and I. Reiner, ``Representation 
theory of finite groups and associative algebras'', Wiley 
Interscience 1962.
       \bibitem{Mathieu} O. Mathieu, in Adv. Studies in Pure Math., 
       26 (2000) 145.
       
       \bibitem{PasquierSaleur} V. Pasquier and H. Saleur, Nucl. 
       Phys. B330 (1990) 523.
 
\bibitem{ChariPressley} V. Chari and A. Pressley, ``A guide to 
quantum groups'', Cambridge.
\bibitem{Razumov} A. V. Razumov and Yu G. Stroganov, J. Phys. A34 
(2001) 3185.

 




%\bibitem{Kag}
%V. Kagalovsky, B. Horovitz, Y. Avishai and J.T. Chalker,
%cond-mat/9812155.















\bibitem{completeness}
L.D. Faddeev and L. Takhtajan, J. Sov. Math. {\bf 24}, (1984),\\
A.N. Kirillov, J. Sov. Math. {\bf 34}, 2298 (1985),\\
F.H.L. Essler, V.E. Korepin and K. Schoutens,
Phys. Rev. Lett. {\bf 67}, 3848 (1991).

\bibitem{eks}
F.H.L. Essler, V.E. Korepin and K. Schoutens, 
Int. Jour. Mod. Phys. {\bf B8}, 3205 (1994),\\
K. Schoutens, Nucl. Phys. {\bf B413}, 675 (1994).

\bibitem{BeFr95}
  G. Bed\"urftig and H. Frahm, J. Phys. A \textbf{28}, 4453 (1995).

\bibitem{bef}
G. Bed\"urftig, F.H.L. Essler and H. Frahm, Phys. Rev. Lett. {\bf 77},
5098 (1996),\\
A. Foerster, J. Links  and A.P. Tonel,
Nucl. Phys. {\bf B552}, 707 (1999).

\bibitem{dopes}
  H. Frahm, M.~P. Pfannm\"uller and A.~M. Tsvelik,
  Phys. Rev. Lett. \textbf{81}, 2116 (1998).





\bibitem{marnie}
M.J. Martins, B. Nienhuis and R. Rietman,
Phys. Rev. Lett. {\bf 81}, 504 (1998).

\bibitem{saleur}
H. Saleur, Nucl. Phys. {\bf B578}, 552 (2000).

%THESE TWO ARE NOW CALLED JUST Marcu
%\bibitem{Marcu80}
%M. Marcu,
%\newblock J. Math. Phys. {\bf 21}, 1277 (1980).
%\bibitem{Marcu80b}
%M. Marcu,
%\newblock J. Math. Phys. {\bf 21}, 1284 (1980).


\end{thebibliography}
\end{document}